# Outgroup Animosity Has Risen for Politicians, Journalists, and a Sample of Partisan Users on Twitter and Reddit

Suyash Fulay[1*], Deb Roy[1]

**Affiliations:**

[1]Massachusetts Institute of Technology

*Corresponding author. Email: sfulay@mit.edu

**Abstract**: Using language models, we analyze a sample of 67 million tweets and 30 million Reddit comments referencing a set of 215 political entities from 2010-2023 from partisan users, journalists, and politicians. Our analysis indicates outgroup animosity has increased consistently in our sample, with newer cohorts of users expressing higher levels of animosity than previous ones. Moreover, a small fraction of users are responsible for a disproportionate share of this negative content. We observe systematic differences in topic-level outgroup affect across political orientations: right-leaning users are twice as likely to exhibit outgroup animosity when discussing immigration, while left-leaning users show heightened outgroup animosity when discussing healthcare. On Twitter, U.S. politicians on the left exhibit more outgroup animosity than partisan users in our sample, but in the past four years, politicians on the right have experienced the sharpest rise in outgroup animosity, surpassing journalists, media, and partisan users. On Reddit, a small number of communities account for a large share of polarizing rhetoric, with the rise and eventual ban of r/TheDonald significantly shaping polarizing discourse on the right.

## Introduction

The rise of online platforms has ushered in a new era of political discussion, offering individuals unprecedented opportunities to engage, express their opinions, and shape public discourse. Politicians, journalists, media organizations, and everyday citizens participate in political discussion online, and in turn influence one another in what has become a significant element of the public sphere (1; 2). However, as social media use has grown so have worries about the state of our political discourse online. Social media use can affect opinions on important issues such as vaccination (3) and exposure to different views can affect the extremity of policy positions (4). While we acknowledge that social media does not provide a representative sample of society at large, its use is undoubtedly widespread. Seventy percent of Americans reported using social media in 2021, with twenty-three percent using Twitter and eighteen percent using Reddit (5). Political talk online is also common in both explicitly political and apolitical online spaces (6; 7). Thus, while inherent limitations exist when studying political discourse on social media (e.g., people may also use it primarily for identity curation (8) and signaling rather than expressing sincere opinions) its widespread nature and potential to impact beliefs and have real-world consequences makes it essential to understand.

In this work, we address two main research questions related to the political polarization of online discourse in the American context that we view as important and outstanding. First, we study how aggregate trends in polarizing speech have evolved across two large social media platforms over the past sixteen years. Second, we examine the various groups that contribute to polarizing discourse. We do this by analyzing a sample of comments made by politicians, journalists, and partisan users that reference a political group or entity over the past fifteen years.

Existing work has found that politicians have generally increased their polarizing and uncivil rhetoric on Twitter (9; 10) and there has been rising incivility in partisan subreddits (11; 12). However, many of these studies are limited to a single platform (e.g., Twitter (9; 10; 13; 14) or Reddit (11; 12; 15) only), a single type of user (e.g. politicians (9; 10; 14)) or a single community (e.g. a single subreddit representing each political group (11; 12; 15)), or are in relation to a single topic or event (16). Given the differential roles of the mass public, meso-level (e.g. media, think tanks, journalists, and influencers), and political elite in our media sphere, and differences in platform affordances, we study polarization on platforms and user groups with a consistent methodology.

We also investigate which topics are most likely to lead to polarizing discourse. Party line differences in opinions about policy or policy importance have been well-documented (17; 18). Thus, in addition to identifying and measuring politically polarizing speech, we examine which topics are most likely to elicit polarizing discourse from users. We believe our findings are of practical use to researchers attempting to identify high impact areas to introduce interventions to reduce polarization online (19).

## Results

Our analysis focuses on Twitter and Reddit, two large social media networks that host a significant amount of political discussion throughout the platforms and serve different groups of users (5). To see a high-level overview of our methodology to identify users, collect and process tweets and Reddit comments, and train and apply a polarization classifier, see **Figure 1**.

Our main findings concerning aggregate trends in polarizing discourse are:

i) In our sample of politicians, journalists, and partisan users, outgroup animosity rose over the past decade.

ii) Right-leaning users were more likely to use polarizing rhetoric on immigration, while left-leaning users were more likely to use polarizing rhetoric on healthcare.

iii) A minority of users contributed a large share of negative political content relative to neutral political content, illustrating that negatively charged politically content was created by a more concentrated group of users in our sample of partisan users.

At the user group level, our analysis shows that:

iv) While every group has increased their level of outgroup derogation, right-leaning politicians increased their use of polarizing rhetoric more rapidly than other groups since 2016 (see **Figure 2**).

v) Recent cohorts of politicians joining Twitter exhibited more outgroup animosity than previous cohorts (see **Figure 3**).

vi) While journalists and other political influencers on Twitter were less likely to use polarizing rhetoric when tweeting about politics, they contributed to a large share of aggregate political discussion (see **Figure 5**).

vii) On Reddit, a small number of communities—such as *r/politics* on the left and *r/TheDonald* on the right—accounted for much of the polarizing discourse. The rise and eventual ban of *r/TheDonald* significantly reduced polarizing rhetoric from right-leaning users in our sample in the short term. However, following the ban, other right-leaning subreddits increased their contribution to polarizing discourse, leading to a renewed rise in polarizing discourse amongst our sample of right-leaning partisan Reddit users.

## Datasets

### Twitter Users

On Twitter, we study four main groups of users: politicians, journalists, and partisan users. We focus on the United States political context, although we do not filter to users solely based in the United States. The politician group primarily consists of elected members of or candidates running for U.S. Congress but also includes governors and the executive branch (20; 21). In total, our sample includes 1,361 politicians. Journalists (1,165 users) were identified by (22) and covered reporters and writers working for U.S. based English language media organizations (e.g., The New York Times) and major news outlets (e.g., CNN and Fox News). Finally, we identify general politically active users by collecting all the followers of the 1,361 politicians identified and filtering to those users that follow at least ten politicians. We then infer the political orientation of users following the method by (21) and use the cutoffs provided by (22) to select a final group of left and right-leaning partisan users. This led to a general group of partisan users of ~1.1 million.

### Reddit Users

To identify politically active Reddit users, we first create a list of political subreddits using measures of partisan-ness provided by (23) and then manually inspect which subreddits display a clear relation to politics. We identify 78 subreddits which are listed in Table S7. Users that left at least ten comments in political subreddits are included in our sample, which led to roughly half a million Reddit users. We apply a similar method to the one on Twitter to identify partisans on Reddit, relying on commenting patterns in the different subreddits to infer users' partisan leaning.

Identifying Content Referencing a Political Group or Entity

An important preprocessing step involved filtering the collected tweets and comments to include only those that referenced a political group or entity (see Table S9 for the list of entities and the Methods section for a description of how these terms were selected). In total, we use 215 political entities to filter the content. Each piece of content is tagged based on whether it references a left- or right-leaning entity. To ensure clearer distinctions, we exclude any comments that mentioned both left- and right-leaning entities. This approach allows us to identify whether a comment was likely referring to a user's ingroup or outgroup.

Thus, on Twitter the total number of tweets collected from all users is close to one billion; however, after filtering to content referencing a political entity, the bulk of our analysis focuses on a subset of 67 million tweets (see Table S5 for a more detailed breakdown of the number of users and posts). On Reddit we collect 1.5 billion comments, of which 30 million contain a reference to a political entity in our list. Those thirty million political comments fall into 38,818 subreddits. We provide a detailed breakdown of the top twenty subreddits by volume of comments in Table S8. While previous work has studied interactions in specific political subreddits (12) it has recently been shown that a significant amount of political discourse on Reddit happens in non-political subreddits (6). We ensure that all comments made by politically active users are collected, even if posted in non-political subreddits. We refer to content that contains a reference to a political group or entity in our list as "political content". Although this term is broad, we adopt this definition for consistency and conciseness throughout the manuscript.

Identifying Polarizing Speech

Polarization has been measured in a variety of ways in the literature, ranging from quantitative methods that measure ideological polarization (24) to qualitative coding of tweets (14) to machine learning classification of tweets (9). Since we are interested in measuring politically polarizing speech at scale across groups and platforms, we adopt a machine learning classification approach that can be applied to millions of tweets or Reddit comments.

We draw on literature from political science, specifically surveys measuring affective polarization. Typically, these surveys use a "thermometer" rating from zero to one hundred to measure how citizens feel about certain political groups or politicians (25). Using this as the theoretical underpinning, we examine content that contains a reference to a political group or entity and examine what fraction of them convey "(dis)like or (dis)trust" (phrasing explicitly drawn from these surveys) to get an empirical measure of polarization that reflects natural, online discourse. We are careful to label only those comments that exhibit negative or positive affect *towards the specific political group or entity* as opposed to general comment valence, to capture partisan feeling.

We classify the comment as displaying positive, negative, or neutral affect towards the reference political group. For example, posts such as "*the [political group] is a lawless party led by an as-yet-unindicted criminal*" or "*i can't stand you [politician name] your green deal sucks like you suck as a politician*" would be classified as having negative affect, as they exhibit clear negative feelings towards a political group or entity. See Table S10 and Table S11 for examples of polarizing speech and their corresponding labels. We fine-tune several language models for this task and ensure that our results are robust to the specific model choice.

We code comments as negative (1), neutral (0), or positive (–1). Our primary measure is the mean affect across all political comments, where higher scores reflect more negative affect. By identifying both the political leaning of content authors and the orientation of the political entities they reference, we can calculate the average affect for left- and right-leaning users. This includes measuring affect toward political ingroups and outgroups—for example, the average affect of comments by left-leaning users referencing right-leaning entities reflects the level of negative outgroup sentiment among left-leaning users. We will primarily use the terms "in(out)group animosity" or "in(out)group derogation" to refer to the mean affect of a group of users when discussing their respective in/outgroups. We will also use the term "polarizing discourse" or "polarizing rhetoric" more generally to refer to content that contains negative affect, irrespective of the referent group.

Aggregate Trends in Outgroup Animosity

Although Reddit and Twitter cater to different audiences and features, we observe consistent trends in outgroup animosity across both platforms. Outgroup animosity has generally increased, although it declined among right-leaning users post-2018 before rising again following the 2020 U.S. presidential election (see **Figure 2**). On Reddit, the temporary decrease can partly be attributed to the rise and subsequent ban of the subreddit *r/TheDonald* (see **Figure 4**) (26). In our sample, each cohort of new users joining both platforms has generally exhibited outgroup animosity at levels equal to or exceeding those of prior cohorts (see **Figure 3**). Additionally, a small subset of users generates a disproportionate amount of outgroup derogatory content. On Twitter, the top decile of users by volume of negative affect tweets produces 44% of all such tweets, while the top 10% of neutral political tweeters accounts for 38% of neutral affect content. A similar pattern is evident on Reddit, where the top decile contributes 52% of negative and 43% of neutral affect comments. Further statistical comparisons, including a Kolmogorov–Smirnov test and Lorenz curve analysis, support these findings (see S6 Distribution of Polarizing Discourse across Users).

We also examine levels of outgroup animosity across political topics. We train a classifier to identify various political topics and measure relative levels of outgroup animosity across groups and platforms. See **Topic Classifier** in Methods and S3.2 Topic Model Performance for more details on the topic classifier. We train a ordinal logistic regression to predict the affect of a comment based on the author's political orientation, the topic, and an interaction variable between the two for the set of comments where a user's comment includes a reference to their political outgroup. We report the exponentiated interaction coefficients, where this coefficient represents how much more likely a right-leaning users is to use outgroup animosity on this topic than a left-leaning user (see **Figure 6**). Right-leaning users are more likely to express outgroup animosity when discussing immigration ($exp(\beta_{reddit})$ = 1.703, $CI$ = [1.625, 1.785]) ($exp(\beta_{twitter})$ = 2.253, $CI$ = [2.146, 2.365]) and international affairs ($exp(\beta_{reddit})$ = 1.129, $CI$ = [1.078, 1.183]) ($exp(\beta_{twitter})$ = 1.349, $CI$ = [1.271, 1.433]), whereas left-leaning users are more likely to express outgroup animosity when discussing health ($exp(\beta_{reddit})$ = 0.814, $CI$ = [0.775, 0.854 ]) ($exp(\beta_{twitter})$ = 0.756, $CI$ = [0.721, 0.792]).

Polarizing Discourse by User Group

After analyzing aggregate levels of outgroup animosity, we also analyze trends in polarizing discourse by group. Given that Twitter operates primarily in a "top-down" manner—where influential users, often with large followings, shape the broader conversation (27)—we examine how a user's influence (measured by follower count) relates to their likelihood to engage in polarizing rhetoric. We also explore how users occupying different roles within the media landscape contribute to polarizing discourse. In contrast, Reddit's community-driven structure offers a distinct dynamic, so we focus on how individual subreddits contribute to the overall levels of polarizing discourse.

*Trends on Twitter: Comparing Politicians, Journalists, and Partisan Users*

Many groups contribute to political discourse, including journalists, the media, politicians, and the public. However, we have not fully understood *which* of these groups plays the biggest role in contributing to polarizing discourse. Are politicians and journalists more likely to blame and attack their outgroup, or is it members of the public? Considering these groups have varying levels of influence, knowing which ones are more likely to use polarizing language can provide a clearer picture of the increasing polarization perceived by many Americans (28).

In the following sections, we report the mean level of polarizing discourse by computing the average affect of political comments and report the 95% confidence interval of this mean. When comparing groups, we report the Mann-Whitney U test statistics and report the mean affect *M*, where a higher mean indicates more negative affect. A limitation of our work is that comparisons across groups may be affected by how we sampled data, defined and labeled polarizing discourse, and trained the classifier and so our results are conditional on these decisions. Overall our analysis indicates that today, politicians exhibit high levels of outgroup animosity ($M = 0.652$, *CI* = [0.646, 0.658]) exceeding journalists ($M = 0.206$, *CI* = [0.202, 0.209]), media ($M = 0.134$, *CI* = [0.123, 0.145]) and partisan users on both Reddit ($M = 0.307$, *CI* = [0.305, 0.308]) and Twitter ($M = 0.478$, *CI* = [0.478, 0.479]). Comparing politicians to all other users in 2022 yields significant differences in levels of outgroup animosity ($U = 5.7$ x $10^{10}$, *p(U)* < .001, $t = 72.45$, *p(t)* < 0.001). See **Figure 2** and S5 Twitter Ordinal Logistic Regressions for ordinal logistic regressions to verify results across multiple base language models.

We also find that this difference between politicians and partisan users has shifted over time. While left-leaning politicians ($M = 0.576$, *CI* = [0.571, 0.580]) have consistently expressed more outgroup animosity than partisan left-leaning users ($M = 0.434$, *CI* = [0.434, 0.434]), with an average gap of ~15%, from 2009-2016 years right-leaning politicians ($M = 0.315$, *CI* = [0.302, 0.327]) and partisan users (M = $0.323$, *CI* = [0.318, 0.327]) exhibited similar levels of outgroup animosity. This similarity in levels of outgroup derogation started to change in 2016, as right-leaning politicians' outgroup animosity began to rise steadily. See Table S23 which uses an ordinal logistic regression to show that right-leaning politicians increased their levels of outgroup derogation faster than other groups. In 2019, levels of outgroup animosity of right-leaning politicians surpassed partisan right-leaning Twitter users for the first time, and it has been consistently higher since ($U = 4.5$ x $10^{10}$, *p(U)* < 0.001, $t = 64.49$, *p(t)* < 0.001). Investigating this increase further, we see that post-2016, each cohort of right-leaning politicians joining the platform exhibited more outgroup animosity than any cohort prior to 2016 ($U = 3.3$ x $10^8$, *p(U)* < 0.001, $t = 25.30$, *p(t)* < 0.001), while on the left the patterns across cohorts are more mixed (see **Figure 3**). From 2020-2023, the gap between politicians and partisan users has reached 13% ($U = 2.8$ x $10^{10}$, *p(U)* < 0.001, $t = 31.43$, *p(t)* < 0.001) and 17% ($U = 6.1$ x $10^{10}$, *p(U)* < 0.001, $t = 43.16$, *p(t)* < 0.001) for left/right leaning users, respectively.

Ingroup animosity has increased slightly over the past fifteen years, albeit at much lower levels than outgroup animosity. Right-leaning politicians again exhibit a different trend, where politicians reduced their ingroup animosity from 2010 to 2020. However, ingroup animosity amongst this group increased from 2020 to 2022.

Finally, journalists ($M = 0.185$, $CI = [0.183, 0.187]$) and media outlets ($M = 0.159$, $CI = [0.152, 0.166]$) generally exhibit low levels of outgroup animosity, although with notable increases over the past sixteen years. We analyze the most influential accounts and find that although journalists use less outgroup animosity than politicians (i.e., when they tweet about an outgroup political entity, they are less likely to use polarizing rhetoric), they contribute to a large share of the total amount of political discussion. In the next section we discuss this in more detail.

*Influencers on Twitter*

While much attention has been devoted to studying the rhetoric of politicians on social media (9; 10; 13; 14), recent interest has expanded to include a broader spectrum of influential figures (24). To gain insights into the behaviors of the most influential users in our study, we identify the top two hundred most followed accounts that have posted at least ten tweets containing political content and manually annotated their roles. See Table S1 through Table S4 for a full list of the influencers. These roles encompass media accounts, journalists, politicians, celebrities (such as actors and singers who engage in political discourse), and various "political influencers" – individuals like political commentators or former government officials who are not journalists by profession. It is worth noting that these influencers often have multiple roles, and we have classified the accounts to the best of our ability. The complete table detailing these influencers is available in the Supplementary Information (SI) for reference.

We assess the contribution of different types of influential users to the overall level of polarizing discourse within this group of influential accounts. This contribution is determined by two factors: the average level of in(out)group derogation expressed by members of each group, and the proportion of total comments made by that group. A group can still contribute substantially to polarizing discourse even if they post infrequently, as long as their comments are consistently polarizing. To quantify this, we calculate each group's average in(out)group derogation, their share of total comments, and then multiply these two values to estimate their overall contribution (see **Figure 5**).

Our analysis shows that politicians ($M_{left} = 0.510$, $CI = [0.432, 0.589]$) ($M_{right} = 0.459$, $CI = 0.378$, $0.541$) exhibit the highest levels of outgroup animosity. Among celebrities, several on the left also display elevated levels of outgroup animosity ($M_{left} = 0.506$, $CI = [0.406, 0.606]$).

Journalists ($M_{left} = 0.184$, $CI = [0.125, 0.243]$) ($M_{right} = 0.319$, $CI = [0.266, 0.372]$) tend to express lower levels of outgroup animosity, but because they tweet far more frequently, they have the second-highest overall contribution to outgroup animosity among influential users. Finally, we observe that several political influencers—particularly on the right—also contribute significantly to outgroup animosity.

*Audience Size and Polarizing Discourse on Twitter*

Although users with large amounts of followers are undoubtedly influential, most accounts belong to people with fewer than one thousand followers. Here, we examine the propensity to use polarizing discourse over users with differently sized audiences. There are various reasons that we may expect polarization levels to differ by audience size. Outgroup animosity has been shown to increase engagement (29), and accounts with more followers may be more driven to use this tactic to keep their audiences engaged. On the other hand, these same accounts face public scrutiny and may be more cognizant of alienating members of their potentially diverse audiences.

We empirically find that follower count and outgroup animosity are negatively related. For accounts with less than 1K followers, outgroup animosity is 0.44 ($CI$ = [0.439, 0.440]) and 0.48 ($CI$ = [0. 484, 0.485]) for the left and right, respectively. This is higher than 0.31 ($CI$ = [0.305, 0.312] and 0.40 ($CI$ = [0.388, 0.408]) for left and right leaning accounts with between 100K and 1M followers, respectively. This indicates that when smaller accounts tweet about politics, they are significantly more likely to exhibit outgroup animosity than larger accounts.

We also investigate whether there are partisan differences in the relationships between audience size and outgroup animosity since a difference across party lines could asymmetrically affect meta-perceptions of polarization (30). Upon first inspection, it appears that right-leaning influencers tend to exhibit more outgroup animosity than left-leaning ones compared to the differences in outgroup animosity amongst normal partisan users. However, further investigation reveals that this trend is largely driven by a single, vocal right-leaning influencer that repeatedly posts highly similar tweets with negative affect about a left-leaning politician. After removing this influencer, we do not find a significant partisan difference in this trend. See S6.6 Followers vs. Polarizing Rhetoric for charts illustrating these trends.

*Polarizing Discourse on Reddit: Contributions by Subreddit and Cohort*

Users on Reddit (Redditors) are required to post in defined "subreddits", making the community structure much more explicit than on Twitter. Thus, we study how different subreddits and cohorts of users entering the platforms have contributed to polarizing discourse on Reddit. In **Figure 4**, we can see on the left, the subreddit *r/politics* dominates the share of polarizing discourse. On the right, outgroup animosity peaked from 2016-2020 ($U$ = 1.1 x $10^{11}$, $p(U) < 0.001$, $t$ = 3.19, $p(t)$ = 0.004), primarily driven by new right-leaning users who joined the subreddit *r/TheDonald.* Then, the ban of *r/TheDonald* in 2020 due to policy violations, including the promotion of violence and hate speech (26) resulted in a decline in outgroup animosity among right-leaning users on the platform, although evidence shows that the users may have simply migrated to other platforms (31). We also see that the subreddit *r/Conservative* increased its contribution to polarizing discourse after the ban of r/*TheDonald*. Conversely, left-leaning users experienced heightened outgroup animosity from 2008-2016 and from 2020-2023 ($U$ = 9 x $10^{11}$, $p(U) < 0.001$, $t$ = 40.30, p*(t)* < 0.001).

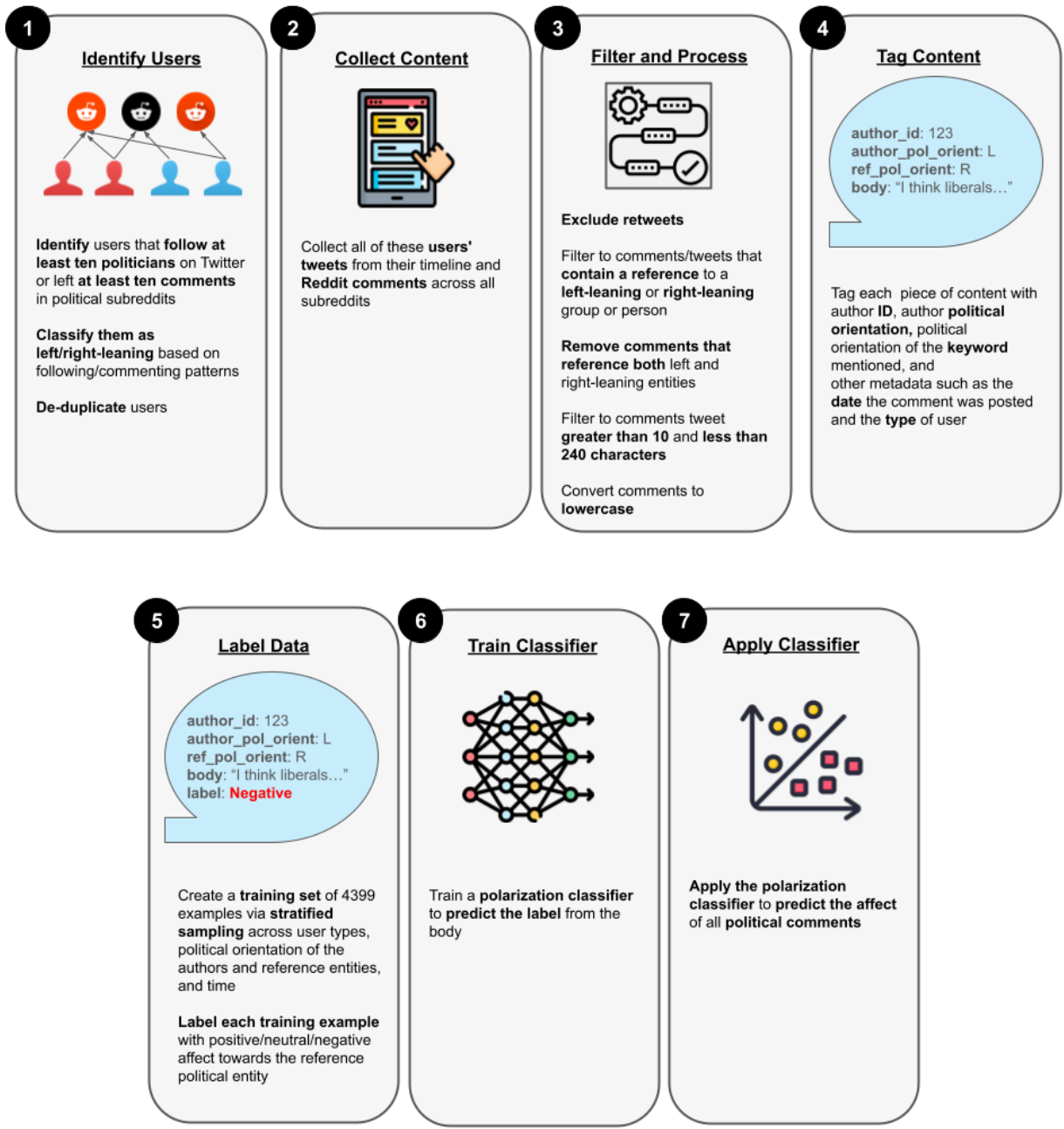

*Figure 1: Overview of data processing pipeline. This figure illustrates the full methodological workflow used in the study. The pipeline includes (1) identifying politically active users on Twitter and Reddit, (2) collecting thee users' tweets and Reddit comments, (3) filtering their posts to retain only those referencing a predefined list of 215 political entities, (4) adding tags to the data to understand author characteristics and the political entities referenced, (5) labeling the affect of the comments, (6) training the classifier, and (7) applying the classifier to the set of political content. Political entities used for filtering (e.g., politician and parties) were manually curated and categorized as left- or right-leaning; these classifications are detailed in the Supplementary Information (SI). Partisan users were identified using methods adapted from prior work, and public datasets were used to label journalists and politicians. Additional preprocessing and classification details can be found in the Methods and SI.*

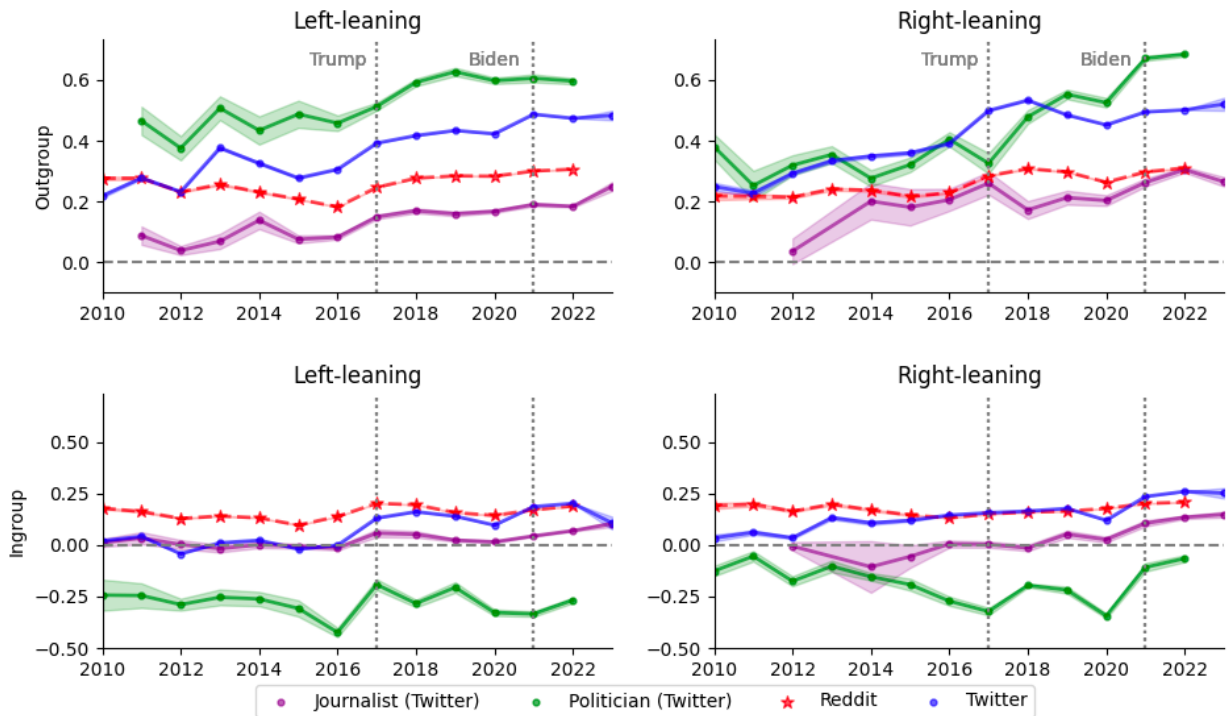

**Figure 2: For each group and their political in/outgroup, the level of polarizing discourse since 2010.** *The y-axis is the average affect towards the specified group (higher indicates more negative affect). The top row shows outgroup animosity (e.g. right-leaning users discussing left-leaning entities) while the bottom row shows ingroup animosity. The left column is for left-leaning users while the right column is for right-leaning users. Data pulled from the User Timeline API. See S7 for chart with Historical Powertrack data and other models. Shaded areas shown are 95% confidence intervals.*

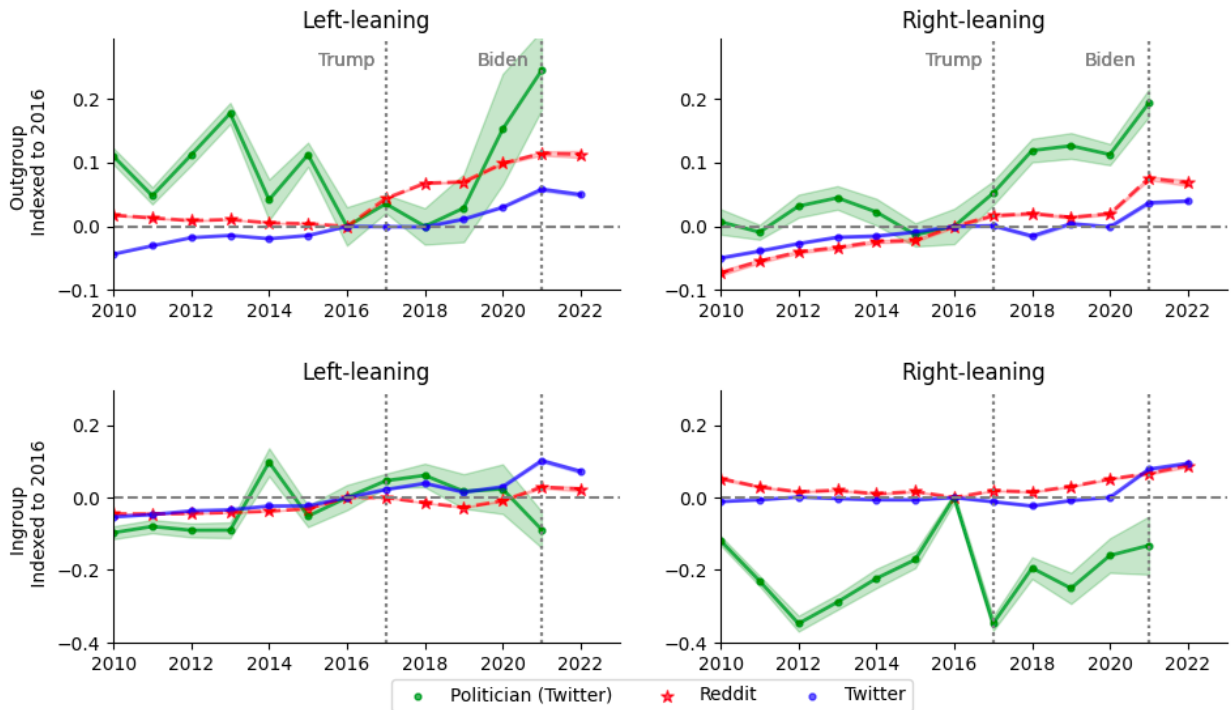

**Figure 3: The level of polarizing discourse by each cohort of users (cohort defined as the year the user entered the platform) indexed to 2016 to better compare trends.** *We see that generally each new cohort of users exhibits more outgroup animosity than previous cohorts, and recent cohorts of politicians are especially polarized relative to previous ones.*

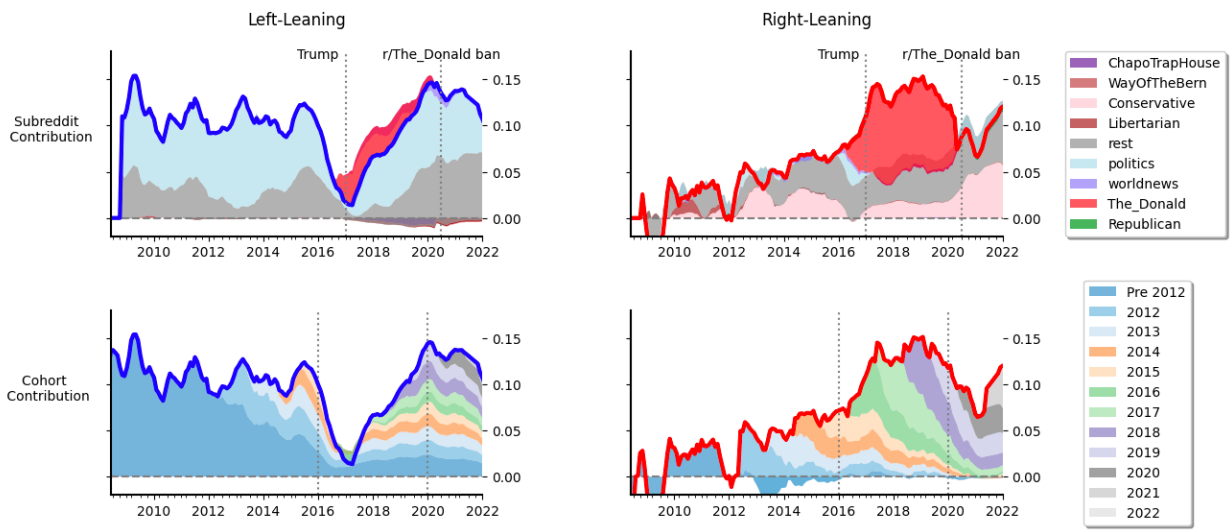

**Figure 4: Each subreddit and each cohort's contribution to the partisan gap (average affect towards outgroup − average affect towards ingroup) on Reddit.** *The top chart shows how much each subreddit contributed to this partisan gap, while the bottom chart shows the contribution of each cohort of users. These patterns illustrate the outsize role particular subreddits play in using polarizing political speech (r/TheDonald and r/politics) as well as the impact of the influx and subsequent ban of users on the right post 2016.*

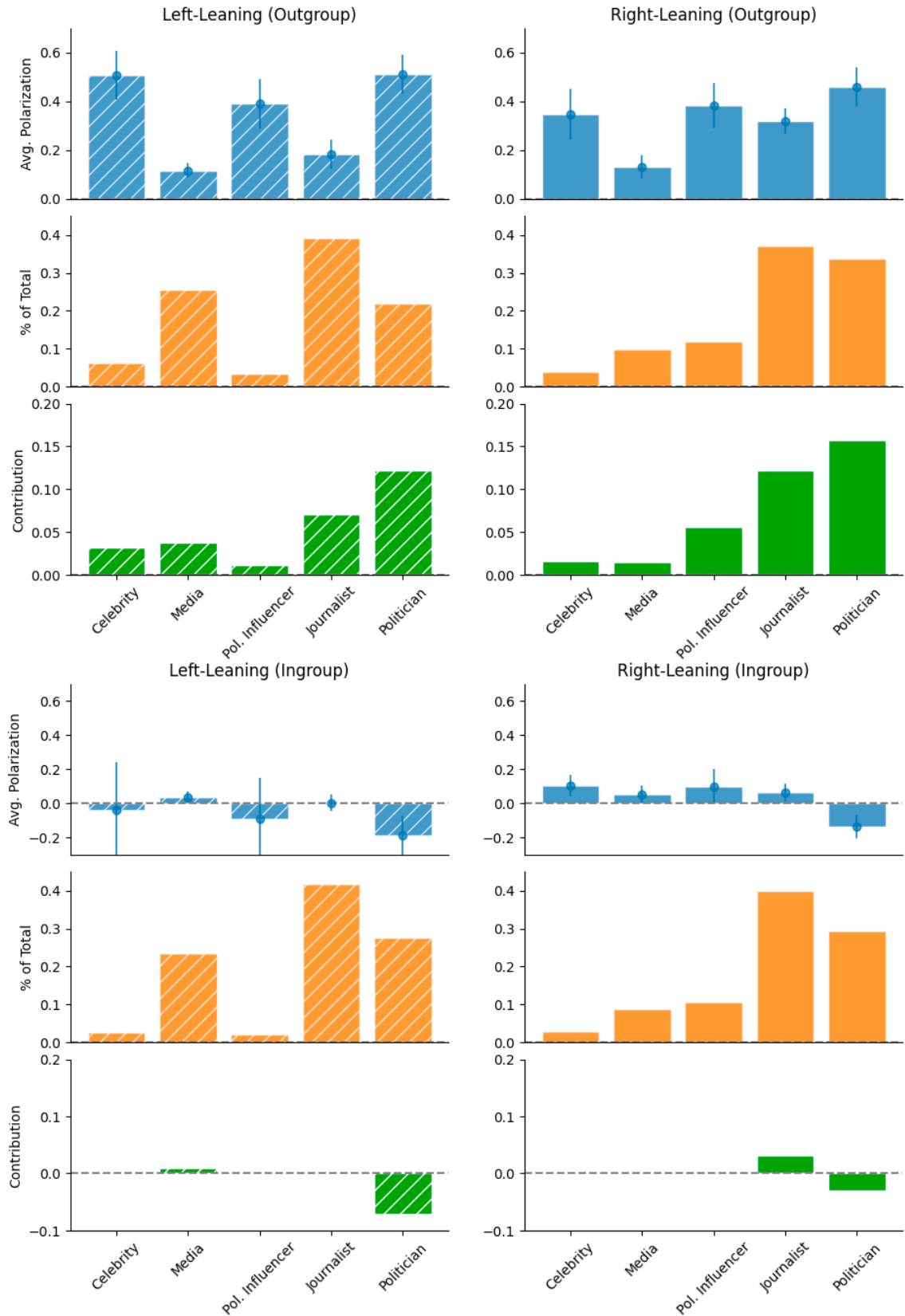

Polarization of Top 200 most Followed Accounts on Twitter

*Figure 5: **Breakdown of polarizing discourse among the top 200 most-followed Twitter accounts posting political content.** For each user type (e.g., celebrities, media, politicians, political influencers, journalists, and politicians), the figure shows: (1) the average level of in-group and out-group derogation, (2) the proportion of total political comments made, and (3) the estimated contribution of each group to the overall level of polarizing discourse (calculated as the product of average derogation and comment share).*

## Topic Coefficients and Interactions on Twitter and Reddit

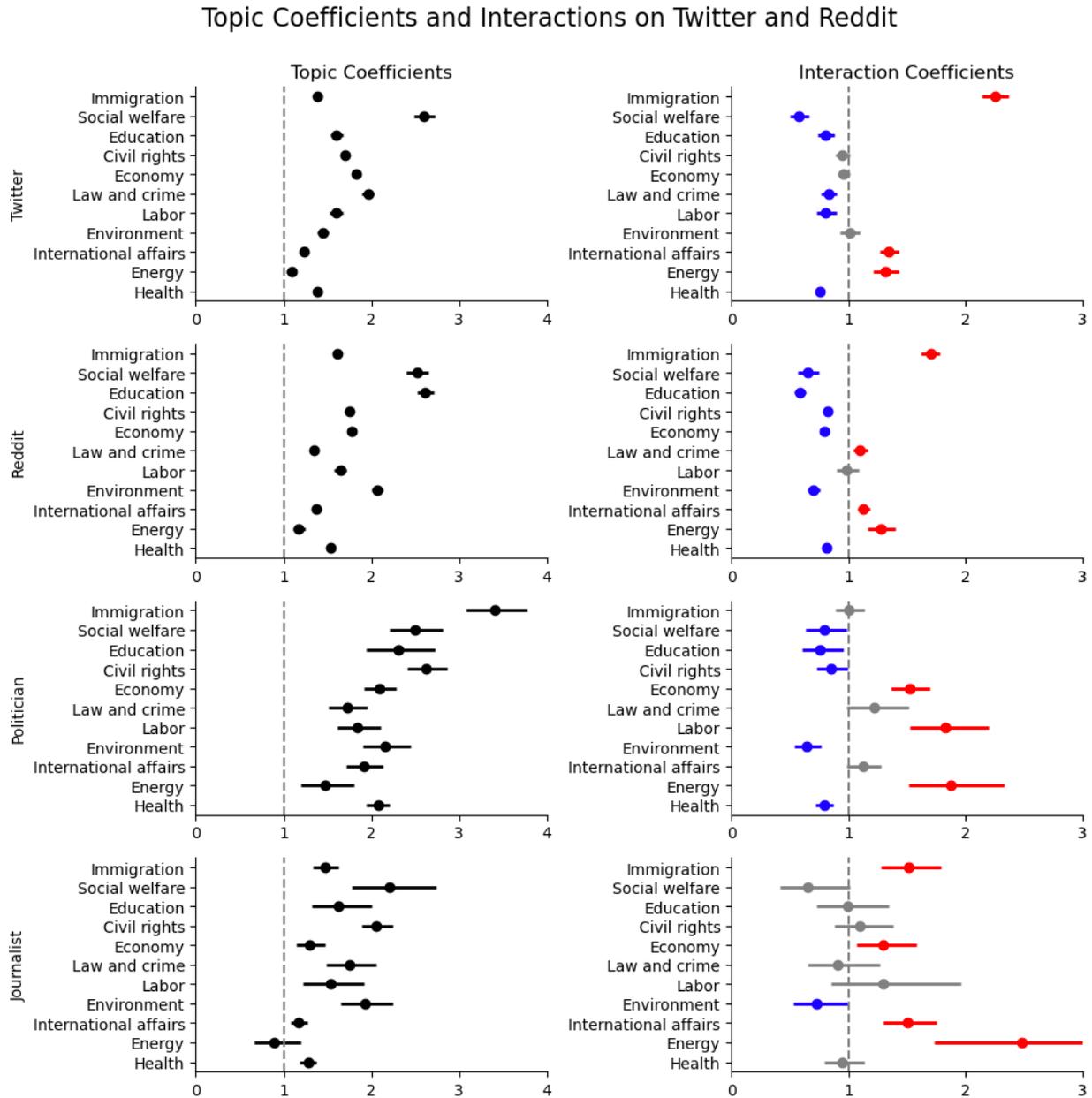

*Figure 6: **The results of an ordinal logistic regression, regressing the affect of a comment on the topic, the author's political leaning, and an interaction of the topic and the political leaning of the user.** The exponentiated coefficients are shown. The topic coefficient represents the increase in the probability that a comment will exhibit negative outgroup affect if it contains the specified topic. The interaction coefficient represents how much more likely a right-leaning users is to use polarizing rhetoric on the specified topic than a left-leaning user. We control for temporal trends when performing this analysis.*

## Discussion

We have conducted a comprehensive analysis of polarizing speech on social media platforms, specifically focusing on Reddit and Twitter over a period of fifteen years and attempting to understand how polarizing discourse has evolved across different groups and platforms.

Our findings indicate a general upward trend in outgroup animosity on both Reddit and Twitter, with newer cohorts of users in our sample displaying more outgroup derogation than their predecessors. This suggests that social media platforms may increasingly reflect and amplify political divides over time. Notably, a small minority of users contribute a significant portion of the negative political content, indicating that a concentrated group of individuals plays a large role in shaping the tone of online discourse. The variation in tendency to use polarizing discourse across topics, with right-leaning users more likely to use outgroup animosity on immigration and left-leaning users more likely to use it on health, highlights how different political issues resonate differently across party lines.

At the user group level, our findings reveal substantial outgroup animosity today amongst U.S. politicians. Most notably, right-leaning politicians substantially increased their levels of outgroup animosity since 2016 and recent cohorts of left and right-leaning politicians exhibit more outgroup animosity than any previous cohort. Since political elites have a significant amount of influence on our political discourse and public opinion it is noteworthy how recent cohorts exhibit more outgroup derogation than previous one (32). Additionally, this trend could be exacerbating the over-perception of polarization in American politics (28). Journalists and other political influencers, although generally less polarizing in their rhetoric, contribute meaningfully to the volume of political discussion and have also tended to increase their outgroup animosity, shaping the overall discourse on Twitter. Use of outgroup animosity tends to decrease with the number of followers an account has. This may suggest that more work could focus on polarizing speech by smaller accounts, as much of the literature focuses on political elite (9; 13; 14). On Reddit, polarizing speech was dominated by *r/politics* on the left and *r/TheDonald* on the right. Content moderation—such as the banning of *r/TheDonald*—has had a measurable impact on the level of polarizing discourse on the right, although we see that other subreddits increased their contribution to overall polarizing discourse.

This analysis is subject to several limitations. First, we study only the American political context and thus cannot draw inferences to trends in other countries, although we view this as important future work. Second, as with any retrospective data collection there are various changes to the data that may have occurred (e.g., platform changes to content moderation policies or algorithmic changes) that may have influenced results (indeed, we see from the removal of *r/TheDonald* the impact that a moderation decision can have on aggregate polarizing discourse). Finally, we acknowledge that our classifiers are imperfect for both polarizing speech detection and topic classification, with potential errors introduced both due to the subjectivity of the tasks and the imperfections in any machine learning model. However, we show stability of the results under various model types (see S7 Robustness Checks for Data Source and Base Model).

At a time when aggregate affective polarization is at an all-time high, we have found that outgroup animosity has also increased amongst a sample of partisan users on Reddit and Twitter as well as amongst journalists and politicians on Twitter. We note that these findings are limited to a sample of social media posts in the U.S. political context, and thus we do not draw conclusions beyond the sample studied. However, given the prevalence and importance of social media it is important that we understand who is contributing most to polarizing discourse, on which platforms and in what contexts. We believe that mapping these trends is a step towards enabling social media users to contextualize the polarizing content they encounter and motivate them to seek out online spaces, whether within these platforms or beyond, that foster constructive political communication.

## Methods

### Identifying Partisan Users

We collect our initial set of political partisans by following the method of Barbera et al. (21) which infers the political leaning of a Twitter user based on their following patterns of a set of political elites. This method constructs a binary "following" matrix where each row corresponds to a user and each column is a politician, and after dimensionality reduction on this matrix one can compute an estimate of the users' political leaning. We filter to users that follow at least ten politicians in our set or have left at least ten comments in a set of 78 explicitly political subreddits.

After applying the method of Barbera et al. to Twitter, we see the expected bimodal distribution of partisan scores (see **Figure S1**). To translate the Barbera method to Reddit, instead of a binary entry in a matrix indicating a follow relationship, we insert the log number of positive scoring comments a user made in each subreddit. Then, the application of Barbera's method would yield conceptually similar results; users that comment in similar subreddits would have similar values across the primary principal component. Indeed, we see a similar bimodal distribution of partisan scores on Reddit (see Figure S2). As expected on both platforms, there are more left leaning than right leaning users, and we use the same cutoffs as previous work (22) of -1.2 and 0 to determine left vs. right leaning users (we flipped the signs as necessary such that on both platforms, a higher theta meant more liberal, and a lower theta meant more conservative). We inspected some users as a sanity check for both Reddit and Twitter. As expected, on Reddit the high theta users primarily comment in *r/politics* and *r/Liberal*, while the low theta users comment primarily in *r/Conservative*. Likewise, on Twitter the users with higher theta primarily followed left-leaning politicians, and vice-versa for users with lower theta.

In summary, to identify partisan users on Reddit and Twitter we:

1. Identified 1,361 politicians on Twitter and 78 political subreddits on Reddit.
2. Filtered to users that followed at least ten politicians on Twitter or left at least ten comments in our list of political subreddits.
3. Used the method by Barbera (21) to infer the partisan leaning of these users.
4. Filtered to users that had partisan scores of greater than zero (classified as left-leaning) and less than -1.2 (classified as right-leaning).

### Collecting Tweets

From these users, we collect tweets from these users in two ways. First, we use the User Timeline API, which allows us to collect the last 3200 tweets of each user. We find t 64.6% of users fall below this threshold, meaning we collect their entire timeline. In total, we pull 988 million tweets from 1.2 million authors. Since for the temporal analysis the 3200-tweet limit could introduce bias, we also pull data from one day each month from 2006 onwards for a sample of users using the Historical Powertrack API, which has no restriction on the number of previous tweets. We additionally pull the tweets of politicians, media, and journalists using account IDs provided by (22) as well as the Ballotpedia dataset. We discard retweets; while previous work has studied engagement with polarizing content, usually from political elites (22) our focus is on the affect of original tweets.

<u>Collecting Reddit Comments</u>

From our sample of Reddit partisans, we collect *all* comments from these users across all subreddits, which leaves us with 1.5 billion comments from roughly half a million users. While previous work has often studied interactions in specific political subreddits (15) it has recently been shown that a significant amount of political discourse on Reddit happens in non-political subreddits (6). Thus, we ensure to collect every comment made by a politically active user, even if that user is posting in non-political subreddits.

*Limitations*

This study has several limitations related to data collection and processing. First, ongoing changes in platform moderation policies—such as the deletion of content or removal of accounts—may have affected the availability and representativeness of the data. While such deletions often target toxic or harmful content, our focus on politically polarizing speech, which is not inherently toxic, suggests that the impact on our findings could be limited. Second, we did not attempt to identify or exclude automated accounts (bots), which may influence the prevalence and nature of political discourse in our sample. The presence of bots could introduce distortions in patterns of polarizing rhetoric. Finally, although the political terms used in this study are derived from the American political context, we did not restrict our dataset to users located within the United States. As a result, our sample may include international users whose engagement with American political language differs in meaning or intent, potentially affecting the interpretation of polarization.

<u>Identifying Political Content and Preprocessing Data</u>

Next, we subset our collected data to a set containing a reference to a political group or entity. The closest methodology to ours by Ballard et al. (9), who study only Congressional tweets, does not take this step. However, since our analysis covers a broader diversity of users (e.g. high visibility politicians and journalists and more general partisan users) one issue is that without filtering the data to ensure some consistency in the type of content being discussed, measures of polarizing political discourse will be heavily skewed by the propensity to discuss politics in the first place. For example, politicians and journalists, given the nature in their profession, are more likely to post about politics than most users and thus have more opportunities to use polarizing discourse (i.e. one cannot use polarizing political rhetoric if one is not discussing politics). This can be seen in Figure S5 which shows that politicians and journalists are more likely to include a tweet that contains a reference to a political group or entity in our list than other partisan users. Additionally, the amount of overall discussion about politics more generally has shifted over time, and thus a metric that is sensitive to these fluctuations may simply capture changes in the volume of overall political discourse. See S4.1 Alternative Metrics for a discussion of the trade-offs involved in filtering content based on political keywords, as well as an analysis of the results when this filtering step is omitted. Additionally, we test our results' stability to these keywords by running a bootstrap type test, randomly dropping 10% of keywords and recreating our main visualizations. We find that our results are robust to these tests (see S4.2 Robustness to Political Keywords). Throughout this work, we will refer to content containing reference to a political group or entity as "political content" (while this term is general, we use it in the interest of brevity and readability).

We take the following preprocessing steps to clean our data.

1. We construct a list of 215 terms that corresponding to political groups or entities. The list of political groups and entities used to filter the data can be found in Table S9, where political group terms were sourced from (33) and influential political figures were sourced by taking the top hundred most followed left-leaning and top hundred most followed right-leaning politicians from our sample of politicians on Twitter. In total, we use 215 political entities used to filter comment and assign a comment as referencing either a left-leaning or right-leaning entity. These terms are then converted to lowercase.
2. We lowercase our existing set of 1.5 billion Reddit comments and 988 million tweets.
3. We remove retweets.
4. We filter our set of comments using the set of political entity keywords, which results in 30 million Reddit comments and 67 million tweets.
5. We remove comments that contain both a reference to a left-leaning and right-leaning entity.
6. We remove tweets and comments with less than ten characters or over 240 characters.

<u>Sampling the Training Set</u>

From our cleaned collection of political tweets and political Reddit comments, we select a subset of 4,399 training examples to label for polarizing speech. We use a stratified sampling approach based on several factors: the political orientation of the comment author, the political orientation of the person or group being referenced, the user type, and the year. Within each stratum, we sample 100 data points. Due to data sparsity in certain groups during earlier years—such as right-leaning journalists, who do not consistently post political content until 2012—we stratify journalist data separately starting in 2012, using the same criteria. We apply a similar stratification process to the Reddit data, sampling 50 examples from each stratum. To see the resulting breakdown across the training set, please refer to the classification reports in Table S12, which captures the support of these strata in the training set. We also provided a table capturing the number of data points per year in Table S6**.**

<u>Labeling Polarizing Speech</u>

Previous work on polarizing speech has typically labeled speech that attempts to create division between the speaker and an outgroup, or speech that attempts to show intra-party loyalty, as "polarizing" speech (9). We adopt a similar approach, coding content based on whether it attempted to create a division between the speaker and their outgroup or show solidarity and support towards their ingroup or members of their ingroup. Annotators were asked to either label the comment as exhibiting explicit "affective polarization", which could include ad-hominem attacks/compliments towards political group or person, or "partisan", where the author exhibits (dis)like or (dis)trust towards a group in the context of governance. We are careful to only label comments that are negative towards or lay blame on a specific group or person. See Table S10 and Table S11 for examples. A team of three annotators labeled 4399 examples of political speech.

This number of examples is in line with other work measuring polarizing speech using machine learning (9) and our reasonable out of sample accuracy and model agnostic results give us confidence that the results are robust. For labeling whether a comment contained either affectively polarizing or partisan speech, the inter-annotator agreement was .78 using Krippendorf's alpha, and for the categories separately it was .66 for affectively polarizing and .68 for partisan speech.

The difference between the two categories can often be hard to distinguish in practice (frequently comments will display features of both, and it is subjective which one ought to be chosen), hence why the agreement for whether the comment contains any type of polarizing rhetoric is higher than which particular category it falls into. Therefore, we follow previous work (9) and take an expansive definition of polarizing rhethoric, building a classifier to predict whether the comment contains any type of polarizing rhetoric (e.g., ad hominem attacks or laying blame on a specific group or person for poor governance).

<u>Training a Polarizing Speech Classifier</u>

Given the importance of the classifier in our analysis, we train three separate classifiers to ensure that our results are not driven by any one model. We use language models available from the HuggingFace library, which are BERT, GPT2, and a version of BERT further fine-tuned on a tweet sentiment task. The best performing individual model is the fine-tuned BERT sentiment classifier, which achieves an accuracy of 81% and we use for the main results and charts. We use five-fold cross validation to measure these metrics out-of-sample. We also replicate our results with all the models individually and find that they agree. See Table S12 for the full classification report on our main model. We also break down performance across different political orientations, political references, user types, and temporal periods, which can be found in Table S13 through Table S16. Finally, we benchmark our model against a Transformer based and keyword-based sentiment classifier and find that it outperforms both (see Figure S3 and Figure S4).

<u>Topic Classifier</u>

We use a set of topics compiled by the Comparative Agendas Project (34) and have three annotators label the same examples outlined above. We only keep topics for which we see at least fifteen examples in the training set. On a set of 800 overlapping examples, the annotators agreed 87% of the time with a Krippendof's alpha of .63. Although the inter-annotator agreement is lower than expected, after discussions with the annotators we note that this may be due in part to the difficulty the annotators had in keeping track of all possible topic labels. For comments with disagreements consensus could often be reached due to an annotator realizing their error. We acknowledge this may make the classifier more unreliable than desired. However, after testing the classifier out of sample it still manages to achieve a 91% out of sample accuracy using five-fold cross validation, and a macro F1 score of .67. See Table S17 for a detailed breakdown of performance over the different topics.

**Data and Code Availability Statement**

Anonymized data will be made available to the degree permitted by the platforms' data sharing agreements. This will be released in conjunction with code necessary to replicate all the visualizations and regression models shown in this manuscript. Please contact Suyash Fulay (sfulay@mit.edu) for data and code.

**Supplementary Materials**

**The file includes:**

## S1 Influencers

We manually annotate the roles of each of the top two-hundred accounts with the most followers in our sample. Several of the accounts were social media influencers, former government officials, or political commentators; we group these into a single "political influencer" category for ease of analysis. However, we make the full list transparent here. We group actors, singers, and comedians into a single "celebrity" category. The tables below show each account, the number of followers they had at the point of data collection, the average affect towards their in/outgroup, the percentage of political tweets they post out of the total number of influencers' political tweets, and then the product of these which is the contribution to overall polarizing discourse amongst the set of influencers.

Influencer Tables

| Screen Name | Followers | Type | Outgroup animosity | % Tweets | Total Contribution |
|---|---|---|---|---|---|
| FoxNews | 23,793,177 | Media | 0.11 | 1.83 | 0.20 |
| DonaldJTrumpJr | 9,585,414 | Influencer | 0.58 | 1.01 | 0.59 |
| seanhannity | 6,323,745 | Journalist | 0.2 | 5.35 | 1.09 |
| benshapiro | 5,516,716 | Journalist | 0.37 | 0.53 | 0.20 |
| IngrahamAngle | 4,530,529 | Journalist | 0.51 | 2.02 | 1.04 |
| RandPaul | 4,510,587 | Politician | 0.52 | 0.72 | 0.38 |
| EricTrump | 4,064,036 | Influencer | 0.47 | 0.24 | 0.11 |
| KellyannePolls | 3,520,211 | Official | 0.25 | 0.73 | 0.18 |
| BillOReilly | 3,292,879 | Journalist | 0.29 | 3.13 | 0.92 |
| kayleighmcenany | 3,198,451 | Commentator | 0.5 | 0.12 | 0.06 |
| JudgeJeanine | 3,156,268 | Journalist | 0.33 | 1.87 | 0.61 |
| SenJohnMcCain | 2,811,878 | Politician | 0.24 | 0.56 | 0.13 |
| greggutfeld | 2,414,702 | Journalist | 0.28 | 0.33 | 0.09 |
| LouDobbs | 2,384,691 | Journalist | 0.56 | 1.66 | 0.93 |
| newtgingrich | 2,371,541 | Journalist | 0.64 | 0.90 | 0.57 |
| TomiLahren | 2,296,126 | Commentator | 0.63 | 0.59 | 0.38 |
| AnnCoulter | 2,109,240 | Journalist | 0.24 | 0.39 | 0.09 |
| NEWSMAX | 2,091,473 | Media | 0.12 | 2.43 | 0.29 |
| TomFitton | 2,086,321 | Journalist | 0.4 | 2.51 | 1.01 |
| kimguilfoyle | 2,064,432 | Journalist | 0.16 | 0.41 | 0.07 |
| michellemalkin | 2,047,882 | Journalist | 0.38 | 0.26 | 0.10 |
| RepMattGaetz | 2,021,100 | Politician | 0.6 | 0.46 | 0.28 |
| scrowder | 2,020,482 | Journalist | 0.47 | 1.51 | 0.70 |
| GovMikeHuckabee | 1,905,393 | Politician | 0.34 | 2.08 | 0.71 |
| JackPosobiec | 1,904,913 | Journalist | 0.19 | 0.24 | 0.05 |
| BreitbartNews | 1,897,308 | Media | 0.23 | 2.61 | 0.61 |
| mattgaetz | 1,820,924 | Politician | 0.19 | 0.20 | 0.04 |
| mtaibbi | 1,788,395 | Journalist | 0.11 | 0.80 | 0.09 |
| DanScavino | 1,657,560 | Official | 0.11 | 0.18 | 0.02 |
| glennbeck | 1,650,365 | Journalist | 0.49 | 1.65 | 0.81 |
| RonDeSantisFL | 1,564,326 | Politician | 0.73 | 0.10 | 0.07 |
| RudyGiuliani | 1,545,532 | Official | 0.47 | 0.81 | 0.38 |
| ksorbs | 1,529,775 | Actor | 0.24 | 0.61 | 0.14 |
| SaraCarterDC | 1,495,693 | Journalist | 0.37 | 0.62 | 0.23 |
| SarahPalinUSA | 1,435,251 | Politician | 0.1 | 3.45 | 0.36 |
| realDailyWire | 1,429,077 | Media | 0.07 | 0.18 | 0.01 |
| w_terrence | 1,412,740 | Comedian | 0.48 | 1.37 | 0.65 |
| brithume | 1,386,431 | Journalist | 0.23 | 0.61 | 0.14 |
| RubinReport | 1,311,042 | Journalist | 0.39 | 0.24 | 0.09 |
| SpeakerBoehner | 1,292,119 | Politician | 0.32 | 3.19 | 1.03 |
| DLoesch | 1,276,655 | Radio | 0.35 | 1.31 | 0.45 |
| DevinNunes | 1,253,363 | Politician | 0.19 | 0.88 | 0.16 |
| MattWalshBlog | 1,243,029 | Commentator | 0.39 | 0.72 | 0.28 |
| RichardGrenell | 1,130,745 | Journalist | 0.33 | 0.59 | 0.19 |
| ericbolling | 1,130,224 | Commentator | 0.33 | 0.39 | 0.13 |
| NEWS_MAKER | 1,044,462 | Influencer | 0.12 | 0.27 | 0.03 |
| nprscottsimon | 1,038,683 | Journalist | 0.08 | 0.16 | 0.01 |
| MZHemingway | 1,036,109 | Journalist | 0.33 | 0.93 | 0.31 |
| JohnStossel | 1,034,068 | Journalist | 0.21 | 0.38 | 0.08 |
| BillKristol | 1,020,382 | Journalist | 0.08 | 0.43 | 0.03 |
| KatiePavlich | 987,605 | Journalist | 0.39 | 1.15 | 0.44 |
| GregAbbott_TX | 972,377 | Politician | 0.55 | 0.07 | 0.04 |
| RyanAFournier | 969,163 | Influencer | 0.36 | 0.73 | 0.26 |
| BuckSexton | 965,464 | Radio | 0.63 | 2.46 | 1.56 |
| MarshaBlackburn | 907,385 | Politician | 0.66 | 3.86 | 2.56 |
| jsolomonReports | 881,142 | Journalist | 0.07 | 0.99 | 0.07 |
| cvpayne | 877,957 | Journalist | 0.26 | 0.20 | 0.05 |
| michaeljknowles | 872,642 | Commentator | 0.2 | 0.59 | 0.12 |
| FoxBusiness | 872,055 | Media | 0.22 | 0.45 | 0.10 |
| theblaze | 865,816 | Media | 0.15 | 1.41 | 0.22 |
| KariLake | 834,304 | Politician | 0.29 | 0.27 | 0.08 |
| LisaMarieBoothe | 831,671 | Journalist | 0.63 | 0.39 | 0.25 |
| ChuckGrassley | 825,220 | Politician | 0.55 | 3.46 | 1.90 |
| AmbJohnBolton | 813,761 | Official | 0.59 | 1.26 | 0.74 |
| KatTimpf | 799,820 | Journalist | 0.15 | 0.17 | 0.03 |
| TimRunsHisMouth | 799,437 | Comedian | 0.38 | 1.57 | 0.61 |
| AllenWest | 795,035 | Journalist | 0.62 | 0.60 | 0.37 |
| ScottBaio | 792,546 | Actor | 0.29 | 0.48 | 0.14 |
| ACTBrigitte | 776,132 | Activist | 0.68 | 2.43 | 1.66 |
| parscale | 724,935 | Official | 0.59 | 0.45 | 0.26 |
| SenatorTimScott | 706,028 | Politician | 0.81 | 1.13 | 0.91 |

| Screen Name | Followers | Type | Outgroup Animosity | % Tweets | Total Contribution |
|---|---|---|---|---|---|
| KatrinaPierson | 698,963 | Activist | 0.46 | 0.09 | 0.04 |
| RepBoebert | 698,958 | Politician | 0.52 | 0.89 | 0.46 |
| RepDanCrenshaw | 696,236 | Politician | 0.63 | 0.27 | 0.17 |
| SenMikeLee | 685,135 | Politician | 0.44 | 0.85 | 0.38 |
| ShannonBream | 670,001 | Journalist | 0.05 | 0.49 | 0.03 |
| MarkDice | 664,094 | Influencer | 0.52 | 0.34 | 0.18 |
| JeffFlake | 663,881 | Politician | 0.03 | 0.52 | 0.01 |
| SenTomCotton | 634,003 | Politician | 0.75 | 2.03 | 1.53 |
| KarlRove | 632,357 | Journalist | 0.27 | 0.68 | 0.18 |
| pnjaban | 614,357 | Activist | 0.31 | 0.17 | 0.05 |
| SenJohnKennedy | 610,835 | Politician | 0.68 | 0.89 | 0.61 |
| TeamCavuto | 608,906 | Journalist | 0.14 | 2.44 | 0.34 |
| RepAndyBiggsAZ | 608,568 | Politician | 0.67 | 0.87 | 0.58 |
| maibortpetit | 607,674 | Influencer | 0.03 | 0.75 | 0.02 |
| BrandonStraka | 599,162 | Influencer | 0.37 | 0.93 | 0.34 |
| RepLizCheney | 591,626 | Politician | 0.74 | 0.85 | 0.63 |
| DavidJHarrisJr | 590,107 | Influencer | 0.37 | 0.36 | 0.13 |
| EpochTimes | 582,567 | Media | 0.11 | 0.59 | 0.07 |
| PAMsLOvE | 578,845 | Influencer | 0.27 | 0.17 | 0.05 |
| Liz_Cheney | 575,343 | Politician | 0.5 | 2.02 | 1.01 |
| SteveScalise | 559,770 | Politician | 0.54 | 0.17 | 0.09 |
| SenatorCollins | 558,555 | Politician | -0.02 | 0.32 | -0.01 |
| jasoninthehouse | 557,170 | Politician | 0.41 | 1.19 | 0.49 |
| ElijahSchaffer | 554,151 | Journalist | 0.15 | 0.09 | 0.01 |
| RepKinzinger | 553,770 | Politician | 0.44 | 0.28 | 0.13 |
| RepDougCollins | 550,656 | Politician | 0.54 | 1.31 | 0.71 |
| RepThomasMassie | 548,528 | Politician | 0.36 | 0.95 | 0.34 |
| SethDillon | 544,614 | Influencer | 0.24 | 0.53 | 0.13 |

*Table S1: Outgroup animosity for right-leaning influencers*

| Screen Name | Followers | Type | Ingroup Animosity | % Tweets | Total Contribution |
|---|---|---|---|---|---|
| FoxNews | 23,793,177 | Media | 0.01 | 1.28 | 0.02 |
| DonaldJTrumpJr | 9,585,414 | Influencer | 0.17 | 0.51 | 0.09 |
| seanhannity | 6,323,754 | Journalist | 0.05 | 4.03 | 0.18 |
| benshapiro | 5,516,716 | Journalist | 0.18 | 0.40 | 0.07 |
| IngrahamAngle | 4,530,529 | Journalist | 0.12 | 0.99 | 0.11 |
| RandPaul | 4,510,595 | Politician | -0.06 | 1.11 | -0.06 |
| EricTrump | 4,064,036 | Influencer | -0.45 | 1.65 | -0.75 |
| KellyannePolls | 3,520,211 | Official | -0.23 | 0.58 | -0.13 |
| BillOReilly | 3,292,879 | Journalist | 0.1 | 2.23 | 0.23 |
| kayleighmcenany | 3,198,451 | Commentator | -0.53 | 0.13 | -0.07 |
| JudgeJeanine | 3,156,268 | Journalist | 0.03 | 3.32 | 0.11 |
| SenJohnMcCain | 2,811,878 | Politician | -0.12 | 1.75 | -0.20 |
| greggutfeld | 2,414,710 | Journalist | 0.17 | 0.26 | 0.04 |
| LouDobbs | 2,384,691 | Journalist | 0.14 | 1.46 | 0.20 |
| newtgingrich | 2,371,541 | Journalist | -0.15 | 0.42 | -0.06 |
| DanaPerino | 2,357,412 | Journalist | 0.0 | 0.10 | 0.00 |
| TomiLahren | 2,296,126 | Commentator | 0.13 | 0.40 | 0.05 |
| AnnCoulter | 2,109,240 | Journalist | 0.01 | 0.81 | 0.01 |
| NEWSMAX | 2,091,473 | Media | 0.0 | 3.71 | 0.01 |
| TomFitton | 2,086,321 | Journalist | 0.22 | 1.20 | 0.26 |
| kimguilfoyle | 2,064,434 | Journalist | -0.37 | 1.57 | -0.57 |
| michellemalkin | 2,047,882 | Journalist | 0.42 | 0.99 | 0.41 |
| RepMattGaetz | 2,021,100 | Politician | 0.06 | 0.90 | 0.05 |
| scrowder | 2,020,482 | Journalist | 0.17 | 0.82 | 0.14 |
| GovMikeHuckabee | 1,905,401 | Politician | 0.16 | 1.76 | 0.28 |
| JackPosobiec | 1,904,913 | Journalist | 0.03 | 0.30 | 0.01 |
| BreitbartNews | 1,897,308 | Media | 0.03 | 1.69 | 0.04 |
| mattgaetz | 1,820,924 | Politician | -0.11 | 0.55 | -0.06 |
| mtaibbi | 1,788,395 | Journalist | 0.1 | 1.82 | 0.18 |
| DanScavino | 1,657,560 | Official | -0.15 | 3.12 | -0.47 |
| glennbeck | 1,650,365 | Journalist | 0.14 | 0.98 | 0.13 |
| RonDeSantisFL | 1,564,326 | Politician | -0.57 | 1.44 | -0.83 |
| RudyGiuliani | 1,545,532 | Official | -0.1 | 0.35 | -0.04 |
| ksorbs | 1,529,775 | Actor | 0.18 | 0.33 | 0.06 |
| SaraCarterDC | 1,495,693 | Journalist | -0.08 | 0.43 | -0.04 |
| SarahPalinUSA | 1,435,251 | Politician | 0.01 | 5.04 | 0.04 |
| realDailyWire | 1,429,077 | Media | 0.04 | 0.21 | 0.01 |
| w_terrence | 1,412,740 | Comedian | 0.03 | 1.43 | 0.04 |
| brithume | 1,386,431 | Journalist | 0.15 | 1.06 | 0.16 |
| RubinReport | 1,311,036 | Journalist | -0.06 | 0.27 | -0.02 |
| SpeakerBoehner | 1,292,119 | Politician | -0.18 | 0.96 | -0.18 |
| DLoesch | 1,276,655 | Radio | 0.09 | 1.22 | 0.11 |
| DevinNunes | 1,253,363 | Politician | -0.06 | 1.55 | -0.09 |
| MattWalshBlog | 1,243,029 | Commentator | 0.22 | 0.78 | 0.18 |
| RichardGrenell | 1,130,745 | Journalist | 0.1 | 0.53 | 0.05 |
| ericbolling | 1,130,224 | Commentator | 0.21 | 0.33 | 0.07 |
| NEWS_MAKER | 1,044,462 | Influencer | 0.12 | 0.28 | 0.04 |
| nprscottsimon | 1,038,683 | Journalist | 0.0 | 0.19 | 0.00 |
| MZHemingway | 1,036,109 | Journalist | 0.25 | 0.99 | 0.25 |
| JohnStossel | 1,034,068 | Journalist | 0.29 | 0.43 | 0.12 |
| BillKristol | 1,020,375 | Journalist | 0.28 | 3.04 | 0.85 |
| KatiePavlich | 987,605 | Journalist | 0.07 | 0.38 | 0.03 |
| RyanAFournier | 969,163 | Influencer | -0.02 | 0.76 | -0.02 |
| BuckSexton | 965,464 | Radio | 0.27 | 0.43 | 0.11 |
| MarshaBlackburn | 907,385 | Politician | 0.07 | 0.37 | 0.03 |
| jsolomonReports | 881,142 | Journalist | 0.02 | 1.80 | 0.04 |
| michaeljknowles | 872,642 | Commentator | 0.18 | 0.90 | 0.16 |
| FoxBusiness | 872,055 | Media | 0.0 | 0.17 | 0.00 |
| theblaze | 865,816 | Media | 0.04 | 0.99 | 0.04 |
| KariLake | 834,301 | Politician | -0.09 | 0.19 | -0.02 |
| LisaMarieBoothe | 831,671 | Journalist | 0.0 | 0.29 | 0.00 |
| ChuckGrassley | 825,220 | Politician | -0.15 | 1.55 | -0.23 |
| AmbJohnBolton | 813,761 | Official | 0.04 | 2.54 | 0.10 |
| Fahrenthold | 809,187 | Journalist | 0.02 | 0.71 | 0.02 |
| TimRunsHisMouth | 799,437 | Comedian | 0.1 | 0.89 | 0.09 |
| AllenWest | 795,035 | Journalist | -0.14 | 1.04 | -0.14 |
| ScottBaio | 792,546 | Actor | 0.11 | 0.33 | 0.04 |
| ACTBrigitte | 776,129 | Activist | -0.09 | 2.55 | -0.22 |
| parscale | 724,935 | Official | -0.13 | 1.62 | -0.21 |
| SenatorTimScott | 706,028 | Politician | -0.23 | 0.23 | -0.05 |
| KatrinaPierson | 698,963 | Activist | 0.1 | 0.18 | 0.02 |
| RepBoebert | 698,958 | Politician | -0.2 | 0.18 | -0.04 |
| RepDanCrenshaw | 696,236 | Politician | 0.0 | 0.11 | 0.00 |
| SenMikeLee | 685,135 | Politician | -0.12 | 1.78 | -0.22 |
| ShannonBream | 670,001 | Journalist | 0.12 | 0.38 | 0.04 |
| MarkDice | 664,094 | Influencer | 0.33 | 0.35 | 0.11 |
| JeffFlake | 663,881 | Politician | -0.11 | 2.45 | -0.27 |
| SenTomCotton | 634,003 | Politician | -0.11 | 0.40 | -0.04 |
| KarlRove | 632,357 | Journalist | -0.06 | 1.12 | -0.07 |
| pnjaban | 614,357 | Activist | 0.08 | 0.78 | 0.06 |
| SenJohnKennedy | 610,852 | Politician | -0.33 | 0.48 | -0.16 |
| TeamCavuto | 608,906 | Journalist | 0.03 | 2.19 | 0.06 |
| RepAndyBiggsAZ | 608,568 | Politician | 0.0 | 0.33 | 0.00 |
| maibortpetit | 607,674 | Influencer | 0.0 | 0.27 | 0.00 |
| BrandonStraka | 599,162 | Influencer | 0.16 | 0.97 | 0.16 |
| RepLizCheney | 591,626 | Politician | -0.09 | 0.67 | -0.06 |
| DavidJHarrisJr | 590,107 | Influencer | 0.17 | 0.37 | 0.06 |
| EpochTimes | 582,567 | Media | 0.06 | 0.55 | 0.04 |
| PAMsLOvE | 578,845 | Influencer | 0.45 | 0.10 | 0.04 |
| Liz_Cheney | 575,343 | Politician | -0.19 | 1.88 | -0.36 |
| SteveScalise | 559,770 | Politician | -0.67 | 0.13 | -0.09 |
| SenatorCollins | 558,555 | Politician | -0.26 | 0.51 | -0.13 |
| jasoninthehouse | 557,161 | Politician | -0.18 | 0.54 | -0.10 |
| ElijahSchaffer | 554,197 | Journalist | 0.08 | 0.11 | 0.01 |
| RepKinzinger | 553,768 | Politician | 0.16 | 1.15 | 0.18 |
| RepDougCollins | 550,596 | Politician | -0.37 | 0.96 | -0.35 |
| RepThomasMassie | 548,528 | Politician | -0.04 | 0.47 | -0.02 |
| SethDillon | 544,614 | Influencer | 0.12 | 0.21 | 0.03 |
| HARRISFAULKNER | 527,129 | Journalist | -0.06 | 0.60 | -0.04 |

*Table S2: Negative ingroup affect for right-leaning influencers*

| Screen Name | Followers | Type | Outgroup Animosity | % Tweets | Total Contribution |
|---|---|---|---|---|---|
| nytimes | 54,986,629 | Media | 0.13 | 0.64 | 0.08 |
| HillaryClinton | 31,595,284 | Politician | 0.51 | 0.37 | 0.19 |
| Reuters | 25,757,772 | Media | 0.03 | 0.21 | 0.01 |
| washingtonpost | 20,040,876 | Media | 0.1 | 1.27 | 0.13 |
| Forbes | 18,780,197 | Media | 0.22 | 0.76 | 0.17 |
| ChrisEvans | 16,182,175 | Actor | 0.67 | 0.17 | 0.11 |
| BernieSanders | 15,552,053 | Politician | 0.74 | 1.41 | 1.05 |
| johnlegend | 13,727,933 | Singer | 0.47 | 0.11 | 0.05 |
| AOC | 13,443,351 | Politician | 0.3 | 0.32 | 0.10 |
| SenSanders | 12,502,471 | Politician | 0.81 | 0.98 | 0.79 |
| maddow | 10,459,893 | Journalist | 0.12 | 0.58 | 0.07 |
| TheDailyShow | 9,512,239 | Media | 0.14 | 1.02 | 0.14 |
| NBCNews | 9,416,428 | Media | 0.06 | 1.01 | 0.06 |
| Sethrogen | 9,368,309 | Actor | 0.2 | 0.58 | 0.12 |
| NewYorker | 8,972,964 | Media | 0.07 | 0.42 | 0.03 |
| CBSNews | 8,916,812 | Media | 0.07 | 0.58 | 0.04 |
| MarkRuffalo | 8,364,416 | Actor | 0.76 | 0.74 | 0.57 |
| SpeakerPelosi | 8,065,399 | Politician | 0.71 | 1.04 | 0.74 |
| chelseahandler | 7,719,580 | Actor | 0.46 | 2.75 | 1.25 |
| SenWarren | 7,033,561 | Politician | 0.82 | 0.87 | 0.72 |
| ewarren | 5,894,853 | Politician | 0.8 | 1.03 | 0.83 |
| MMFlint | 5,883,969 | Journalist | 0.35 | 2.49 | 0.88 |
| CNBC | 5,092,091 | Media | 0.0 | 0.13 | 0.00 |
| MSNBC | 4,992,529 | Media | 0.27 | 7.55 | 2.02 |
| USATODAY | 4,955,901 | Media | 0.13 | 0.37 | 0.05 |
| CoryBooker | 4,839,236 | Politician | 0.4 | 0.14 | 0.06 |
| VanityFair | 4,770,643 | Media | 0.1 | 1.25 | 0.13 |
| politico | 4,642,306 | Media | 0.1 | 0.88 | 0.08 |
| thehill | 4,470,046 | Media | 0.06 | 2.12 | 0.13 |
| soyfdelrincon | 3,813,775 | Journalist | 0.0 | 0.15 | 0.00 |
| Jemima_Khan | 3,758,111 | Journalist | 0.21 | 0.10 | 0.02 |
| PeteButtigieg | 3,677,799 | Politician | 0.34 | 0.52 | 0.18 |
| Newsweek | 3,596,708 | Media | 0.07 | 2.40 | 0.17 |
| Alyssa_Milano | 3,541,392 | Actor | 0.33 | 0.30 | 0.10 |
| NateSilver538 | 3,479,182 | Journalist | 0.05 | 1.21 | 0.06 |
| GeorgeTakei | 3,456,722 | Actor | 0.65 | 0.56 | 0.36 |
| JohnKerry | 3,363,459 | Politician | 0.05 | 0.39 | 0.02 |
| RepAdamSchiff | 3,250,357 | Politician | 0.58 | 1.43 | 0.83 |
| amanpour | 3,198,734 | Journalist | 0.17 | 2.20 | 0.38 |
| jaketapper | 3,184,663 | Journalist | 0.08 | 0.27 | 0.02 |
| IlhanMN | 3,000,249 | Politician | 0.54 | 0.20 | 0.11 |
| staceyabrams | 2,882,261 | Politician | 0.6 | 0.18 | 0.11 |
| ChelseaClinton | 2,858,009 | Commentator | 0.51 | 0.33 | 0.17 |
| Lawrence | 2,837,029 | Other | 0.27 | 1.05 | 0.28 |
| carlaangola | 2,835,597 | Journalist | 0.0 | 0.47 | 0.00 |
| mishacollins | 2,814,706 | Actor | 0.49 | 0.25 | 0.12 |
| marwilliamson | 2,768,425 | Politician | 0.31 | 0.20 | 0.06 |
| JoeNBC | 2,765,504 | Journalist | 0.39 | 2.42 | 0.95 |
| ariannahuff | 2,755,868 | Journalist | 0.2 | 0.11 | 0.02 |
| nowthisnews | 2,708,524 | Media | 0.13 | 0.70 | 0.09 |
| DanRather | 2,707,248 | Journalist | 0.31 | 0.79 | 0.25 |
| MikeBloomberg | 2,674,989 | Politician | 0.25 | 0.53 | 0.13 |
| ezraklein | 2,656,110 | Journalist | 0.33 | 1.43 | 0.47 |
| BetoORourke | 2,549,244 | Politician | 0.46 | 0.67 | 0.31 |
| alroker | 2,412,702 | Journalist | 0.05 | 0.15 | 0.01 |
| chrislhayes | 2,402,852 | Journalist | 0.15 | 0.66 | 0.10 |
| JuddApatow | 2,400,506 | Producer | 0.37 | 0.13 | 0.05 |
| GStephanopoulos | 2,374,523 | Journalist | 0.01 | 2.99 | 0.04 |
| kingsthings | 2,327,966 | Journalist | 0.02 | 0.70 | 0.01 |
| billyeichner | 2,318,140 | Actor | 0.66 | 0.54 | 0.36 |
| Acosta | 2,290,501 | Journalist | 0.14 | 1.38 | 0.19 |
| PBS | 2,241,371 | Media | 0.23 | 0.15 | 0.04 |
| mariashriver | 2,193,577 | Journalist | 0.18 | 0.39 | 0.07 |
| TheAtlantic | 2,120,662 | Media | 0.21 | 0.92 | 0.20 |
| TulsiGabbard | 2,110,579 | Politician | 0.69 | 0.11 | 0.08 |
| chucktodd | 2,060,309 | Journalist | 0.08 | 1.17 | 0.10 |
| JoyAnnReid | 2,057,213 | Journalist | 0.34 | 1.79 | 0.61 |
| GavinNewsom | 2,052,630 | Politician | 0.41 | 0.19 | 0.08 |
| amyklobuchar | 2,035,552 | Politician | 0.29 | 0.39 | 0.11 |
| ggreenwald | 1,999,391 | Journalist | 0.09 | 0.46 | 0.04 |
| ananavarro | 1,993,938 | Commentator | 0.54 | 1.91 | 1.03 |
| mitchellreports | 1,928,460 | Journalist | 0.1 | 5.16 | 0.49 |
| NYMag | 1,793,859 | Media | 0.15 | 0.74 | 0.11 |
| PreetBharara | 1,775,705 | Official | 0.04 | 0.48 | 0.02 |
| johncusack | 1,757,770 | Actor | 0.53 | 0.13 | 0.07 |
| AndrewYang | 1,726,576 | Politician | 0.05 | 0.59 | 0.03 |
| wolfblitzer | 1,725,899 | Journalist | 0.13 | 1.86 | 0.25 |
| katiecouric | 1,699,996 | Journalist | 0.2 | 1.00 | 0.20 |
| maggieNYT | 1,663,152 | Journalist | 0.03 | 3.97 | 0.13 |
| tedlieu | 1,632,354 | Politician | 0.34 | 2.39 | 0.81 |
| RepMaxineWaters | 1,629,426 | Politician | 0.63 | 2.88 | 1.82 |
| MaryLTrump | 1,602,341 | Commentator | 0.58 | 0.27 | 0.15 |
| RBReich | 1,579,444 | Influencer | 0.54 | 0.29 | 0.15 |
| TheRickWilson | 1,554,991 | Official | 0.22 | 0.68 | 0.15 |
| SenGillibrand | 1,550,941 | Politician | 0.63 | 1.05 | 0.66 |
| SenFeinstein | 1,453,136 | Politician | 0.57 | 1.19 | 0.67 |
| davidaxelrod | 1,445,139 | Official | 0.25 | 1.44 | 0.37 |
| RashidaTlaib | 1,435,156 | Politician | 0.71 | 0.20 | 0.14 |
| RepSwalwell | 1,429,334 | Politician | 0.6 | 1.81 | 1.08 |
| AnnCurry | 1,410,240 | Journalist | 0.0 | 0.22 | 0.00 |
| jrpsaki | 1,389,015 | Commentator | 0.14 | 2.56 | 0.35 |
| tribelaw | 1,365,362 | Academic | 0.36 | 1.18 | 0.42 |
| FiveThirtyEight | 1,305,785 | Media | 0.05 | 2.02 | 0.11 |
| hasanthehun | 1,295,320 | Influencer | 0.29 | 1.86 | 0.53 |
| Ilhan | 1,289,923 | Politician | 0.65 | 0.14 | 0.09 |
| SenateDems | 1,278,787 | Gov | 0.0 | 0.08 | 0.00 |
| RepKatiePorter | 1,263,709 | Politician | 0.5 | 0.75 | 0.38 |
| thenation | 1,251,189 | Media | 0.18 | 0.08 | 0.01 |
| RexChapman | 1,243,909 | Influencer | 0.38 | 0.11 | 0.04 |
| ForeignPolicy | 1,237,340 | Media | 0.04 | 0.18 | 0.01 |

*Table S3: Outgroup animosity for left-leaning influencers*

| Screen Name | Followers | Type | Ingroup Animosity | % Tweets | Total Contribution |
|---|---|---|---|---|---|
| BarackObama | 133,304,210 | Politician | -0.47 | 0.89 | -0.42 |
| nytimes | 54,986,253 | Media | 0.0 | 0.91 | 0.00 |
| HillaryClinton | 31,595,284 | Politician | -0.3 | 0.64 | -0.19 |
| Reuters | 25,757,772 | Media | -0.03 | 0.84 | -0.02 |
| washingtonpost | 20,040,736 | Media | 0.02 | 1.32 | 0.03 |
| Forbes | 18,780,197 | Media | 0.03 | 0.62 | 0.02 |
| BernieSanders | 15,552,053 | Politician | -0.22 | 1.63 | -0.36 |
| johnlegend | 13,727,933 | Singer | -0.15 | 0.21 | -0.03 |
| AOC | 13,443,351 | Politician | 0.02 | 0.48 | 0.01 |
| SenSanders | 12,502,426 | Politician | -0.04 | 0.71 | -0.03 |
| TheDailyShow | 9,512,239 | Media | 0.11 | 0.67 | 0.07 |
| NBCNews | 9,416,420 | Media | 0.03 | 1.74 | 0.05 |
| NewYorker | 8,972,964 | Media | -0.03 | 0.68 | -0.02 |
| CBSNews | 8,916,876 | Media | 0.01 | 1.53 | 0.01 |
| MarkRuffalo | 8,364,416 | Actor | 0.2 | 0.60 | 0.12 |
| SpeakerPelosi | 8,065,407 | Politician | -0.54 | 3.14 | -1.71 |
| chelseahandler | 7,719,580 | Actor | 0.11 | 0.58 | 0.06 |
| BDUTT | 7,145,818 | Journalist | -0.19 | 0.22 | -0.04 |
| SenWarren | 7,033,561 | Politician | -0.08 | 2.04 | -0.16 |
| DeptofDefense | 6,526,663 | Gov | -0.06 | 0.18 | -0.01 |
| ewarren | 5,894,853 | Politician | -0.34 | 2.26 | -0.77 |
| MMFlint | 5,883,969 | Journalist | 0.13 | 1.84 | 0.23 |
| CNBC | 5,092,127 | Media | 0.0 | 0.47 | 0.00 |
| MSNBC | 4,992,529 | Media | 0.14 | 2.52 | 0.36 |
| USATODAY | 4,955,894 | Media | -0.01 | 0.79 | -0.01 |
| CoryBooker | 4,839,236 | Politician | -0.5 | 0.43 | -0.21 |
| VanityFair | 4,770,643 | Media | 0.17 | 0.50 | 0.09 |
| politico | 4,642,306 | Media | 0.11 | 0.91 | 0.10 |
| thehill | 4,470,046 | Media | -0.03 | 1.71 | -0.05 |
| jorgeramosnews | 3,797,986 | Journalist | 0.03 | 0.31 | 0.01 |
| Jemima_Khan | 3,758,111 | Journalist | -0.06 | 0.18 | -0.01 |
| PeteButtigieg | 3,677,799 | Politician | -0.44 | 0.80 | -0.35 |
| Newsweek | 3,596,708 | Media | 0.07 | 1.88 | 0.13 |
| Alyssa_Milano | 3,541,392 | Actor | -0.07 | 0.30 | -0.02 |
| NateSilver538 | 3,479,183 | Journalist | 0.04 | 3.34 | 0.12 |
| GeorgeTakei | 3,456,722 | Actor | -0.73 | 0.12 | -0.09 |
| SenSchumer | 3,437,691 | Politician | -0.24 | 0.63 | -0.15 |
| JohnKerry | 3,363,459 | Politician | -0.7 | 0.76 | -0.53 |
| RepAdamSchiff | 3,250,357 | Politician | 0.37 | 0.37 | 0.14 |
| amanpour | 3,198,734 | Journalist | 0.02 | 2.74 | 0.05 |
| jaketapper | 3,184,663 | Journalist | 0.14 | 0.22 | 0.03 |
| IlhanMN | 3,000,249 | Politician | -0.07 | 0.32 | -0.02 |
| staceyabrams | 2,882,261 | Politician | -0.47 | 0.34 | -0.16 |
| ChelseaClinton | 2,858,009 | Commentator | -0.33 | 0.67 | -0.22 |
| Lawrence | 2,837,029 | Other | 0.03 | 0.36 | 0.01 |
| carlaangola | 2,835,597 | Journalist | 0.0 | 0.90 | 0.00 |
| marwilliamson | 2,768,425 | Politician | 0.17 | 0.62 | 0.11 |
| JoeNBC | 2,765,504 | Journalist | 0.12 | 1.40 | 0.17 |
| ariannahuff | 2,755,868 | Journalist | 0.0 | 0.32 | 0.00 |
| nowthisnews | 2,708,524 | Media | 0.01 | 2.50 | 0.03 |
| DanRather | 2,707,248 | Journalist | 0.1 | 0.42 | 0.04 |
| MikeBloomberg | 2,674,989 | Politician | -0.52 | 0.67 | -0.35 |
| ezraklein | 2,656,110 | Journalist | 0.09 | 2.57 | 0.23 |
| alroker | 2,412,702 | Journalist | -0.11 | 0.20 | -0.02 |
| chrislhayes | 2,402,852 | Journalist | 0.11 | 0.59 | 0.06 |
| GStephanopoulos | 2,374,523 | Journalist | 0.01 | 2.92 | 0.04 |
| kingsthings | 2,327,966 | Journalist | -0.34 | 0.31 | -0.11 |
| billyeichner | 2,318,140 | Actor | -0.15 | 0.59 | -0.09 |
| Acosta | 2,290,501 | Journalist | 0.04 | 0.76 | 0.03 |
| PBS | 2,241,371 | Media | 0.08 | 0.27 | 0.02 |
| mariashriver | 2,193,577 | Journalist | -0.26 | 0.75 | -0.19 |
| TheAtlantic | 2,120,662 | Media | -0.02 | 0.49 | -0.01 |
| TulsiGabbard | 2,110,579 | Politician | 0.75 | 0.60 | 0.45 |
| chucktodd | 2,060,309 | Journalist | 0.1 | 0.78 | 0.07 |
| JoyAnnReid | 2,057,213 | Journalist | 0.1 | 0.62 | 0.06 |
| GavinNewsom | 2,052,614 | Politician | -0.57 | 0.15 | -0.09 |
| amyklobuchar | 2,035,552 | Politician | -0.38 | 1.40 | -0.53 |
| ggreenwald | 1,999,391 | Journalist | 0.23 | 2.98 | 0.68 |
| ananavarro | 1,993,938 | Commentator | -0.12 | 0.86 | -0.11 |
| NickKristof | 1,963,891 | Journalist | 0.14 | 0.23 | 0.03 |
| mitchellreports | 1,928,460 | Journalist | 0.0 | 4.60 | -0.01 |
| NYMag | 1,793,859 | Media | 0.2 | 0.79 | 0.16 |
| PreetBharara | 1,775,705 | Official | -0.15 | 0.42 | -0.06 |
| janemarielynch | 1,757,952 | Actor | 0.5 | 0.17 | 0.09 |
| AndrewYang | 1,726,576 | Politician | 0.13 | 0.57 | 0.07 |
| wolfblitzer | 1,725,899 | Journalist | 0.0 | 2.97 | 0.00 |
| VOANews | 1,704,368 | Media | 0.03 | 0.31 | 0.01 |
| katiecouric | 1,699,996 | Journalist | 0.08 | 0.53 | 0.04 |
| maggieNYT | 1,663,152 | Journalist | 0.07 | 0.32 | 0.02 |
| tedlieu | 1,632,354 | Politician | 0.01 | 1.01 | 0.01 |
| RepMaxineWaters | 1,629,407 | Politician | -0.05 | 1.41 | -0.06 |
| MaryLTrump | 1,602,341 | Commentator | 0.12 | 0.17 | 0.02 |
| NYCMayor | 1,598,840 | Gov | -0.75 | 0.13 | -0.10 |
| SenGillibrand | 1,550,941 | Politician | -0.42 | 1.21 | -0.50 |
| SenFeinstein | 1,453,136 | Politician | -0.56 | 2.52 | -1.41 |
| davidaxelrod | 1,445,139 | Official | 0.01 | 0.81 | 0.01 |
| RashidaTlaib | 1,435,156 | Politician | 0.0 | 0.39 | 0.00 |
| RepSwalwell | 1,429,334 | Politician | 0.16 | 0.52 | 0.09 |
| AnnCurry | 1,410,240 | Journalist | 0.01 | 1.38 | 0.01 |
| jrpsaki | 1,389,015 | Commentator | -0.17 | 5.74 | -0.95 |
| tribelaw | 1,365,362 | Academic | -0.07 | 0.15 | -0.01 |
| SinghLions | 1,341,022 | Influencer | -0.31 | 0.55 | -0.17 |
| HouseDemocrats | 1,308,240 | Gov | -0.38 | 0.36 | -0.14 |
| FiveThirtyEight | 1,305,785 | Media | -0.01 | 1.84 | -0.02 |
| glamourmag | 1,303,748 | Media | -0.07 | 0.15 | -0.01 |
| hasanthehun | 1,295,320 | Influencer | 0.11 | 1.30 | 0.15 |
| Ilhan | 1,289,923 | Politician | -0.21 | 0.35 | -0.07 |

| SenateDems | 1,278,787 | Gov | -0.03 | 0.39 | -0.01 |
| orengo_james | 1,275,386 | Politician | -0.25 | 0.13 | -0.03 |
| RepKatiePorter | 1,263,709 | Politician | 0.08 | 0.54 | 0.04 |

*Table S4: Negative ingroup affect for left-leaning influencers*

## S2 Data

<u>Total number of comments, tweets, and authors</u>

We show the total number of posts collected, the subset which contain a reference to a political entity (one of the 215 in our list) and the total number of authors of these posts.

| **Group** | **Total Posts** | **Political Posts** | **Total Authors** |
|---|---|---|---|
| Twitter | 988,344,411 | 66,791,701 | 1,188,080 |
| Politicians | 2,197,330 | 197,295 | 1,361 |
| Journalist | 2,129,798 | 442,967 | 1,165 |
| Media | 251,065 | 33,503 | 93 |
| Reddit | 1,514,559,439 | 29,513,175 | 467,438 |

*Table S5: Total number of tweets, comments, and authors*

| Year | Value |
|------|-------|
| 2010 | 116 |
| 2011 | 118 |
| 2012 | 354 |
| 2013 | 283 |
| 2014 | 360 |
| 2015 | 369 |
| 2016 | 362 |
| 2017 | 343 |
| 2018 | 350 |
| 2019 | 374 |
| 2020 | 300 |
| 2021 | 381 |
| 2022 | 353 |
| 2023 | 155 |

*Table S6: Number of data points in training set by year*

S2.2 Partisan Scores on Reddit and Twitter

We show below here the distribution of scores measuring the partisanship of the Twitter and Reddit users in our sample computed from the method from (21). These scores were computed based on the following patterns of political elite on Twitter and on subreddit commenting patterns on Reddit. Users with scores greater than zero were classified as left-leaning, while users below -1.2 were classified as right-leaning.

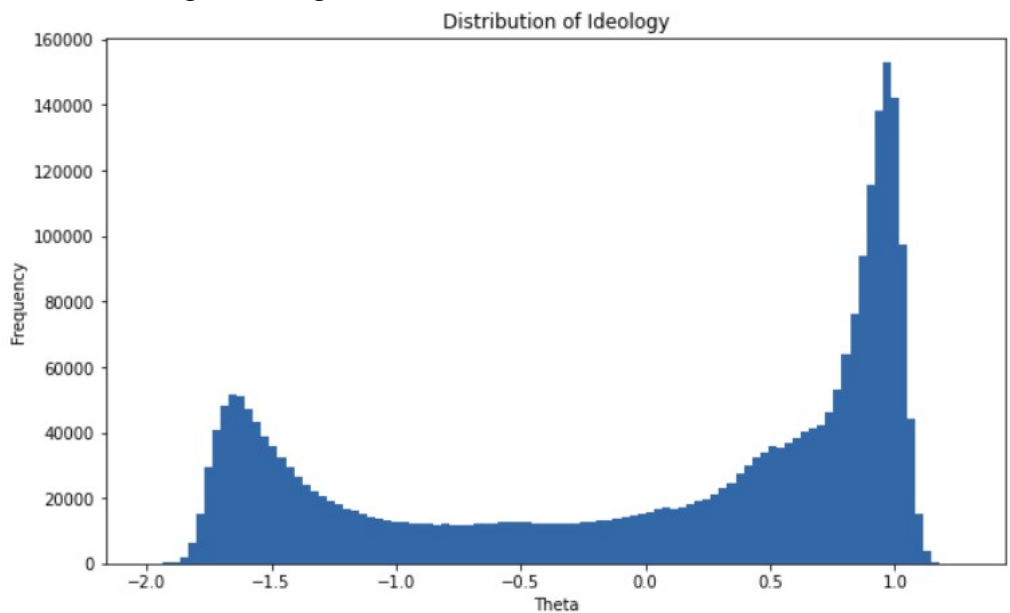

*Figure S1: Distribution of Twitter partisan scores*

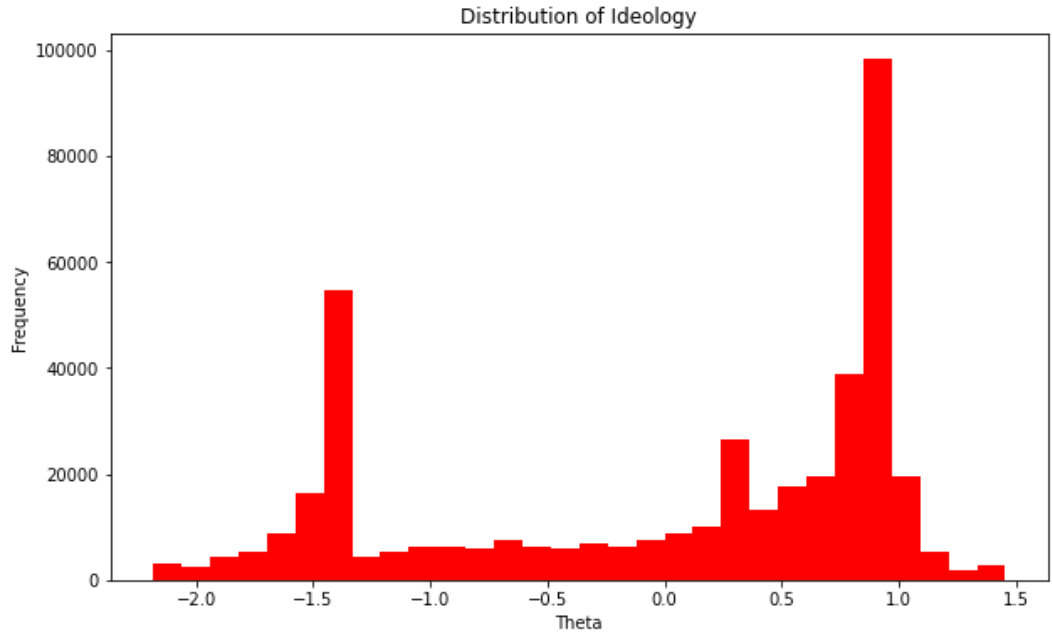

*Figure S2: Distribution of Reddit partisan scores*

S2.3 Subreddits

These were the subreddits we used to find "politically active" users and infer their political orientation using Barbera's method (21) on Reddit.

| Left | Right |
|------|-------|
| AntiTrumpAlliance, PutinsB****, socialanarchism, antifascism,<br><br>Delete_The_Donald, justicedemocrats, Egalitarianism, CornbreadLiberals, Drumpf, neoprogs, EnoughHillHate, N***Hunting, TinyTrumps, Trump_Watch, NeverTrump,<br><br>StillSandersForPres, BernTheConvention,<br><br>HillaryForAmerica, S***The_DonaldSays, SandersForPresident, Impeach_Trump, alltheleft, WayOfTheBern,<br><br>Enough_Sanders_Spam, EnoughTrumpSpam, liberalgunowners, hillaryclinton, F***thealtright, democrats, esist, centerleftpolitics, Political_Revolution,<br><br>MarchAgainstTrump, BlueMidterm2018,<br><br>Trumpgret, OurPresident,<br><br>Kossacks_for_Sanders, Liberal, TrumpCriticizesTrump, progressive, anarcho_socialism, chapotraphouse2, ChapoTrapHouse, AOC, obama, bidenbro, TinyhandsInc | randpaul, BenCarson, HillaryMeltdown, new_right, PresidentKushner, TheRightBoycott, AltRightChristian,<br><br>JohnKasich, Sorosforprison, Romney,<br><br>BobbyJindal, EnoughObamaSpam, TedCruz,<br><br>CarlyFiorina, TeaParty, MartinOMalley, ChrisChristie, presidentbannon, Mr_Trump, republicans, RickSantorum,<br><br>EnoughPaulSpam, trump, HillaryForPrison, ronpaul, tucker_carlson, The_Donald,<br><br>ConservativesOnly, Conservative,<br><br>Republican, conservatives |

*Table S7: Subreddits used to find politically active Reddit users and infer their partisanship*

After collecting all comments left by users in our sample, we collect the political comments left by those users from 38,818 subreddits. However, the distribution of comments by subreddits follows a power law distribution, with a few large subreddits dominating political discussion (such as r/politics and r/The_Donald). Below, we show the top twenty subreddits by number of political comments for left and right-leaning users and compute the percentage of all political comments contributed by users in that subreddit. We also compute the percentage of comments that contain a reference to a political entity out of all the comments in the subreddit to get a sense of how "political" the discussion in the subreddit is. In seventeen percent of subreddits in our sample, more than five percent of the comments contain a reference to a political entity (indicating high levels of political discussion in those subreddits). Twenty-two percent of subreddits contain between one and five percent political comments, and sixty percent contain less than one percent (indicating low levels of political discussion).

| Subreddit | # Political Comments | Total Comments | # Political Comments / Total Comments | % All Political Comments |
|---|---|---|---|---|
| politics | 7,215,371 | 68,382,543 | 11% | 27% |
| The_Donald | 2,509,869 | 33,738,019 | 7% | 9% |
| worldnews | 812,462 | 24,868,187 | 3% | 3% |
| Conservative | 682,694 | 7,376,589 | 9% | 3% |
| news | 642,870 | 21,730,984 | 3% | 2% |
| AskReddit | 552,080 | 78,446,702 | 1% | 2% |
| PoliticalHumor | 544,022 | 5,914,005 | 9% | 2% |
| SandersForPresident | 484,457 | 5,564,400 | 9% | 2% |
| neoliberal | 483,628 | 8,674,826 | 6% | 2% |
| ChapoTrapHouse | 383,323 | 7,114,333 | 5% | 1% |
| pics | 317,885 | 18,044,027 | 2% | 1% |
| conspiracy | 291,480 | 7,685,055 | 4% | 1% |
| PoliticalDiscussion | 282,094 | 2,885,286 | 10% | 1% |
| Enough_Sanders_Spam | 229,407 | 2,297,626 | 10% | 1% |
| WayOfTheBern | 193,312 | 2,556,913 | 8% | 1% |
| Libertarian | 179,639 | 4,169,073 | 4% | 1% |
| EnoughTrumpSpam | 169,127 | 1,600,278 | 11% | 1% |
| PublicFreakout | 152,042 | 6,337,561 | 2% | 1% |
| PoliticalCompassMemes | 139,804 | 4,084,629 | 3% | 1% |
| hillaryclinton | 138,628 | 1,155,687 | 12% | 1% |

*Table S8: Top 20 subreddits by number of political comments on Reddit*

S2.4 List of political keywords

We show below the list of political keywords use to both filter the data to a set of "political content" and determine the entity being referenced in the text. We filtered content that contained a reference to both a left and right-leaning entity.

| Pol Orient | Type | Name |
|---|---|---|
| L | group | antifa, dems, liberals, liberal, leftist, leftists, libtard, libtards, libs, democratic party,left-wing, dem, left wingers, left winger, libruls, progressive, progressives, democrat, democrats |
| L | name | obama, clinton, biden, sanders, pelosi, jimmy carter, al gore, warren, michael bloomberg, john kerry,kamala harris, aoc, ocasio-cortez, charles schumer, cory booker, andrew cuomo, madeleine albright, john edwards, maxine waters, anthony weiner,michael dukakis, julian castro, dianne feinstein, al franken, joe biden, bernie sanders, nancy pelosi, elizabeth warren, chuck schumer, adam schiff,ilhan omar, gavin newsom, amy klobuchar, tulsi gabbard, ted lieu, kirsten gillibrand, rashida tlaib, eric swalwell, katie porter, john lewis,chris murphy, ayanna pressley, claire mccaskill, beto orourke, martin omalley |
| L | twitter | @barackobama, @hillaryclinton, @billclinton, @joebiden, @berniesanders, @speakerpelosi, @algore, @ewarren, @mikebloomberg, @johnkerry,@senkamalaharris, @repaoc, @senschumer, @corybooker, @nygovcuomo, @madeleine, @juliancastro, @diannefeinstein, @alfranken,@petebuttigieg, @ilhan, @nycmayor, @adamschiff, @sensanders, @senwarren, @repadamschiff, @ilhanmn, @gavinnewsom, @amyklobuchar, @tulsigabbard, @tedlieu, @sengillibrand, @senfeinstein, @rashidatlaib, @repswalwell, @repkatieporter, @repjohnlewis, @chrismurphyct, @ayannapressley,@teampelosi, @clairecmc |
| R | group | trumpster, trumpeters, republican party, far right, right wing, far-right, republican, republicans, alt-right, repubs,maga, repub, right-wing, right winger, right wingers, conservative, conservatives, alt\-right, gop |
| R | name | trump, schwarzenegger, pence, palin, romney, cheney, ted cruz, jeb bush, ben carson, paul ryan,mcconnell, gingrich, marco rubio, henry kissinger, bob dole, mike huckabee, chris christie, rand paul, dan quayle, lindsey graham,jared kushner, jeff sessions, donald rumsfeld, rick santorum, ron desantis, jim jordan, john mccain, mitt romney, matt gaetz, kevin mccarthy,marjorie taylor greene, trey gowdy, john boehner, devin nunes, josh hawley, greg abbott, marsha blackburn, chuck grassley, dan crenshaw, tim scott,adam kinzinger, jeff flake, mike lee, tom cotton |
| R | twitter | @realdonaldtrump, @schwarzenegger, @sarahpalinusa, @mittromney, @sentedcruz, @georgehwbush, @secretarycarson, @speakerryan, @senatemajldr, @newtgingrich, @marcorubio, @senatordole, @govmikehuckabee, @chrischristie, @randpaul, @lindseygrahamsc, @jaredkushner, @jeffsessions, @rumsfeldoffice, @ricksantorum,@secretaryperry, @barrygoldwater, @speakerboehner, @nikkihaley, @tedcruz, @govrondesantis, @jim_jordan, @senjohnmccain, @repmattgaetz, @gopleader,@mattgaetz, @repmtg, @rondesantisfl, @tgowdysc, @devinnunes, @hawleymo, @gregabbott_tx, @marshablackburn, @chuckgrassley, @repdancrenshaw,@senatortimscott, @adamkinzinger, @jeffflake, @senmikelee, @sentomcotton |

*Table S9: Political groups and entities used to filter data and identify whether the content refers to a left or right-leaning entity*

S2.5 Polarizing Speech Examples

We show below different types of polarizing speech we labeled. These were eventually combined into one category due to high subjectivity in which category the speech would fall into (so if content was coded as either affectively polarizing or partisan, we considered it to be polarizing speech).

*Affectively Polarizing Speech*

Affectively polarizing speech encompasses ad-hominem attacks as well as comments that express distrust that the group/party is doing what's right (for themselves and the country). This is grounded in the feeling thermometer and trust questions that surveys use, as well as the general definition of affective polarization that is "dislike and distrust to those from the out party".

| Comment | Affect |
|---------|--------|
| i can't stand you **@aoc** your green deal sucks like you suck as a politician | Negative |
| i'm not sure people grasp how far to the right **gop** is going and has gone. i'm not sure people understand that their freedom, democracy is on the line. | Negative |
| @wikileaks @hillaryclinton haha!!!!! go wikileaks! good thing **hillary** can laugh at the mess she's made!" | Negative |
| at least they're admitting that blm was burning towns down and committing violence because **obama** told them it was | Negative |
| yes, the **gop** is a lawless party led by an as yet-unindicted criminal. | Negative |
| @premr that was credit card reform measure signed into law by president **obama**.... | Neutral |
| **obama** was not the most leftist president in history. contextually (contrasting the politics of the time they were in office), the most leftist president in history was probably thomas jefferson. | Neutral |
| i don't see why **trump** would pardon him. has he said anything on the matter? this isn't even related to anything about trump either. | Neutral |
| Happy birthday **joe biden**! | Positive |
| i absolutely adore **maxine waters**. she doesn't pull any punches. | Positive |
| **conservative** women are level headed and actually use their head when they process information | Positive |

*Table S10: Examples of affective polarizing speech*

*Partisan Speech*

These comments are not necessarily ad-hominem attacks. However, they often blame negative outcomes on a group/person or criticize their governance. The comment explicitly blames/praises a certain group/person or clearly implying a positive/negative sentiment and often attempts to criticize their competence.

| Comment | Affect |
|---|---|
| there is one major political party in the world that doesn't believe in climate change. **the republican party**. as long as the party that continue to deny the problem is in power, we won't be able to make the changes necessary to solve it. #climatechange | Negative |
| the way to keep guns out of our schools is not to put guns in them. **trumps** suggestion today is absolutely unfathomable and must be stopped. | Negative |
| if **gop** senators spent more time doing their job, they'd realize america is broke and a health entitlement program is stupid! | Negative |
| the fact that the **biden administration** is spending $86 million to pay for hotel rooms for illegal immigrants is absurd. let's take care of our homeless veterans and hurting americans first. | Negative |
| over and over again, **obama's** foreign policy actions have proven reckless and not in our nation's best interest. | Negative |
| i mean, it makes sense. they are a publicly traded company subject to sec and congressional oversight. since all 3 houses are controlled by **liberals**, only makes sense to silence the opposing party. | Neutral |
| i don't refute that point. my point is that **obama** has taken more total below-the-belt punches than his predecessors, and that you'd have to be an, extremist, not a moderate, to believe/claim otherwise. | Neutral |
| not sure what to think of this..."hacker group claims to have **romney's** tax returns" http.//t.co/rdlbsrrf via @mashable | Neutral |
| **sarah palin** suggests she will get involved in kentucky senate race #politics | Neutral |
| thanks **@realdonaldtrump** for approving sd's disaster declaration for severe storms and flooding. this will help local units of | Positive |
| government & certain non-profit organizations to rebuild critical infrastructure and facilities that have been severely damaged. | |
| so glad that @debrarodman73, @vvfordelegate, @adamsfordel, @pwcdanica, @chrishurstva and so many more **progressive** leaders are headed to the house of delegates! @vahousedems | Positive |
| @ingrahamangle i can't think of a more important reason to support president **trump** in his re-election. | Positive |

*Table S11: Examples of partisan speech*

**S3 Models**

S3.1 Polarization Model

We fine tune the *cardiffnlp/twitter-roberta-base-sentiment, bert-base-uncased,* and *gpt2* classifiers from Huggingface on our 4399 labeled examples using five-fold cross validation to get a robust estimate of out of sample performance. For data cleaning, we keep all comments that are under 240 characters and greater than ten characters and lowercase all text. We chose to do this since we are comparing text across different platforms and users, and we wanted to standardize the data as much as possible. However, we acknowledge that this may lead to lost information for the model. We fine tune all the models for three epochs, with a batch size of eight and performance evaluation every 200 gradient updates. The only hyperparameter tuning we do is on the learning rate, which we adjust to 1.25e-5 for *bertbase-uncased* and *gpt2* and 2.5e-5 for *cardiffnlp/twitter-roberta-base-sentiment* to ensure smooth training. We then use the weights of the model with the lowest evaluation cross entropy loss. We see that fine-tuning the base models allows us to reach the best performance across our outcomes, as measured by the macro and weighted F1 scores.

*S3.1.2 Main Polarization Classifier Performance*

Below, we show the classification report from our classifier used in the main text of the manuscript.

| Class | Precision | Recall | F1-Score | Support |
|---|---|---|---|---|
| Negative Affect | 0.63 | 0.62 | 0.63 | 866 |
| Neutral Affect | 0.88 | 0.87 | 0.88 | 3315 |
| Positive Affect | 0.54 | 0.61 | 0.57 | 218 |
| Accuracy | | | 0.81 | 4399 |
| Macro Avg | 0.68 | 0.70 | 0.69 | 4399 |
| Weighted Avg | 0.81 | 0.81 | 0.81 | 4399 |

*Table S12: The classification report from our main model to measure polarizing discourse across all data*

*S3.1.2 Performance across Time and User-types*

We also estimate the model performance separately across the political orientations of users, the political entity reference groups, user types, and time periods. to ensure that any biases in performance would not be materially impacting our results. We find similar performance when grouping by these various subsets of data, though the model is slightly better at predicting polarization of content by left-leaning users over right-leaning users as well as politicians and politically active users over journalist. However, we also note that we run several robustness checks such as varying the base models and political keywords and find that the results are consistent.

| Political Orientation | Precision | Recall | F1-Score | Support |
|---|---|---|---|---|
| **Right (R)** | | | | |
| Negative Affect | 0.62 | 0.63 | 0.62 | 489 |
| Neutral Affect | 0.87 | 0.87 | 0.87 | 1748 |
| Positive Affect | 0.52 | 0.55 | 0.53 | 110 |
| Macro Avg | 0.67 | 0.68 | 0.68 | 2347 |
| Weighted Avg | 0.80 | 0.80 | 0.80 | 2347 |
| **Accuracy** | 0.80 | | | |
| **Left (L)** | | | | |
| Negative Affect | 0.63 | 0.62 | 0.63 | 377 |
| Neutral Affect | 0.89 | 0.88 | 0.88 | 1567 |
| Positive Affect | 0.56 | 0.67 | 0.61 | 108 |
| Macro Avg | 0.69 | 0.72 | 0.71 | 2052 |
| Weighted Avg | 0.82 | 0.82 | 0.82 | 2052 |
| **Accuracy** | 0.82 | | | |

*Table S13: Classifier performance across users of different political orientations*

| Reference | Precision | Recall | F1-Score | Support |
|---|---|---|---|---|
| **Left (L)** | | | | |
| Negative Affect | 0.65 | 0.63 | 0.64 | 494 |
| Neutral Affect | 0.88 | 0.88 | 0.88 | 1807 |
| Positive Affect | 0.51 | 0.64 | 0.57 | 106 |
| Macro Avg | 0.68 | 0.71 | 0.70 | 2407 |
| Weighted Avg | 0.82 | 0.81 | 0.82 | 2407 |
| **Accuracy** | 0.81 | | | |
| **Right (R)** | | | | |
| Negative Affect | 0.60 | 0.62 | 0.61 | 372 |
| Neutral Affect | 0.88 | 0.87 | 0.88 | 1508 |
| Positive Affect | 0.58 | 0.57 | 0.57 | 112 |
| Macro Avg | 0.68 | 0.69 | 0.69 | 1992 |
| Weighted Avg | 0.81 | 0.81 | 0.81 | 1992 |
| **Accuracy** | 0.81 | | | |

*Table S14: Classifier performance across content with left versus right-leaning political entities mentioned*

| Year Bin | Precision | Recall | F1-Score | Support |
|---|---|---|---|---|
| **2011-2015** | | | | |
| Negative Affect | 0.65 | 0.59 | 0.62 | 259 |
| Neutral Affect | 0.89 | 0.89 | 0.89 | 1148 |
| Positive Affect | 0.54 | 0.64 | 0.58 | 77 |
| Macro Avg | 0.69 | 0.71 | 0.70 | 1484 |
| Weighted Avg | 0.83 | 0.83 | 0.83 | 1484 |
| **Accuracy** | | | 0.83 | |
| **2016-2020** | | | | |
| Negative Affect | 0.62 | 0.62 | 0.62 | 369 |
| Neutral Affect | 0.86 | 0.87 | 0.87 | 1262 |
| Positive Affect | 0.60 | 0.56 | 0.58 | 98 |
| Macro Avg | 0.69 | 0.68 | 0.69 | 1729 |
| Weighted Avg | 0.80 | 0.80 | 0.80 | 1729 |
| **Accuracy** | | | 0.80 | |
| **2021-2023** | | | | |
| Negative Affect | 0.63 | 0.69 | 0.66 | 211 |
| Neutral Affect | 0.88 | 0.83 | 0.85 | 638 |
| Positive Affect | 0.47 | 0.68 | 0.55 | 40 |
| Macro Avg | 0.66 | 0.73 | 0.69 | 889 |
| Weighted Avg | 0.80 | 0.79 | 0.79 | 889 |
| **Accuracy** | | | 0.79 | |

*Table S15: Classifier performance across different periods of time*

| User Type | Precision | Recall | F1-Score | Support |
|---|---|---|---|---|
| **Journalists** | | | | |
| Negative Affect | 0.49 | 0.57 | 0.53 | 68 |
| Neutral Affect | 0.94 | 0.92 | 0.93 | 579 |
| Positive Affect | 0.43 | 0.55 | 0.48 | 11 |
| Macro Avg | 0.62 | 0.68 | 0.64 | 658 |
| Weighted Avg | 0.88 | 0.87 | 0.88 | 658 |
| **Accuracy** | | | 0.87 | |
| **Politicians** | | | | |
| Negative Affect | 0.70 | 0.67 | 0.69 | 242 |
| Neutral Affect | 0.84 | 0.82 | 0.83 | 728 |
| Positive Affect | 0.59 | 0.69 | 0.64 | 150 |
| Macro Avg | 0.71 | 0.73 | 0.72 | 1120 |
| Weighted Avg | 0.77 | 0.77 | 0.77 | 1120 |
| **Accuracy** | | | 0.77 | |
| **Politically Active Users** | | | | |
| Negative Affect | 0.64 | 0.66 | 0.65 | 381 |
| Neutral Affect | 0.88 | 0.87 | 0.87 | 1208 |
| Positive Affect | 0.43 | 0.47 | 0.45 | 43 |
| Macro Avg | 0.65 | 0.66 | 0.66 | 1632 |
| Weighted Avg | 0.81 | 0.81 | 0.81 | 1632 |
| **Accuracy** | | | 0.81 | |

*Table S16: Classifier performance across different types of users*

### S3.1.2 Performance comparison across Classifiers

To benchmark our main classifier's performance, we also report our models' performance against an untrained BERT sentiment classifier as well as the VADER model, which only uses keywords to create a sentiment score. To ensure our results are not dependent on the base model, we train a base BERT model and base GPT-2 model on our labeled dataset and ensure our results are robust to this choice (see **S7 Robustness Checks for Data Source and Base Model**). Below, we report the results for the main model used in the manuscript (Senti (FT)), as well as the GPT-2 and BERT models used for robustness checks. We compare their performance to a baseline sentiment classifier and VADER, which infers sentiment based on keywords, finding that our fine-tuned models outperform both.

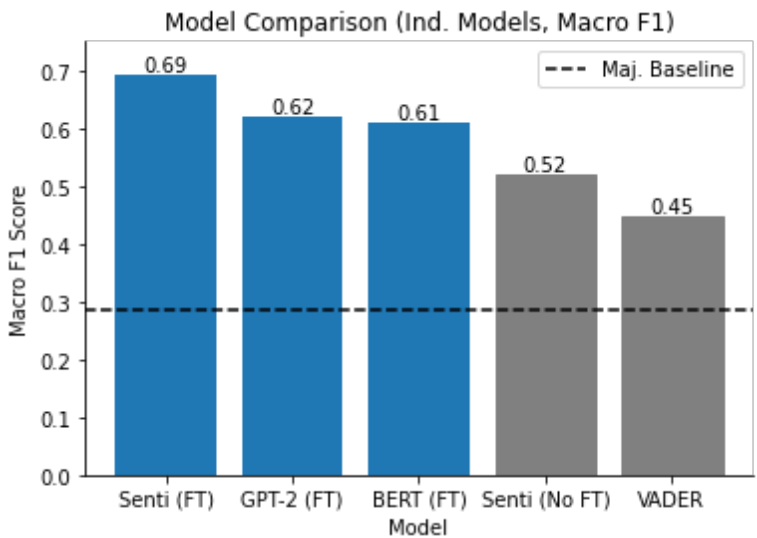

*Figure S3: Macro F1 performance of our main models versus baseline sentiment classifiers*

Additionally, we report the precision and recall breakdown by class for each classifier. We make two observations; first, the untuned sentiment model has high recall but low precision, likely due to the data distribution being different between the pre-training data (predicting sentiment) and the task of predicting polarizing rhethoric. However, after fine-tuning the model we see a more reasonable balance between precision and recall is found that maximizes the F1-score. Additionally, since the fine-tuned sentiment model has better recall on polarizing comments than the original BERT or GPT models, we see that this model is more likely to predict a comment as being polarizing. However, while the models display different sensitivities to predicting whether a comment is polarizing, we show that all our main findings are robust to the choice of model.

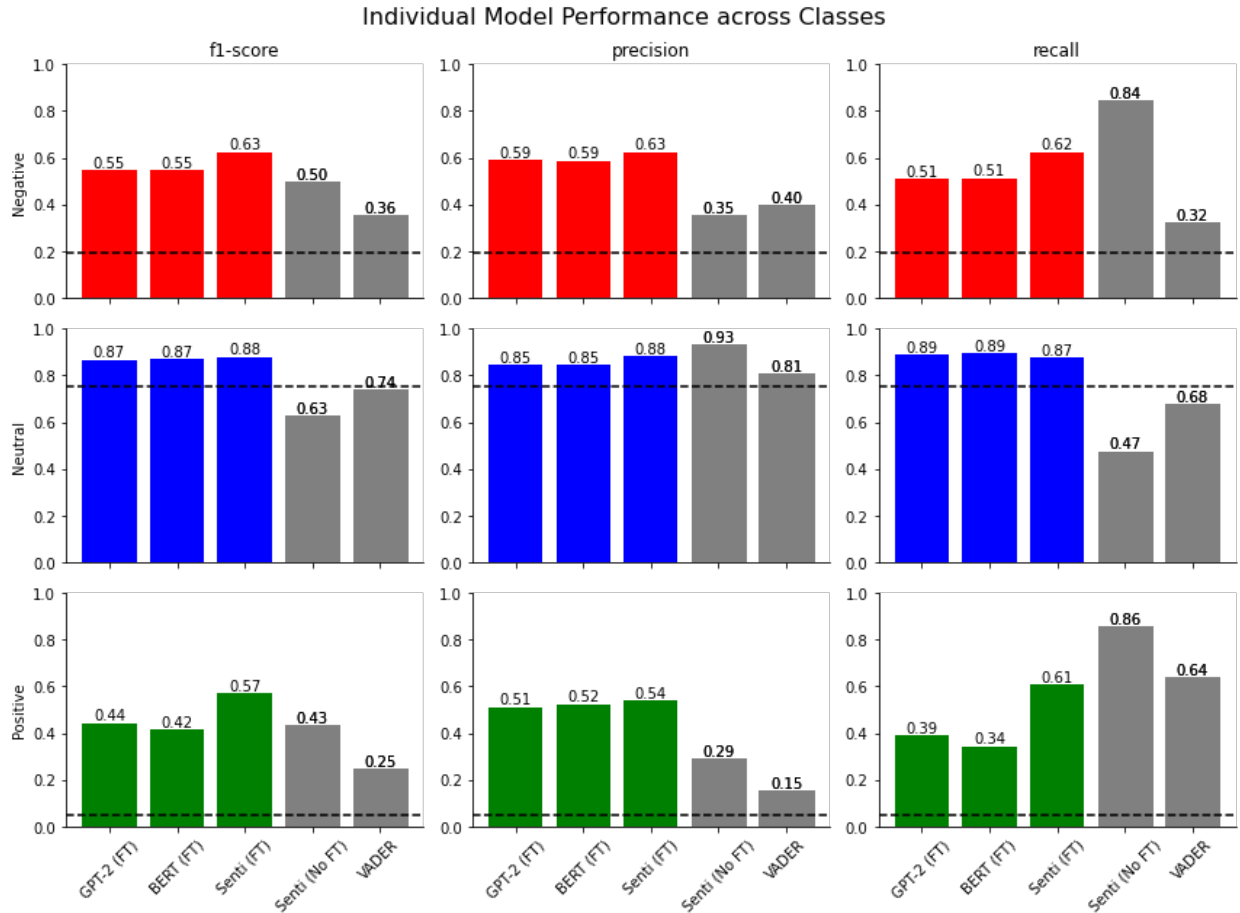

*Figure S4: Precision, recall, and F1 breakdown of our classifiers versus baseline sentiment classifiers*

### S3.2 Topic Model Performance

Although most comments did not contain any specific topic, we can still reasonably classify the topics when they do occur. Our overall accuracy is 91% when including tweets with no topic and 66% when only considering tweets that contained a given topic. Given the large number of topics and their relative sparsity of occurrence, the classifier's performance is modest. However, this seems roughly in line with other work that attempts to use topic classifiers on political social media data (22) .

| Topic | Precision | Recall | F1 | Support |
|---|---|---|---|---|
| None | 0.95 | 0.95 | 0.95 | 0.85 |
| International affairs | 0.65 | 0.70 | 0.67 | 0.03 |
| Economy | 0.60 | 0.62 | 0.61 | 0.02 |
| Health | 0.58 | 0.59 | 0.59 | 0.02 |
| Immigration | 0.77 | 0.78 | 0.78 | 0.02 |
| Civil rights | 0.61 | 0.57 | 0.59 | 0.02 |
| Labor | 0.70 | 0.82 | 0.75 | 0.01 |
| Environment | 0.79 | 0.79 | 0.79 | 0.01 |
| Education | 0.78 | 0.78 | 0.78 | 0.01 |
| Law and crime | 0.55 | 0.27 | 0.36 | 0.01 |
| Social welfare | 0.64 | 0.41 | 0.50 | 0.01 |
| Energy | 0.65 | 0.69 | 0.67 | 0.00 |

*Table S17: Topic model performance*

**S4 Metrics**

S4.1 Alternative Metrics

When assessing the degree of polarizing discourse in online discussions, different measurement approaches can lead to varying interpretations. One alternative to our chosen method involves creating a metric that measures the overall likelihood of a comment displaying politically polarizing rhetoric, relative to all comments—both political and apolitical. In this framework, apolitical comments are treated as "neutral."

However, this approach has several limitations. First, it is highly sensitive to fluctuations in the volume of political content on the platform. For example, an observed increase in polarizing rhetoric may simply reflect a rise in the proportion of political discussion, rather than a genuine increase in polarization. Similarly, differences between groups may reflect how often each group engages in political conversation, rather than differences in how polarizing their content is.

Second, this metric often yields relatively small values because political content makes up only a portion of all online discourse, and polarizing content is only a subset of that. This can make the results harder to interpret. That said, one could argue that this approach offers a more holistic view of platform dynamics: even if both neutral and polarizing political content increase, the overall rise in politically charged rhetoric still signals a shift in the nature of discourse.

Given these tradeoffs, we separate our analysis into two parts: first, we examine how prevalent political content is across different groups and platforms. Then, we explore how our metric behaves when we allow it to vary in tandem with the volume of political content.

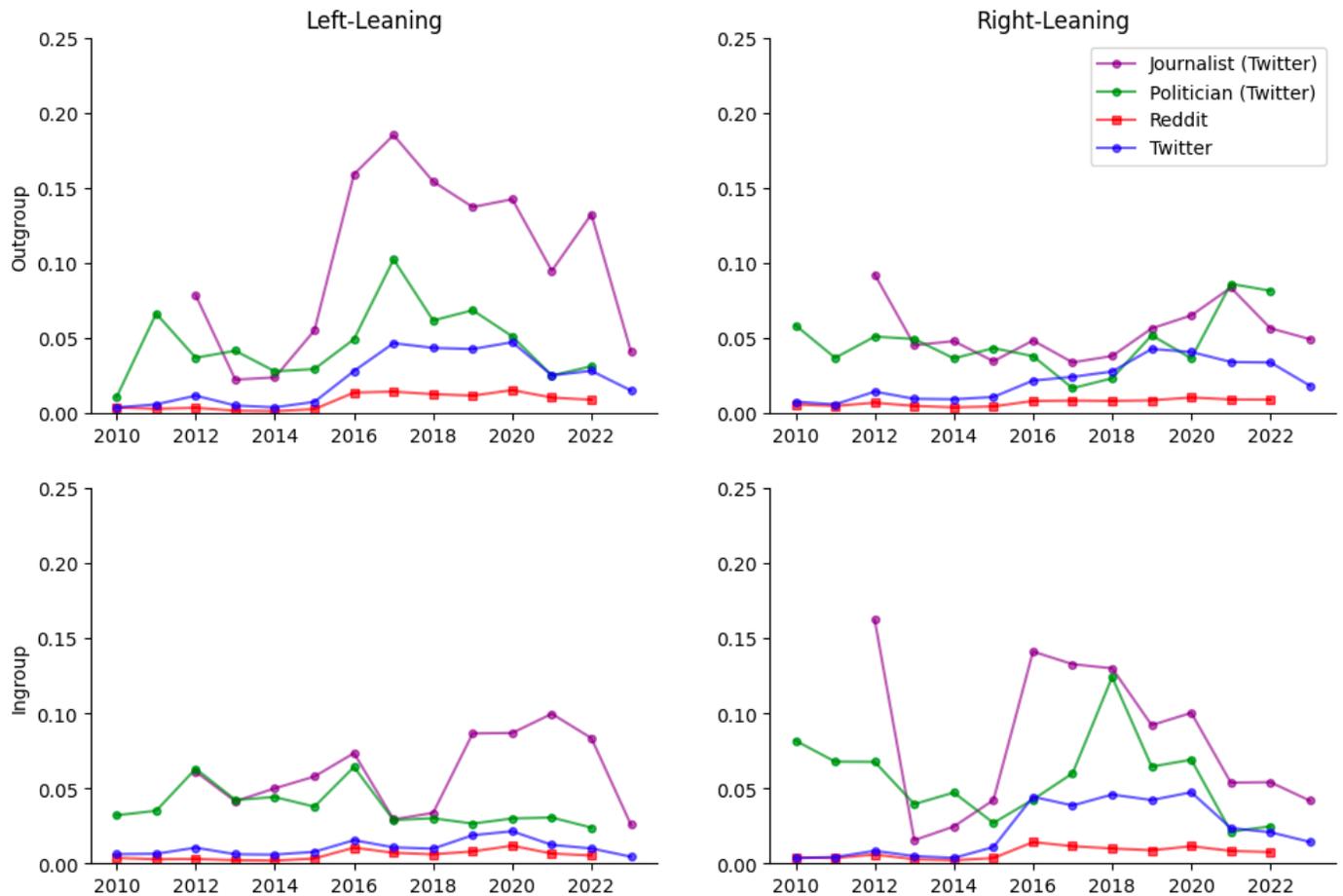

*Figure S5: Percentage of all statements containing a reference to a political group or entity. These are the fraction that we consider "political comments". As we can see, this varies substantially across groups and time.*

**Figure S5** shows the percentage of political comments relative to all comments over time. As expected, journalists and politicians produce the highest proportion of political content in our dataset. We also observe distinct temporal patterns. For instance, left-leaning users showed an increase in references to their political outgroups shortly after the 2016 election—likely in response to Donald Trump's victory—followed by a gradual decline. In contrast, right-leaning politicians peaked in their outgroup mentions during 2021 and 2022. Reddit users, on the other hand, generally exhibit a lower proportion of political speech in their overall commenting behavior.

It's important to stress that our goal here is not to draw definitive conclusions about any group's inherent tendency to engage in political speech. Rather, we aim to illustrate the varying levels of political discourse observed across groups and platforms, and how these variations could affect metrics that are sensitive to such fluctuations.

There are also key limitations to consider. Differences in platform behavior—such as estimating political leaning on Reddit and Twitter using a similar method, despite their distinct user cultures—can influence the observed patterns. Additionally, how we define what counts as "political" may shape these trends.

To reduce the risk of our metric being overly influenced by the total volume of political speech, we instead use a measure that is conditional on the content being political. We also believe this approach offers a more intuitive interpretation. For example, in 2015, about 30% of tweets from right-leaning politicians that mentioned the political left contained negative outgroup sentiment. By 2022, that figure had risen to over 60%. In practical terms, this means that in 2015, if a user saw a tweet from a right-leaning politician about the left, there was a 30% chance it conveyed negative sentiment. By 2022, that probability had nearly doubled.

We present an alternative approach—illustrated in **Figure S6**—that treats apolitical comments as neutral. This method calculates the level of polarizing discourse by taking the difference between the number of negative and positive comments and dividing it by the total number of comments, including both political and apolitical ones.

This analysis reveals two main trends. First, temporal changes in polarizing discourse are largely driven by shifts in the proportion of political content relative to all content. Second, while journalists frequently discuss politics, their rhetoric tends to be less polarized than that of politicians, who consistently show the highest levels of outgroup derogation.

We also find that the rise in polarizing discourse among right-leaning politicians after 2016 is due to both an increase in the proportion of political content and a greater likelihood that this content is polarizing. In contrast, left-leaning users show a decline in political speech over time—a trend that may reflect the heightened volume of tweets about President Trump during and after the 2016 election.

It's important to note the limitations of comparing platforms, especially when our metric is influenced by the volume of political discussion. For example, Reddit appears significantly less polarized than Twitter using this measure, but this may be due to differences in how users were sampled. Many Reddit comments likely occur in apolitical subreddits, reducing the chance of referencing political entities and thereby lowering measured polarizing discourse.

Given these considerations, we see that filtering the data using a set of political keywords helps reduce potential biases introduced by differences in how often various groups or platforms engage in political discussion over time.

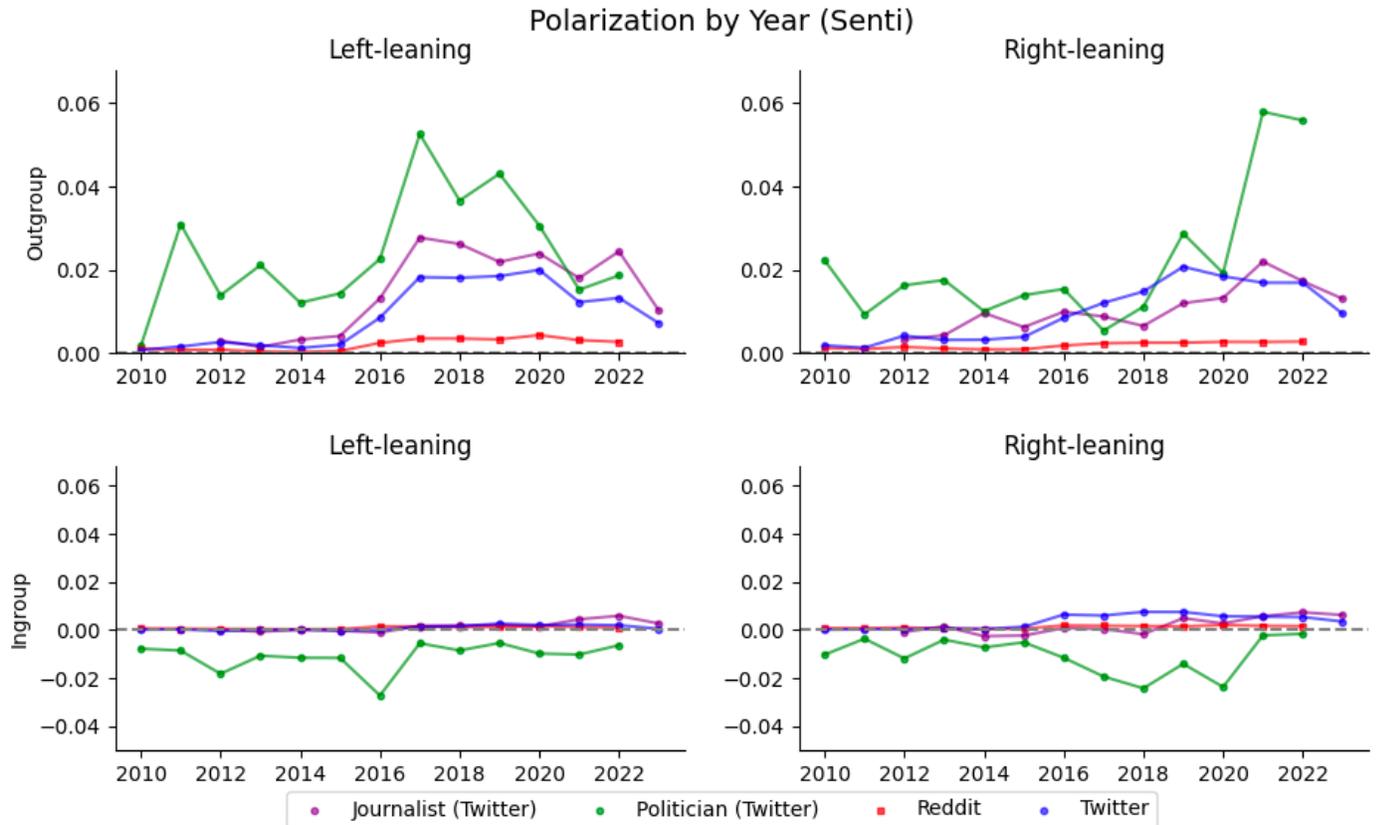

*Figure S6: An alternative measure of polarizing discourse where we compute the difference between the number of negative affect comments and the number of positive affect comments and divide by the total number of comments (political and apolitical). This metric will be sensitive to the overall level of political discourse, and we can see that it closely mirrors trends in the previous figure that capture the propensity to have political content for different user groups and over time.*

Finally, we note that any metric we choose will have imperfections; although we motivate our work with types of measures often used in the political science literature to measure polarization, it is difficult to find a perfect mapping from the motivating idea (e.g., a thermometer scale rating given by survey participants) and the practical implementation.

S4.2 Robustness to Political Keywords

We also test the robustness of our approach to choice of political keywords. To simulate any potential missing keywords, we randomly drop ten percent of the keywords and re-create our main chart five times. We find that all our main results hold; rising outgroup derogation over time, with a significant increase by right-leaning politicians starting in 2016.

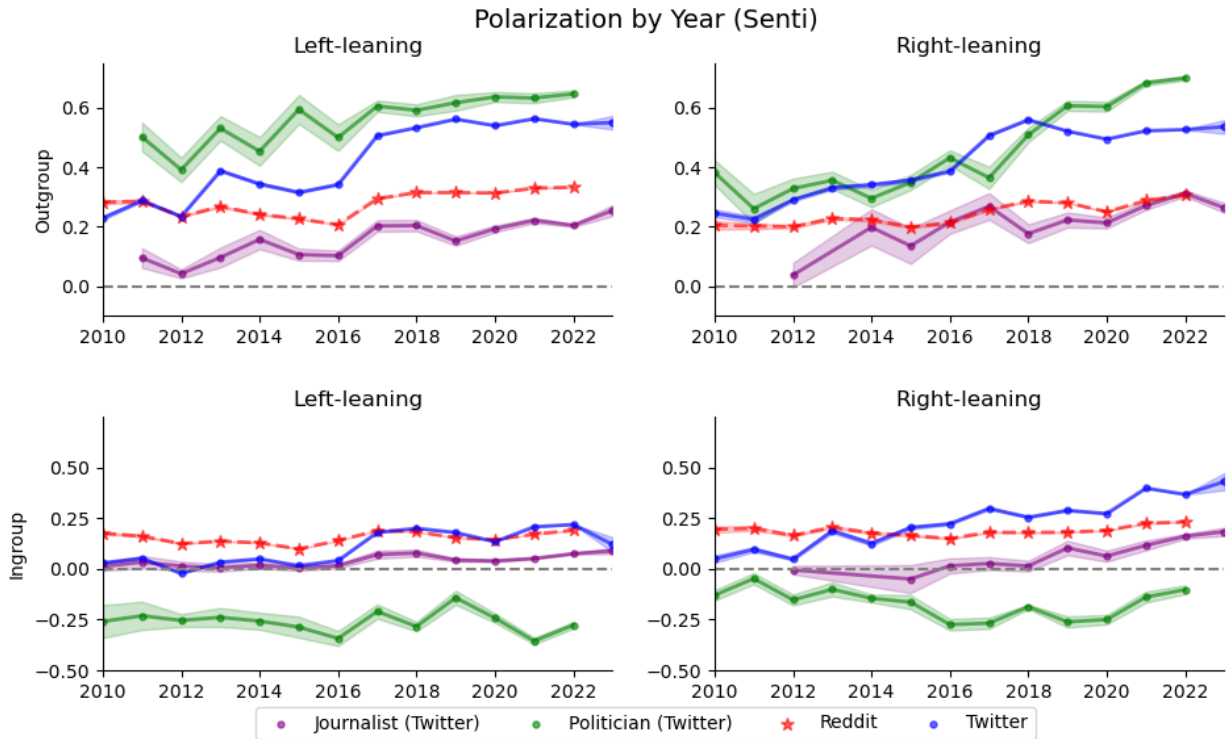

*Figure S7: First run dropping 10% of political keywords and re-creating our main visualization*

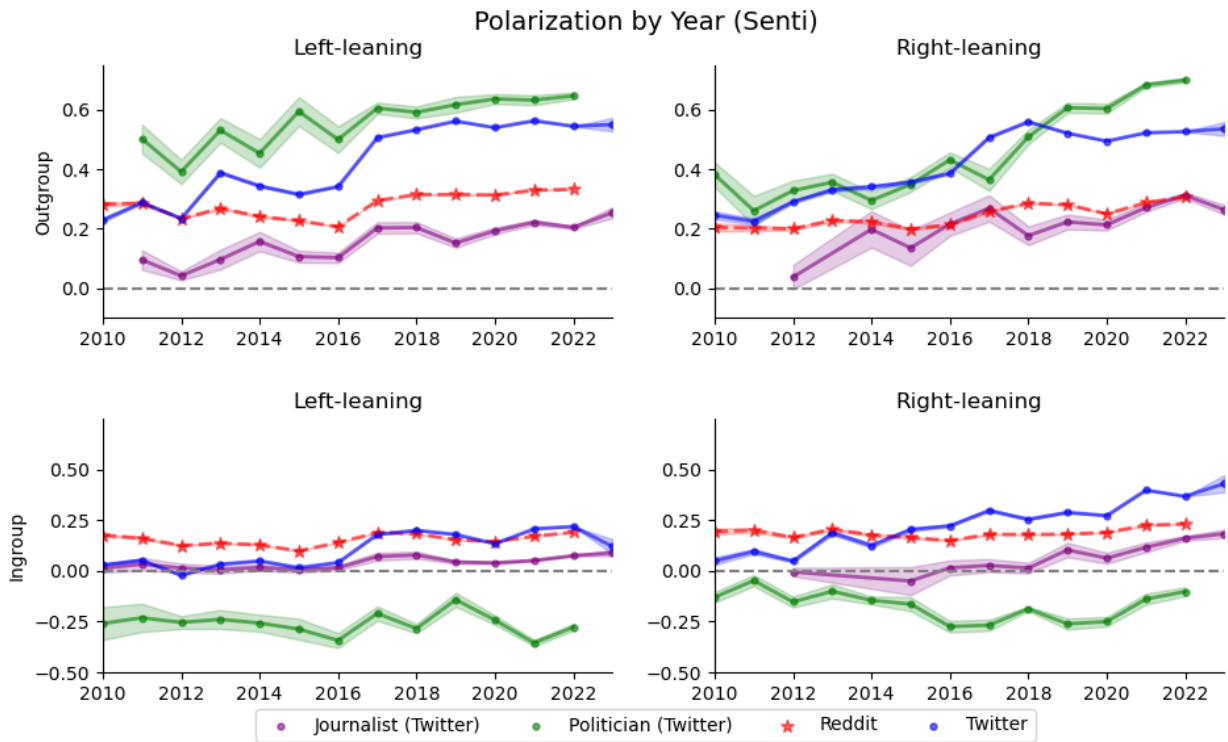

*Figure S8: Second run dropping 10% of political keywords and re-creating our main visualization*

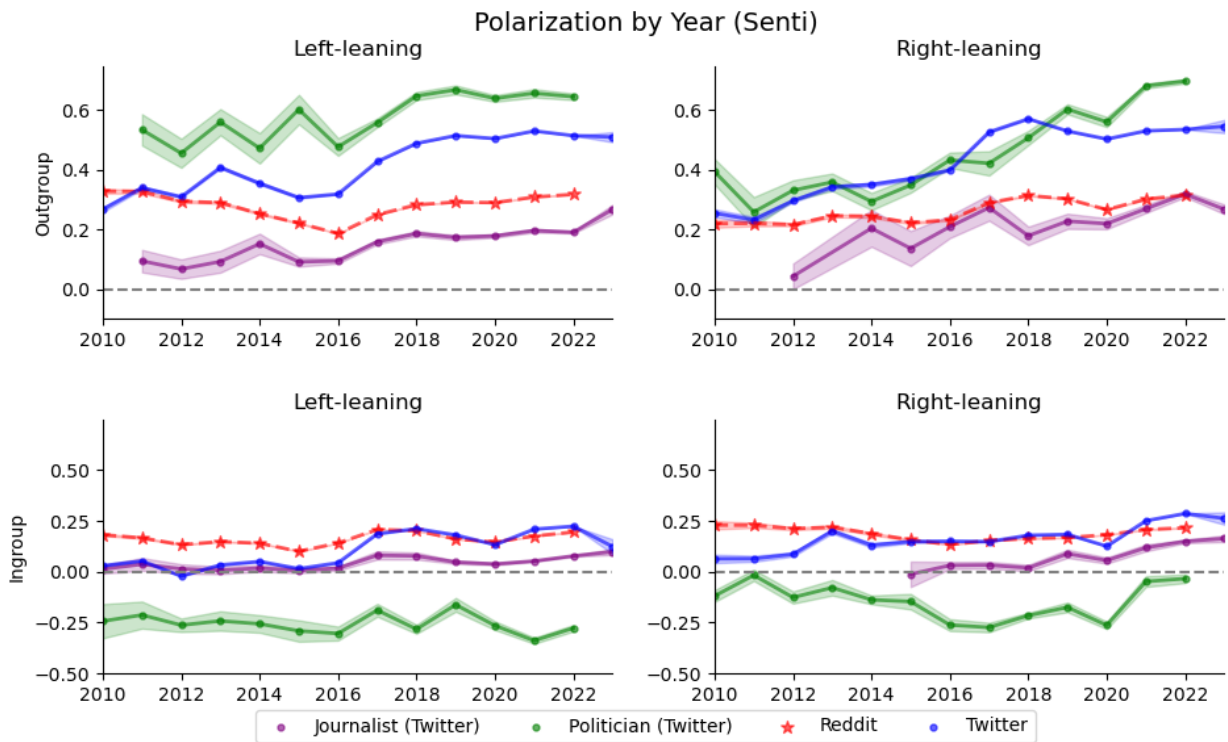

*Figure S9: Third run dropping 10% of political keywords and re-creating our main visualization*

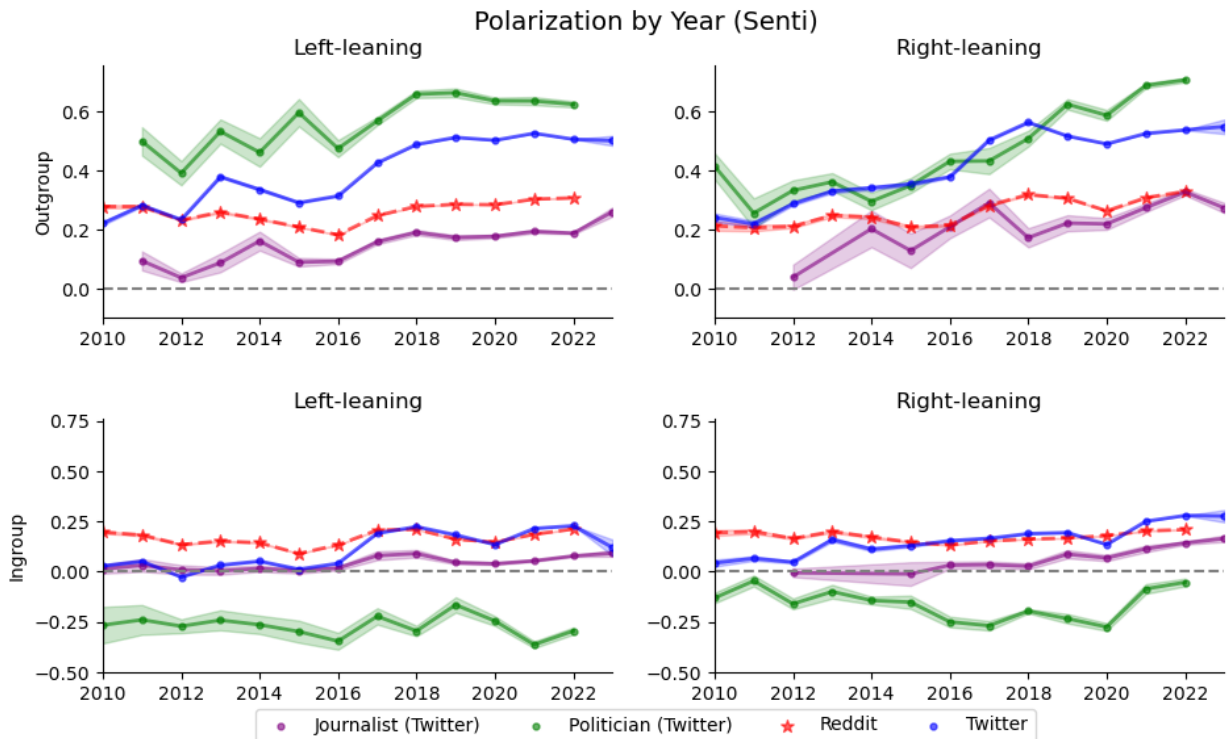

*Figure S10: Fourth run dropping 10% of political keywords and re-creating our main visualization*

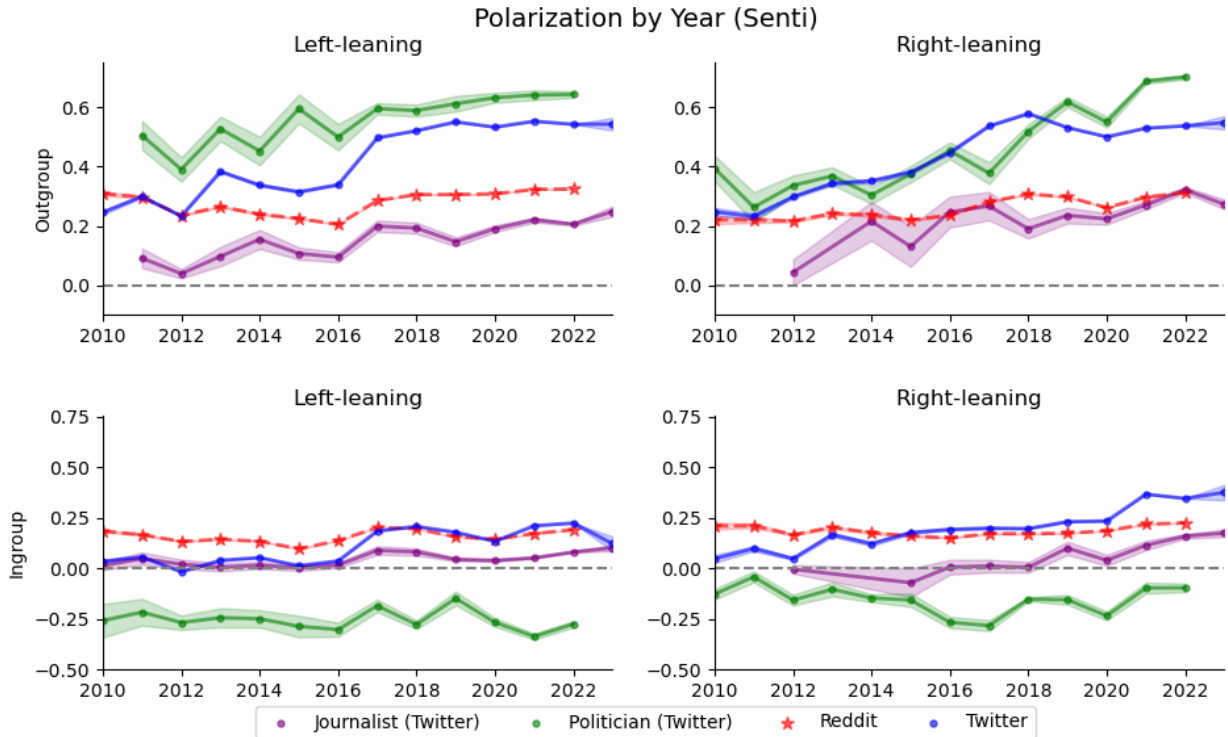

*Figure S11: Fifth run dropping 10% of political keywords and re-creating our main visualization*

## S4.3 Validation with External Metrics

### S4.3.1 DW-Nominate versus Politician Polarization

To validate our metric of polarizing discourse against existing measures, we first follow Ballard et al. (9) and Yu et al. (35), who hypothesize (and show) that ideological extremity is associated with polarizing tweeting behavior amongst U.S. politicians. Ballard et al. (9) claim that "ideological polarization and partisan animosity are distinct concepts (25), but related". To test this relationship, we follow their approach and regress the outgroup animosity of a politician against their DW-Nominate score, a measure of political extremity based on the co-sponsorship of bills. We also find that polarizing speech tends to increase with ideological extremity, confirming our metric follows expected trends and previous findings in the literature.

|  | BERT Left | BERT Right | GPT-2 Left | GPT-2 Right | Senti Left | Senti Right |
|---|---|---|---|---|---|---|
| DW-Nominate | 0.347*** | 0.175** | 0.270*** | 0.093+ | 0.480*** | 0.278*** |
|  | (0.001) | (0.007) | (0.001) | (0.073) | (0.001) | (0.001) |
| Num.Obs. | 371 | 434 | 371 | 434 | 371 | 434 |
| R2 | 0.041 | 0.017 | 0.047 | 0.007 | 0.077 | 0.028 |
| + p < 0.1, * p < 0.05, ** p < 0.01, *** p < 0.001 | | | | | | |

*Table S18: We regress the mean outgroup affect of politicians versus their DW-Nominate score (a measure of ideological extremity based on voting patterns) across our three fine-tuned models. As expected, more ideologically extreme politicians exhibit more outgroup derogation.*

### S4.3.2 Comparison to Following Patterns

We initially used the approach by Barbera (21) based on Twitter following patterns to infer the political orientation of users. In their work, they validated their measure against various external measures (e.g. the DW-NOMINATE score as well as the political orientation of users matched to their voting records) and found that their measure predicted these external metrics well. We test whether our measure of polarizing discourse (based on the content of tweets) is positively correlated to a metric of partisanship based on structural patterns (i.e. who the users follow). We indeed find that the measures are positively related; users that are inferred to be more partisan based on who they follow are also more likely to exhibit outgroup derogation.

|          | BERT                | GPT-2               | Senti               |
|----------|---------------------|---------------------|---------------------|
| theta    | 0.056***<br>(0.001) | 0.050***<br>(0.001) | 0.106***<br>(0.001) |
| Num.Obs. | 349762              | 349762              | 349762              |
| + p < 0.1, * p < 0.05, ** p < 0.01, *** p < 0.001 | | | |

*Table S19: We regress the level of outgroup derogation for each right-leaning Twitter user based on their tweets against their inferred level of partisanship based on their following patterns of elite politicians. Twitter users that are inferred to be more partisan based on their following patterns are also more likely to use outgroup derogation.*

|          | BERT                | GPT-2               | Senti               |
|----------|---------------------|---------------------|---------------------|
| theta    | 0.052***<br>(0.001) | 0.023***<br>(0.001) | 0.073***<br>(0.001) |
| Num.Obs. | 1092238             | 1092238             | 1092238             |
| + p < 0.1, * p < 0.05, ** p < 0.01, *** p < 0.001 | | | |

*Table S20: We regress the level of outgroup derogation for each left-leaning Twitter user based on their tweets against their inferred level of partisanship based on their following patterns of elite politicians. Twitter users that are inferred to be more partisan based on their following patterns are also more likely to use outgroup derogation.*

### S4.2.3 State Voting Patterns versus Polarizing Discourse

In addition to examining the relationship between politicians' ideological extremity and polarizing rhetoric on Twitter, we also investigate, at the state level, whether tweets authored by users in more ideologically extreme states tend to exhibit higher levels of outgroup animosity. To achieve this, we geolocate approximately two hundred thousand users based on self-reported location information in their Twitter bios. Subsequently, we calculate the outgroup derogation for each state using this data. We then compare this state-level metric of outgroup animosity with a measure of partisan leaning by state provided by the site 538 (36). Their metric, derived from voting patterns by state, reflects the degree to which states lean further left or right than the national average, offering insights into the relationship between state-level ideological positioning and Twitter discourse polarization.

We note that there are several limitations with this comparison. First, ideological extremity and polarizing speech are distinct concepts and there may be differences in how they manifest. Additionally, social media users are not a representative sample of U.S. voters and there may also be a bias in types of users that self-report their location.

Using this measure of ideological extremity by state, we make two comparisons. First, regress the level of outgroup derogation on Twitter amongst the users in each state against the absolute value of the ideological extremity based on the voting patterns of the state. When including all states, these are positively related (i.e., more ideologically extreme states have higher levels of outgroup derogation). However, further inspection reveals that Washington, D.C. and Wyoming are outliers with high ideological polarization and high levels of polarizing discourse on Twitter, respectively. After removing these from the analysis, the relationship weakens considerably and is not statistically significant (see **Figure S12**).

We then investigate the relationship between the original measure of partisan leaning (without taking the absolute value) and polarizing discourse. The results indicate that tweets authored by users from right-leaning states tend to exhibit higher levels of outgroup derogation compared to those from left-leaning states, a trend that persists even after excluding Washington D.C. and Wyoming from the analysis (see **Figure S13).** While the exact reason for this may involve the idiosyncrasies of Twitter as a platform, it could also be related to other forms of asymmetric polarization (37).

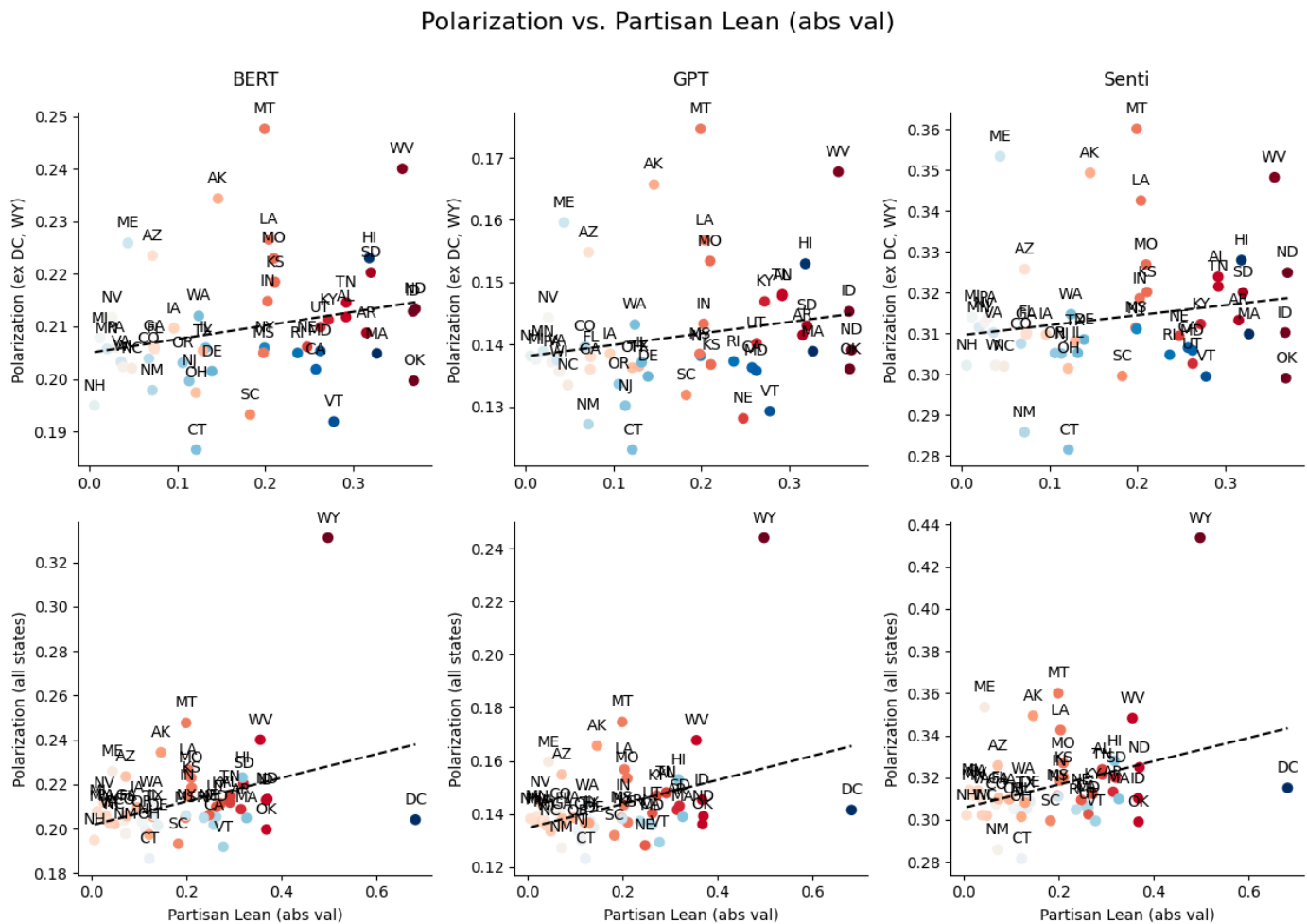

*Figure S12: We measure the average level of outgroup derogation of users located in each state in our sample. We then regress this against the absolute value of the ideological extremity of the state based on their voting patterns. We*

*find that the more ideologically extreme a state is, the more likely they are to use outgroup animosity. Since Wyoming and Washington, D.C. have extreme values that may skew the analysis, the top three charts show this relationship without these states.*

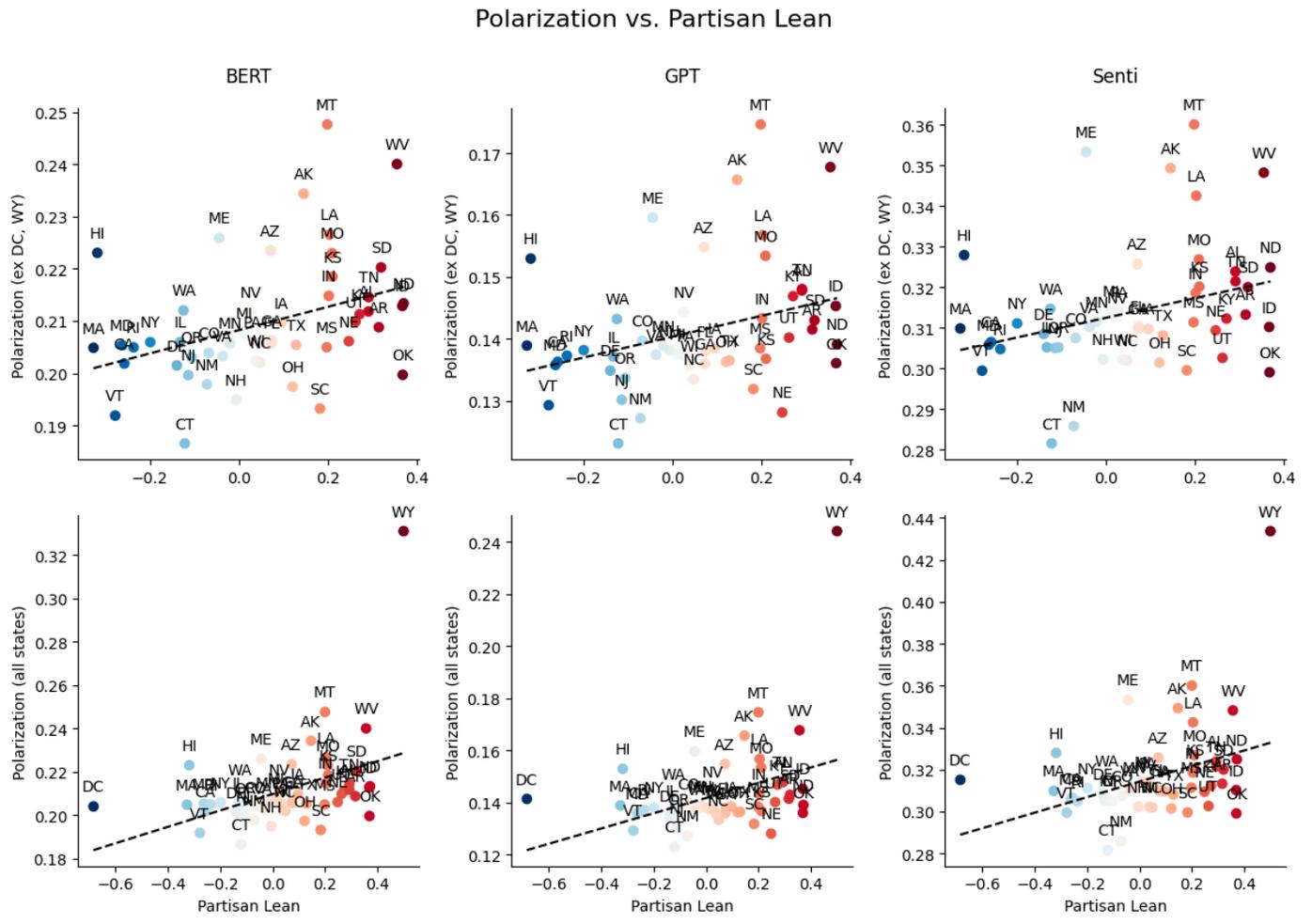

Figure S13: *We measure the average level of outgroup derogation of users located in each state in our sample. We then regress a measure of ideological extremity where a higher value indicates the state is more right-leaning and a lower value indicates the state is more left-leaning. We find that users in right-leaning states tend to use more outgroup derogation than left-leaning states.*

|  | BERT (ex WY, DC) | BERT (all) | GPT (ex WY, DC) | GPT (all) | Senti (ex WY, DC) | Senti (all) |
|---|---|---|---|---|---|---|
| Partisan (abs) | 0.026+ | 0.053* | 0.018 | 0.045* | 0.024 | 0.055* |
|  | (0.091) | (0.011) | (0.175) | (0.011) | (0.232) | (0.017) |
| Num.Obs. | 49 | 51 | 49 | 51 | 49 | 51 |
| R2 | 0.060 | 0.125 | 0.039 | 0.124 | 0.030 | 0.111 |

+ p < 0.1, * p < 0.05, ** p < 0.01, *** p < 0.001

Table S21: *We regress the level of outgroup derogation by state against the absolute value of that state's ideological extremity and find that more ideologically extreme states have users with higher levels of outgroup derogation.*

|  | BERT (ex WY, DC) | BERT (all) | GPT (ex WY, DC) | GPT (all) | Senti (ex WY, DC) | Senti (all) |
|---|---|---|---|---|---|---|
| Partisan | 0.022** | 0.038** | 0.017* | 0.029** | 0.024* | 0.037** |
|  | (0.007) | (0.002) | (0.023) | (0.005) | (0.029) | (0.006) |
| Num.Obs. | 49 | 51 | 49 | 51 | 49 | 51 |
| R2 | 0.142 | 0.180 | 0.106 | 0.147 | 0.097 | 0.143 |

+ p < 0.1, * p < 0.05, ** p < 0.01, *** p < 0.001

*Table S22: We regress the level of outgroup derogation by state against that state's ideological extremity, where higher values are more right-leaning and lower values are more left-leaning. We find that right-leaning states exhibit more outgroup derogation.*

## S5 Twitter Ordinal Logistic Regressions

### S5.1 Group Trends

We compare how different groups have evolved in their use of polarizing discourse by using an ordinal logistic regression where we regress the affect of a comment on an indicator variable corresponding to the period the comment was posted, as well as an interaction between the period and the type of user.

$$logit(Pr(Y \le j)) = \theta_j - (\beta_1 \cdot Period + \beta_2 \cdot UserType + \beta_3 \cdot (Period \times UserType))$$

We focus on $\beta_3$, which indicates whether a certain user type (politician, journalist, or partisan user) has a higher likelihood of using polarizing rhetoric in a given period of time. We run separate models corresponding to the political orientation of the users as well as the reference group (ingroup or outgroup). The following regressions are all performed using the Historical Powertrack data, as it has the least probability of a temporal bias, and we also control for time using a covariate representing the year when necessary. These results confirm our finding that right-leaning politicians used higher levels of outgroup derogation than other groups in recent years.

| | variable | BERT | Senti | GPT-2 | Joint |
|---|---|---|---|---|---|
| 1 | Journalist x Biden | -0.82*** | -0.98*** | -0.95*** | -1.29*** |
| 2 | Media x Biden | -0.96*** | -1.2*** | -1.55*** | -2.01*** |
| 3 | Politician x Biden | 0.3*** | 0.4*** | 0.23** | 0.49*** |
| 4 | Journalist x Trump | -1.13*** | -1.05*** | -1.31*** | -1.92*** |
| 5 | Media x Trump | -1.27*** | -1.31*** | -1.79*** | -2.2*** |
| 6 | Politician x Trump | -0.15 | 0.03 | -0.18 | -0.12 |
| 7 | Journalist x Obama (2nd) | -0.98*** | -0.91*** | -1.18*** | -1.6*** |
| 8 | Media x Obama (2nd) | -1.68*** | -1.41*** | -2.07*** | -2.86*** |
| 9 | Politician x Obama (2nd) | -0.28** | -0.27** | -0.29** | -0.42** |
| 10 | Journalist x Obama (1st) | -0.67*** | -0.69*** | -0.73*** | -1.09*** |
| 11 | Media x Obama (1st) | -1.74*** | -1.05*** | -1.86*** | -2.38*** |
| 12 | Politician x Obama (1st) | -0.74*** | -0.36** | -0.67*** | -0.87*** |
| 13 | -1\|0 | -6.94*** | -5.22*** | -7.22*** | -8.42*** |
| 14 | 0\|1 | 1.4*** | 0.41*** | 1.67*** | 1.62*** |
| 15 | Obama (1st) | -0.41*** | -0.5*** | -0.38*** | -0.66*** |
| 16 | Obama (2nd) | 0.03 | -0.25*** | 0.06* | -0.08* |
| 17 | Trump | -0.06** | -0.14*** | 0.01 | -0.08** |

*Table S23: For comments where right-leaning Twitter users are discussing their outgroup, we regress the affect of each tweet on the type of user, the presidential term in which the tweet was made, and an interaction between the two. We highlight the coefficients of politicians in the red boxes. In the Obama era, right-leaning politicians were generally used less likely to use outgroup derogation than the average right-leaning user (indicated by the negative coefficient). Then, in the Trump era they were roughly equal, indicated by non-significant coefficients. Finally, in the Biden era, politicians are more likely to use outgroup derogation than partisan users as indicated by the positive coefficient.*

| | variable | BERT | Senti | GPT-2 | Joint |
|---|---|---|---|---|---|
| 1 | Journalist x Biden | -0.9*** | -0.94*** | -1.17*** | -1.69*** |
| 2 | Media x Biden | -0.99*** | -1.25*** | -1.57*** | -2.1*** |
| 3 | Politician x Biden | 0.33*** | 0.37*** | 0.19* | 0.41** |
| 4 | Journalist x Trump | -0.92*** | -0.94*** | -1.13*** | -1.5*** |
| 5 | Media x Trump | -0.9*** | -1.04*** | -1.67*** | -2.1*** |
| 6 | Politician x Trump | 0.53*** | 0.42*** | 0.31*** | 0.57*** |
| 7 | Journalist x Obama (2nd) | -1.19*** | -1.26*** | -1.44*** | -1.88*** |
| 8 | Media x Obama (2nd) | -1.09*** | -1.07*** | -1.57*** | -2.07*** |
| 9 | Politician x Obama (2nd) | 0.26** | 0.35*** | 0.2* | 0.42*** |
| 10 | Journalist x Obama (1st) | -1.28*** | -1.13*** | -1.32*** | -1.85*** |
| 11 | Media x Obama (1st) | -1.52*** | -1.3*** | -1.85*** | -2.26*** |
| 12 | Politician x Obama (1st) | 0.08 | 0.27* | 0.26 | 0.1 |
| 13 | -1\|0 | -6.83*** | -5.29*** | -6.82*** | -8.25*** |
| 14 | 0\|1 | 1.59*** | 0.52*** | 1.87*** | 1.85*** |
| 15 | Obama (1st) | -0.35*** | -0.46*** | -0.26*** | -0.49*** |
| 16 | Obama (2nd) | -0.17*** | -0.28*** | -0.18*** | -0.32*** |
| 17 | Trump | -0.16*** | -0.18*** | -0.07*** | -0.18*** |

*Table S24: For comments where left-leaning Twitter users are discussing their outgroup, we regress the affect of each tweet on the type of user, the presidential term in which the tweet was made, and an interaction between the two. We highlight the coefficients of politicians in the red boxes. The positive coefficients on politicians indicates that left-leaning politicians have consistently used more outgroup derogation than partisan left-leaning users. However, the largest gap seems to appear during the Trump era, and the difference was not significant in the first Obama term*

We then focus on confirming our findings that right-leaning politicians have become more negative towards their outgroups at a faster rate than other groups. We regress the affect of a comment on an interaction between the year (represented as the number of years from the earliest comment) and the user type. A higher coefficient essentially corresponds to a higher rate of growth in outgroup animosity of the group relative to other groups. On the right, outgroup animosity by politicians also grew faster than the other right-leaning groups.

$$logit(\Pr(Y \leq j)) = \theta_j - (\beta_1 \cdot Year + \beta_2 \cdot UserType + \beta_3 \cdot (Year \times UserType))$$

|   | variable | BERT | Senti | GPT-2 | Joint |
|---|----------|------|-------|-------|-------|
| 1 | Politician x year | 0.11*** | 0.09*** | 0.1*** | 0.15*** |
| 2 | Journalist x year | 0.01 | -0.02* | 0.01 | 0.01 |
| 3 | Media x year | 0.11*** | 0.01 | 0.06** | 0.09** |
| 4 | -1\|0 | -6.89*** | -4.79*** | -7.22*** | -8.17*** |
| 5 | 0\|1 | 1.44*** | 0.83*** | 1.66*** | 1.84*** |
| 6 | Journalist | -1.08*** | -0.82*** | -1.19*** | -1.65*** |
| 7 | Media | -2.28*** | -1.4*** | -2.36*** | -3.15*** |
| 8 | Politician | -1.11*** | -0.75*** | -1*** | -1.45*** |
| 9 | year | 0 | 0.03*** | 0 | 0.01** |

*Table S25: For right-leaning partisan users, we regress the affect of the comment on the type of user, year, and interaction between the year and type of user. A positive coefficient indicates that the rate of growth in outgroup derogation of right-leaning politicians and media exceeded that of partisan users. However, we note that the overall level of outgroup derogation of media is low relative to all other groups.*

|   | variable | BERT | Senti | GPT-2 | Joint |
|---|----------|------|-------|-------|-------|
| 1 | Politician x year | 0.02 | 0.01 | 0 | 0.02 |
| 2 | Journalist x year | 0.06*** | 0.04*** | 0.05*** | 0.06*** |
| 3 | Media x year | 0.05** | 0.01 | 0.01 | 0.01 |
| 4 | -1\|0 | -6.52*** | -4.77*** | -6.57*** | -7.73*** |
| 5 | 0\|1 | 1.9*** | 1.03*** | 2.12*** | 2.37*** |
| 6 | Journalist | -1.5*** | -1.42*** | -1.67*** | -2.18*** |
| 7 | Media | -1.38*** | -1.13*** | -1.73*** | -2.16*** |
| 8 | Politician | 0.24* | 0.28** | 0.23 | 0.33* |
| 9 | year | 0.02*** | 0.04*** | 0.02*** | 0.04*** |

*Table S26: For left-leaning partisan users, we regress the affect of the comment on the type of user, year, and interaction between the year and type of user. We see that left-leaning journalists increased their polarizing rhetoric faster than any group, albeit at lower levels overall.*

## S6 Distribution of Polarizing Discourse across Users

In addition to examining aggregate measures of polarizing discourse, we also analyze how users contribute differently to negative and neutral affect political content. As established in prior research, user activity on platforms like Reddit and Twitter tends to follow a power-law distribution—a small subset of users generates a disproportionate share of the content. However, we observe that these distributions vary depending on the affective tone of the content.

In **Figure S14**, we present a Lorenz curve to visualize these differences. The x-axis represents user deciles, ranked by the percentage of political comments or tweets they post, while the y-axis shows the cumulative percentage of total content contributed by users in each decile or higher. To ensure uniform group sizes, we randomly broke ties so that each decile contains the same number of users.

Our findings show that on Twitter, the top 10% of users by volume of neutral affect political content contribute 39% of all such comments for left-leaning users, and 35% for right-leaning users. In contrast, when it comes to negative affect political content, the top decile of users contribute 46% (left-leaning) and 40% (right-leaning) of all posts. These results suggest that a more concentrated group of users is responsible for spreading negative affect content compared to neutral content.

To confirm that the distributions are indeed statistically significantly different, we discretize the contributions into percentiles and run the one-sided Kolmogorov–Smirnov test that determines whether the CDF of negative comments is higher than that of neutral comments. For all users except left-leaning politicians and media on Twitter, the distribution of negative comments is indeed more skewed than neutral comments ($p < .05$).

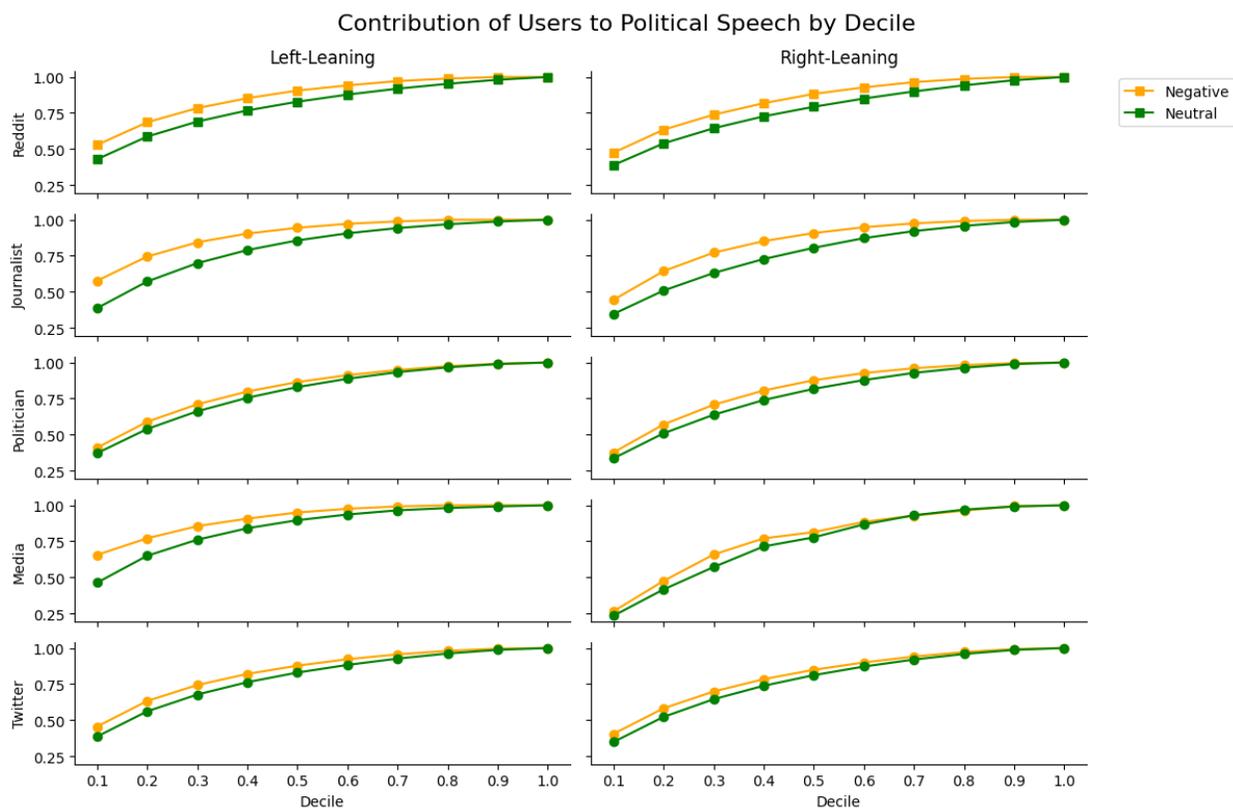

*Figure S14: We plot the Lorenz curves of neutral and negative affect political content. For each decile of users, ranked by the amount of content posted, we compute the percentage of neutral/negative content contributed by that decile of users or higher. If one curve is higher than other, this indicates a more unequal distribution, with the top deciles contributing more. In this case, we see that the top tweeters of negative political content contribute more to the overall amount of negative affect comments than the top tweeters of neutral political content. This suggests negative political content is more concentrated in a smaller set of users.*

| Type | Pol Orient | K-S Statistic | p-value |
| --- | --- | --- | --- |
| Journalist | L | 0.31 | 0 |
| Reddit | L | 0.29 | 0 |
| Reddit | R | 0.29 | 0 |
| Journalist | R | 0.24 | 0.003 |
| Politician | R | 0.19 | 0.027 |
| Twitter | L | 0.17 | 0.056 |
| Twitter | R | 0.17 | 0.056 |
| Media | L | 0.09 | 0.446 |
| Politician | L | 0.08 | 0.529 |
| Media | R | 0.04 | 0.853 |

*Table S33: We report the K-S statistic and corresponding p-value for each group and political orientation, which compares whether the difference between the CDFs of negative versus neutral affect political content is statistically significantly different. For all groups except the media and left-leaning politicians, we find the differences are significant. We discretized the distribution into percentiles to compare CDFs.*

We also plot the same Lorenz curves as above but now compare the two platforms. On the top chart we examine the distribution of neutral affect political content across users on the two platforms, while on the bottom we examine the distribution of negative affect political content. The red curve (Reddit) above the blue curve (Twitter) indicates that the distribution is more skewed on Reddit than on Twitter.

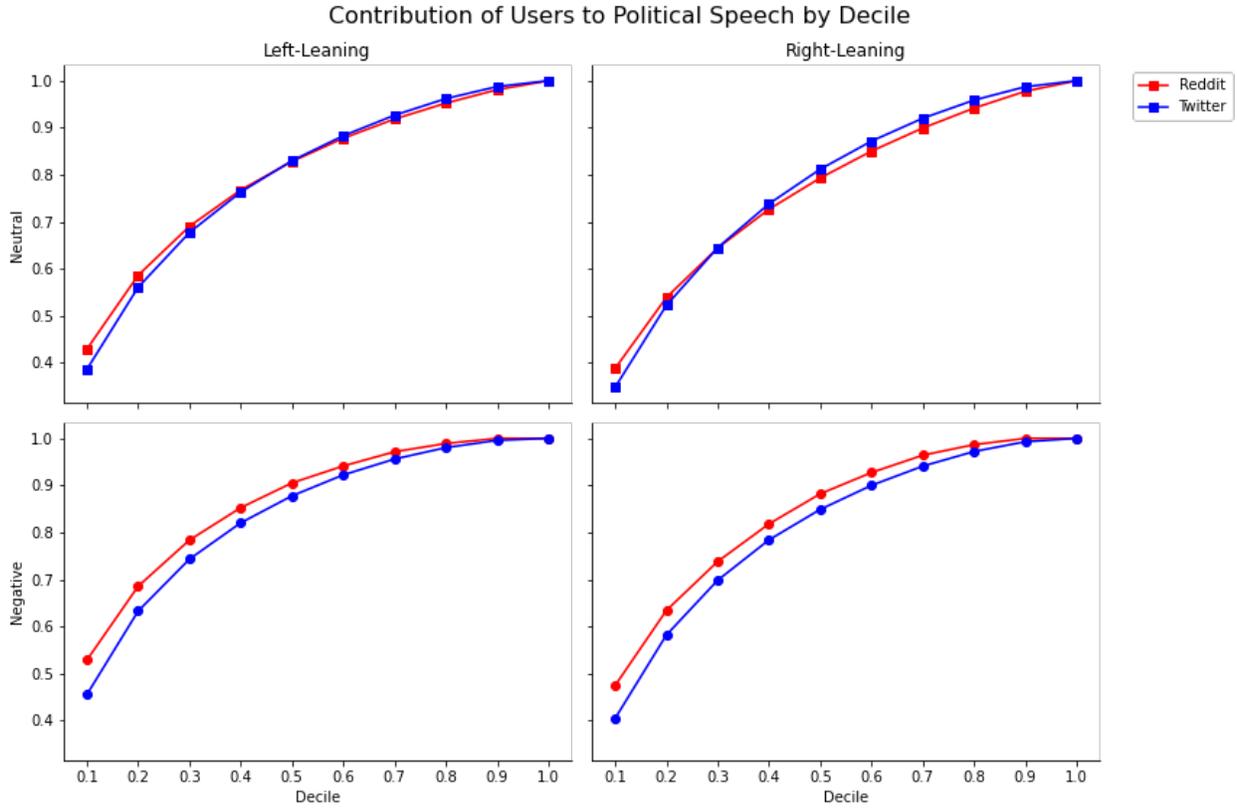

*Figure S15: We plot the Lorenz curves of neutral and negative affect political content on Reddit versus Twitter. For each decile of users, ranked by the amount of content posted, we compute the percentage of neutral/negative content contributed by that decile of users or higher. If one curve is higher than other, this indicates a more unequal distribution, with the top deciles contributing more. We see that negative content on Reddit is more concentrated in the top decile of users than on Twitter.*

## S7 Robustness Checks for Data Source and Base Model

### S7.1 Historical Powertrack Groups/Platforms

Using data from the Historical Powertrack API (a different API to source data from Twitter as opposed to pulling from users' timelines) yields the same results as using timeline data. We show below a version of our main figure recreated with this data.

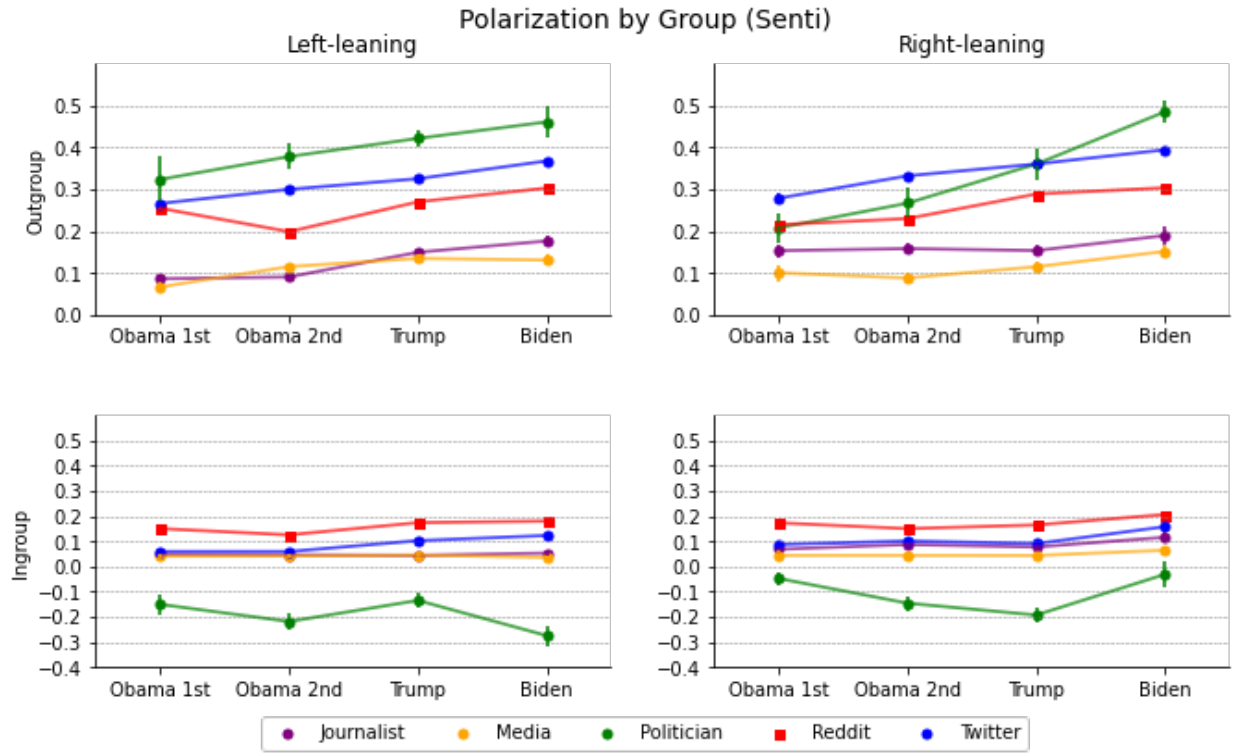

*Figure S22: Using only our sample of Historical Powertrack data, we plot changes in polarizing discourse over time across groups and find that the relative patterns are consistent, with rising outgroup animosity and recent acceleration amongst right-leaning politicians.*

### S7.2 Recreating Plots with Different Base Classifier Models

We re-create all our main charts using different base models to ensure robustness to the underlying model used. We see that the high-level results stay consistent; outgroup animosity is rising for all groups and right-leaning politicians have accelerated the most since 2016. We see that levels of outgroup derogation of left-leaning politicians peaks in 2019 and then declines, although our most accurate model (the fine-tuned sentiment model) shows less of a decline than fine-tuned base BERT and GPT models. When investigating this difference, we find that the higher recall of the fine-tuned sentiment model makes it more likely to find speech that is polarizing by does not include explicit personal attacks. Thus, the greater decline we see for the fine-tuned BERT/GPT models could be due to the shifting nature of the types of polarizing speech that left-leaning politicians are using, although in all cases we find that while polarizing speech by right-leaning politicians has been increasing since around 2016 and is at its peak today, the peak for left-leaning politicians occurred in 2019.

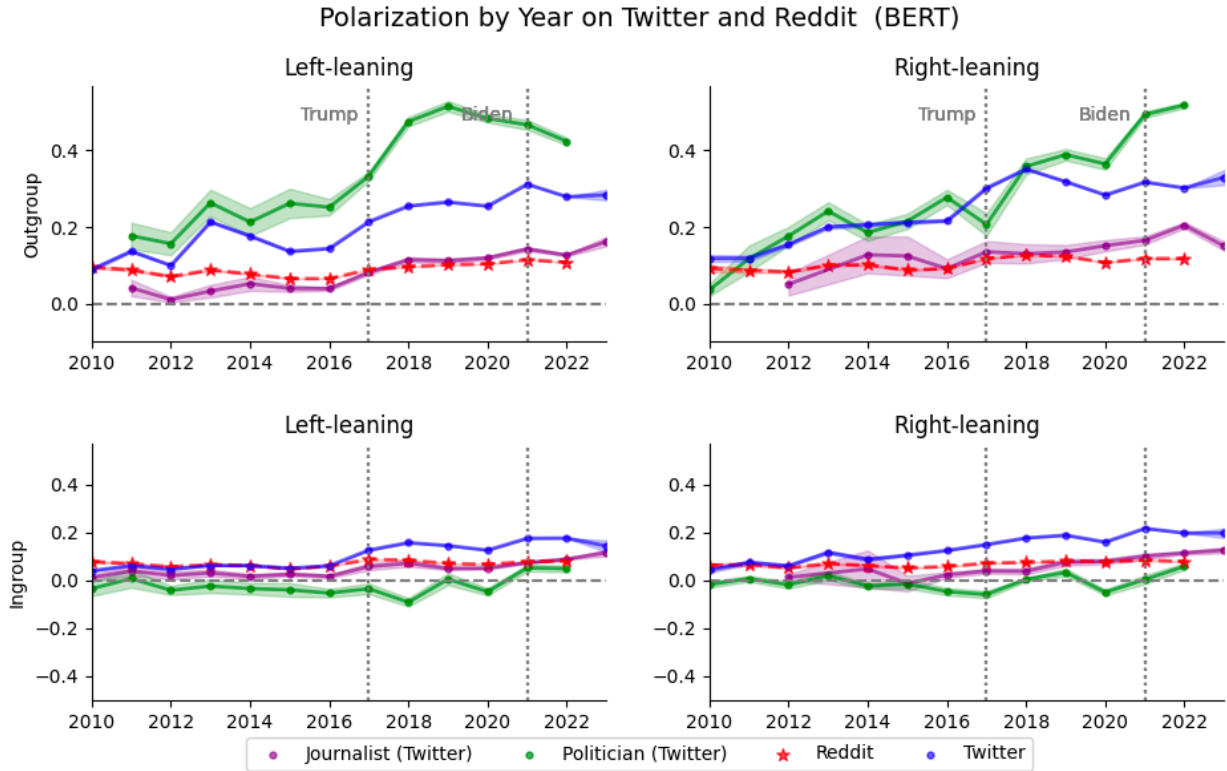

*Figure S16: Recreating our main visualization of polarizing discourse over time using BERT as the base model*

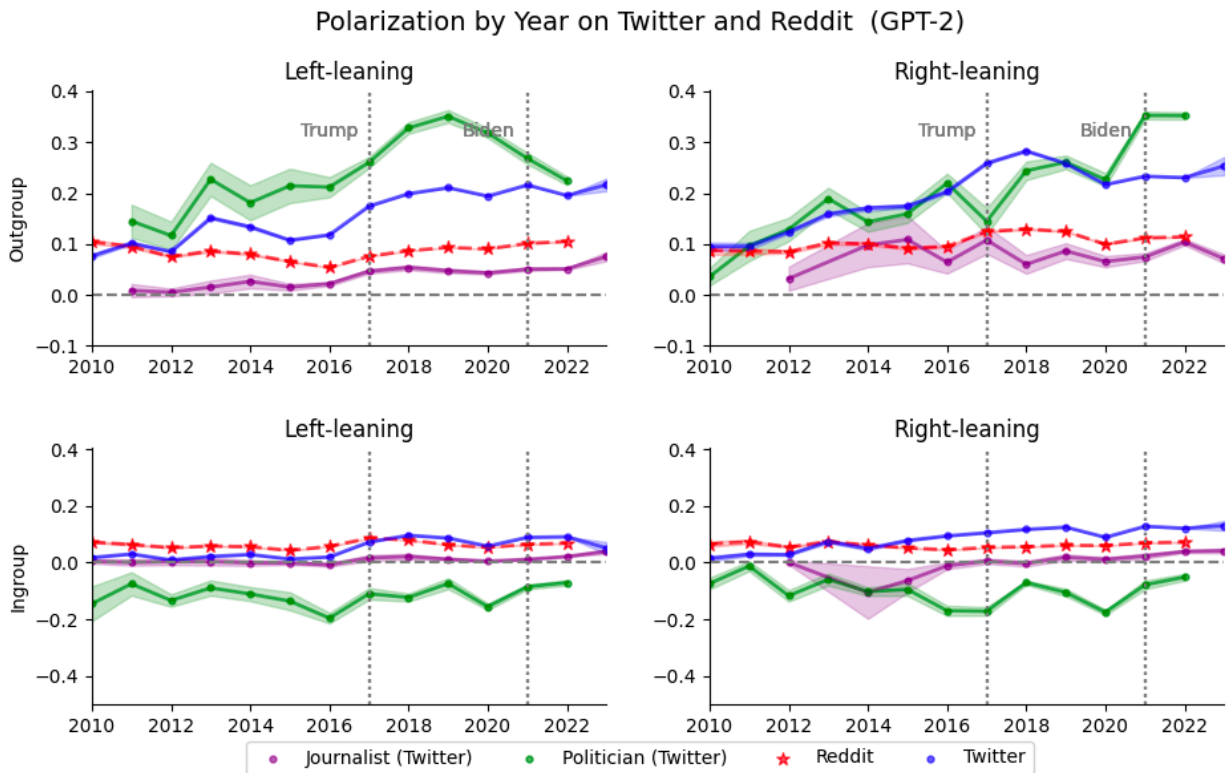

*Figure S17: Recreating our main visualization of polarizing discourse over time using GPT-2 as the base model*

## S7.3 Subreddit and Cohort Contributions

We recreate our subreddit contribution charts with BERT and GPT-2 as our base models.

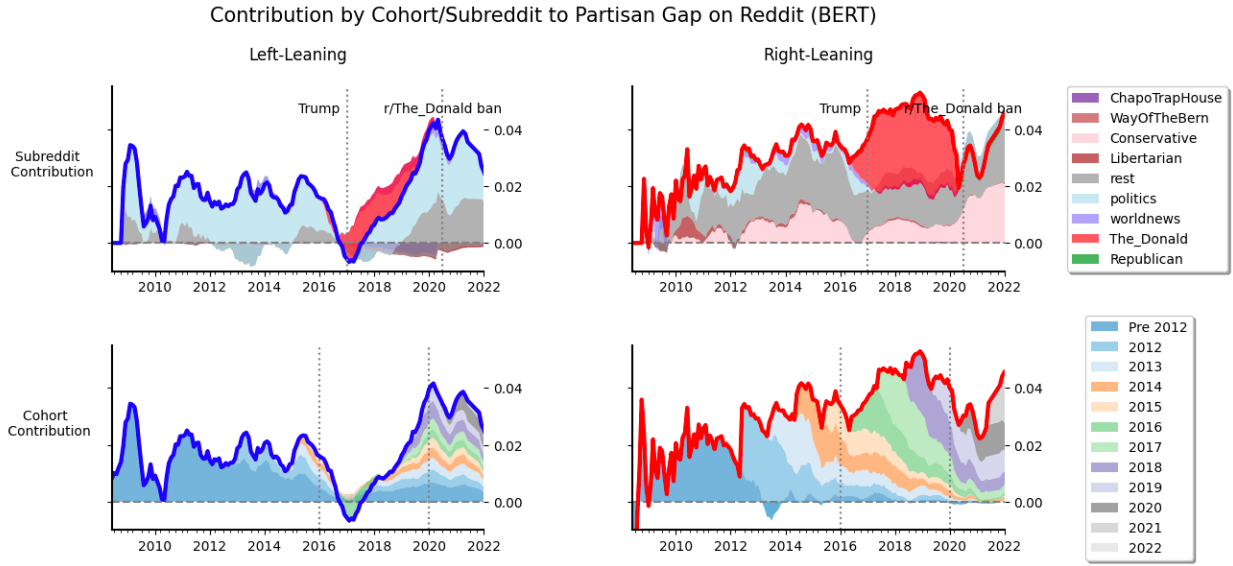

*Figure S18: Recreating our subreddit contribution chart with BERT as the base model*

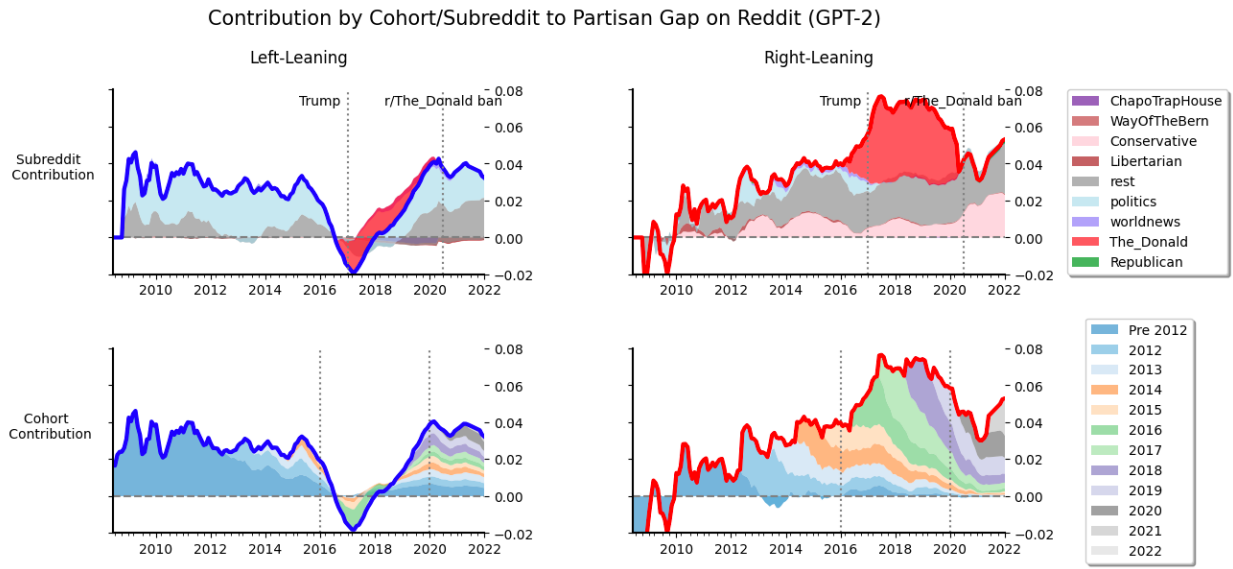

*Figure S19: Recreating our subreddit contribution chart with GPT-2 as the base model*

## S7.5 Influencer Polarizing Discourse

We recreate our charts breaking down polarizing discourse by influencers when using BERT and GPT-2 as our base models.

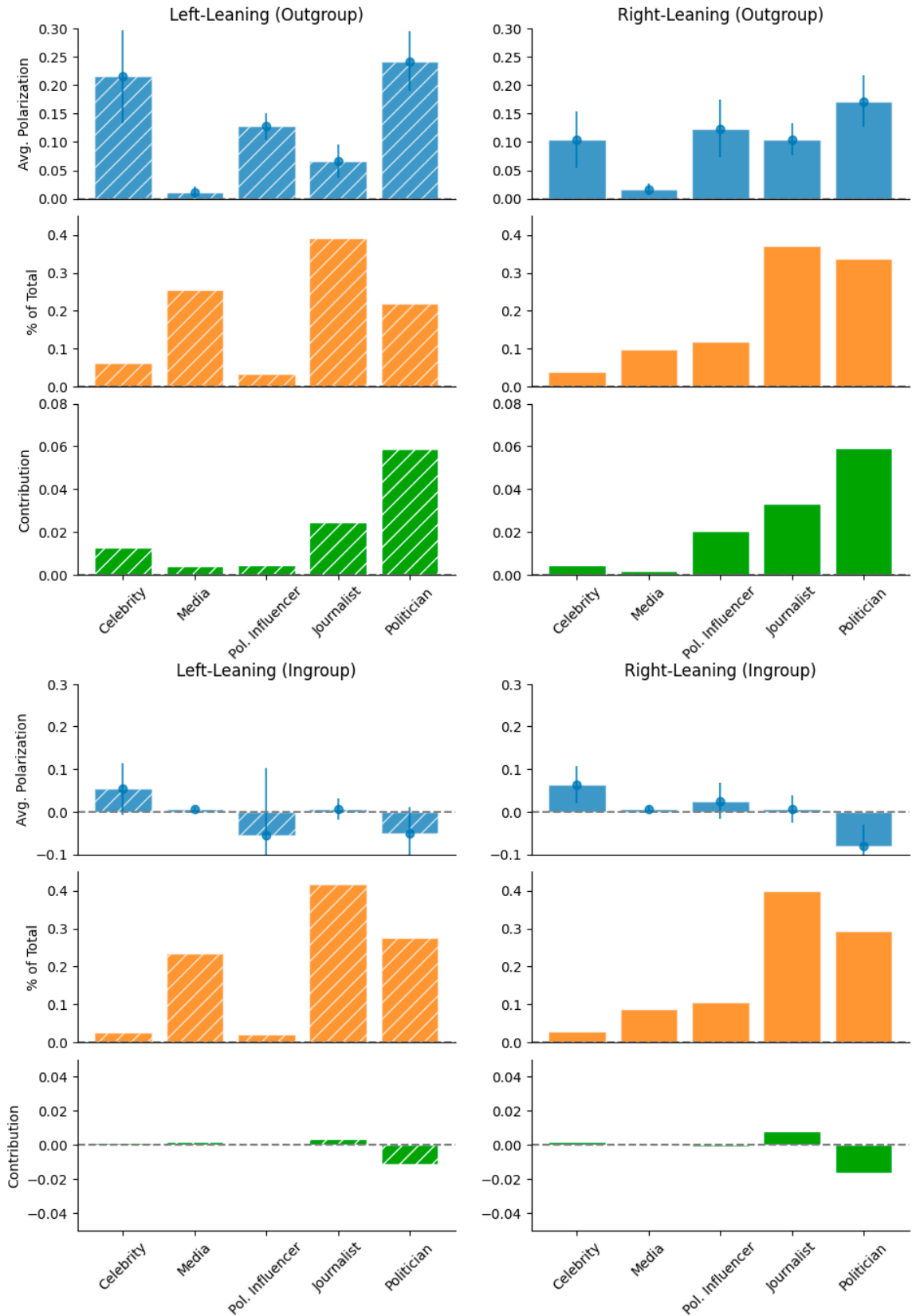

Polarization of Top 200 most Followed Accounts on Twitter (GPT-2)

*Figure S20: Recreating our chart of influencer polarization using GPT-2 as the base model*

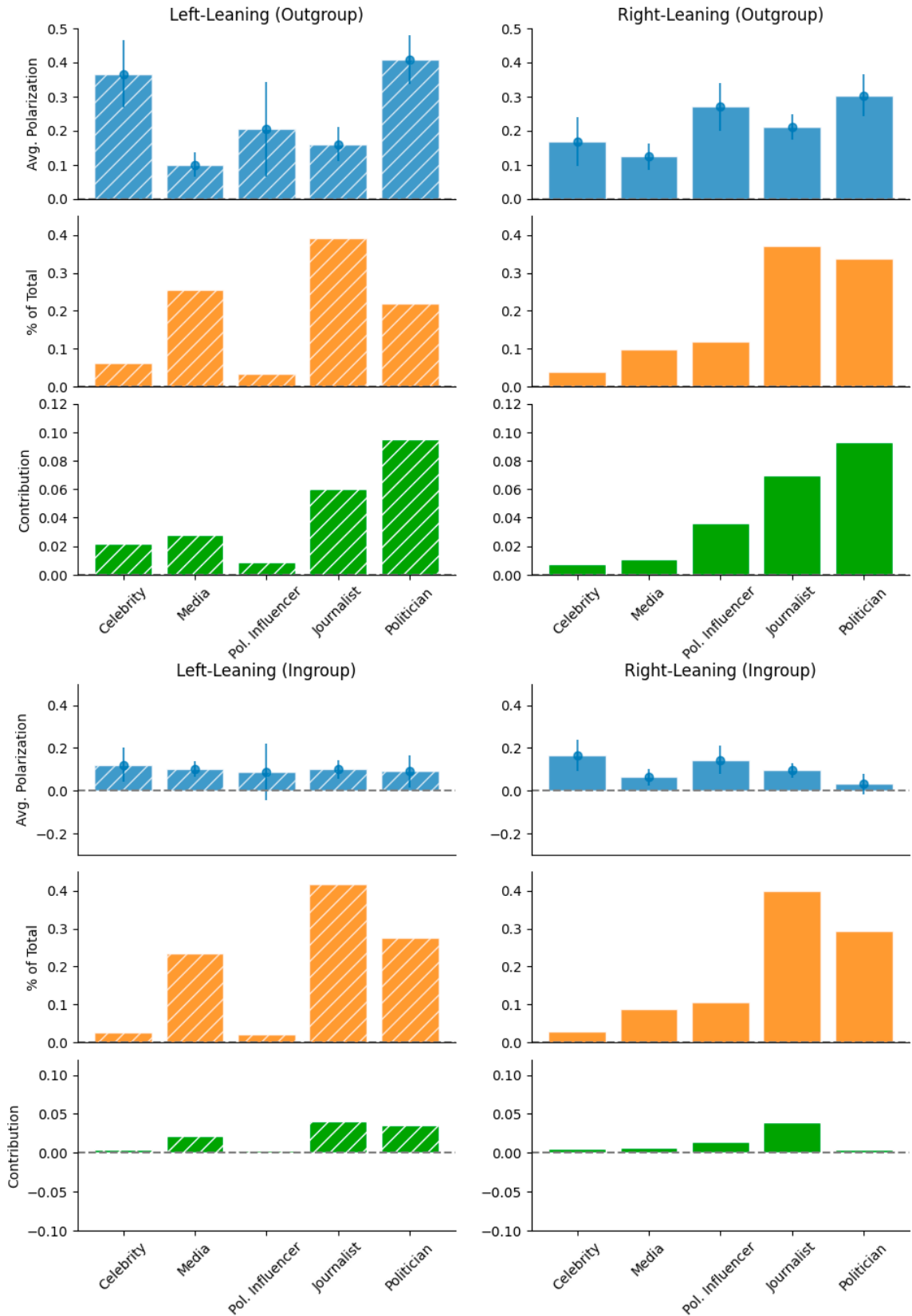

Polarization of Top 200 most Followed Accounts on Twitter (BERT)

*Figure S21: Recreating our chart of influencer polarization using BERT as the base model*

S6.6 Followers vs. Polarizing Rhetoric

Figures **Figure S22** to **Figure S24** illustrate the shifts in polarizing discourse across varying follower counts. To compute the number of followers, we take the latest number of followers that an account has, as this can shift over time. Overall, we observe a consistent negative relationship for both left- and right-leaning users. Notably, accounts with over one million followers appear to show significantly higher outgroup animosity among right-leaning users. However, a closer look reveals that this spike is largely driven by a single influencer who repeatedly posts the same highly negative tweet about Joe Biden. This heavily skews the average. After removing this outlier, the polarization trends between left- and right-leaning users align more closely.

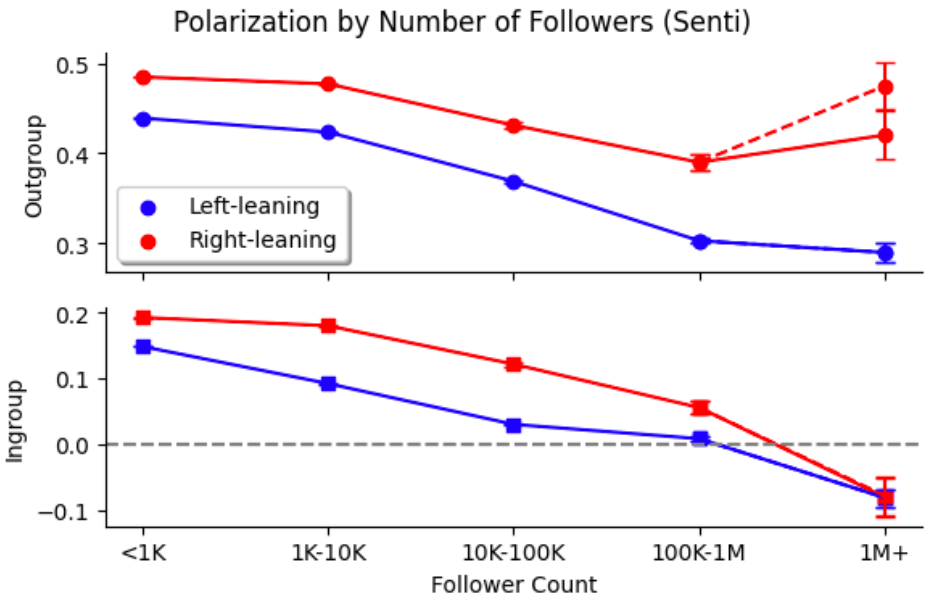

*Figure S22: The level of polarizing discourse across accounts with different amounts of followers using our main model*

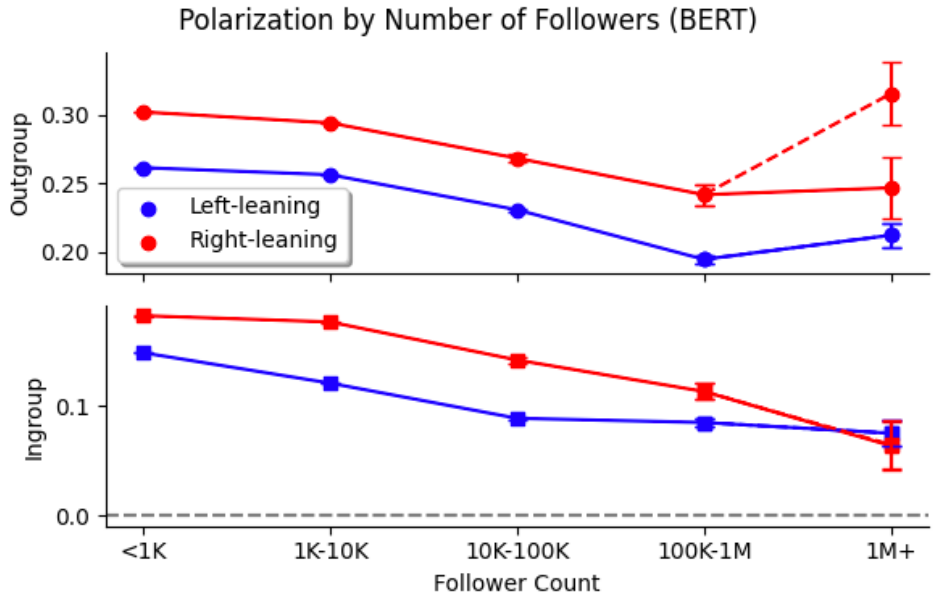

*Figure S23: The level of polarizing discourse across accounts with different amounts of followers using a BERT base model*

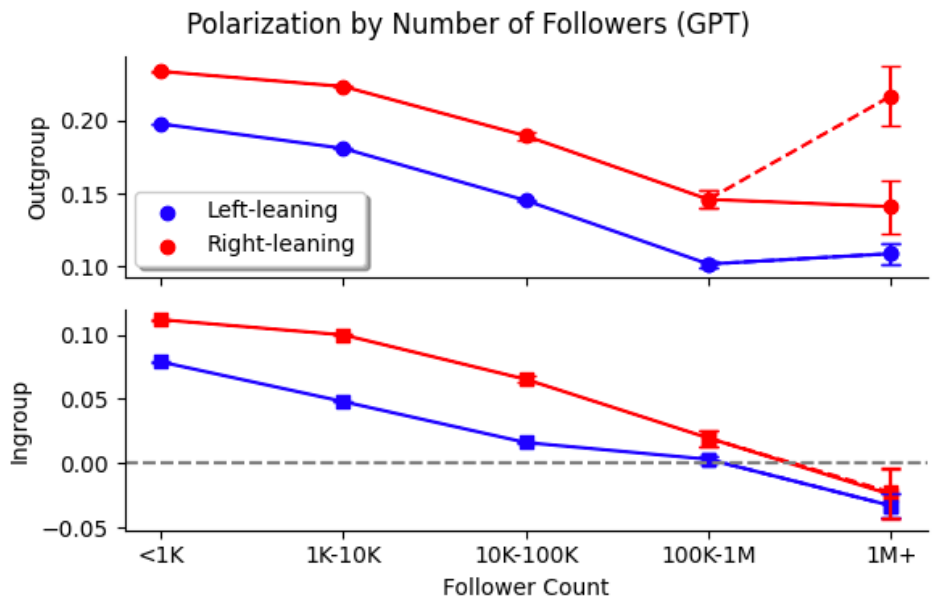

*Figure S24: The level of polarizing discourse across accounts with different amounts of followers using GPT-2 as the base model*

## S6.7 Topics

We re-create our charts showing polarizing discourse on different topics using BERT and GPT-2 as our base models.

# Topic Coefficients and Interactions on Twitter and Reddit (BERT)

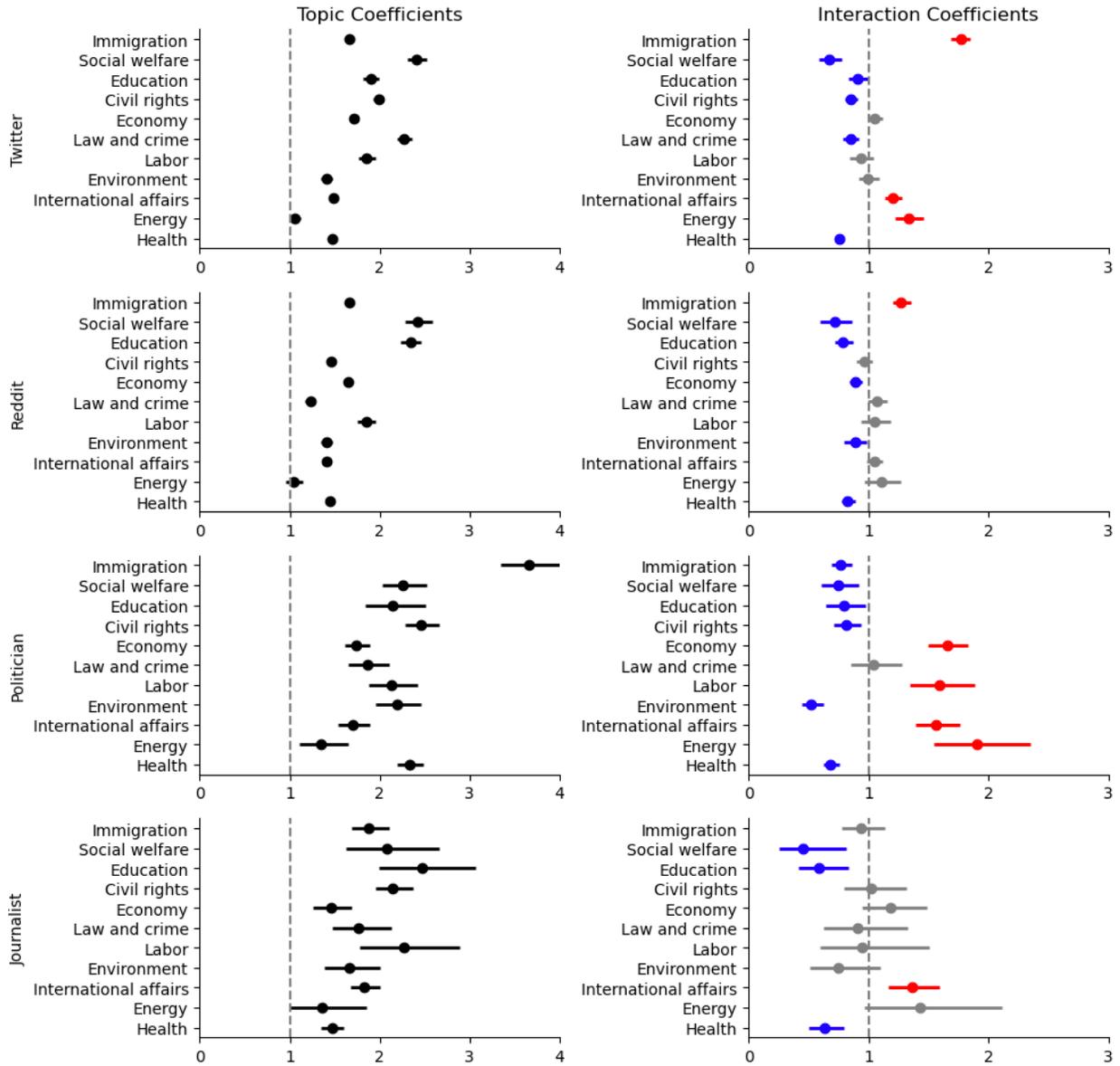

*Figure S25: Re-creating our topic chart with BERT as the base model*

# Topic Coefficients and Interactions on Twitter and Reddit (GPT-2)

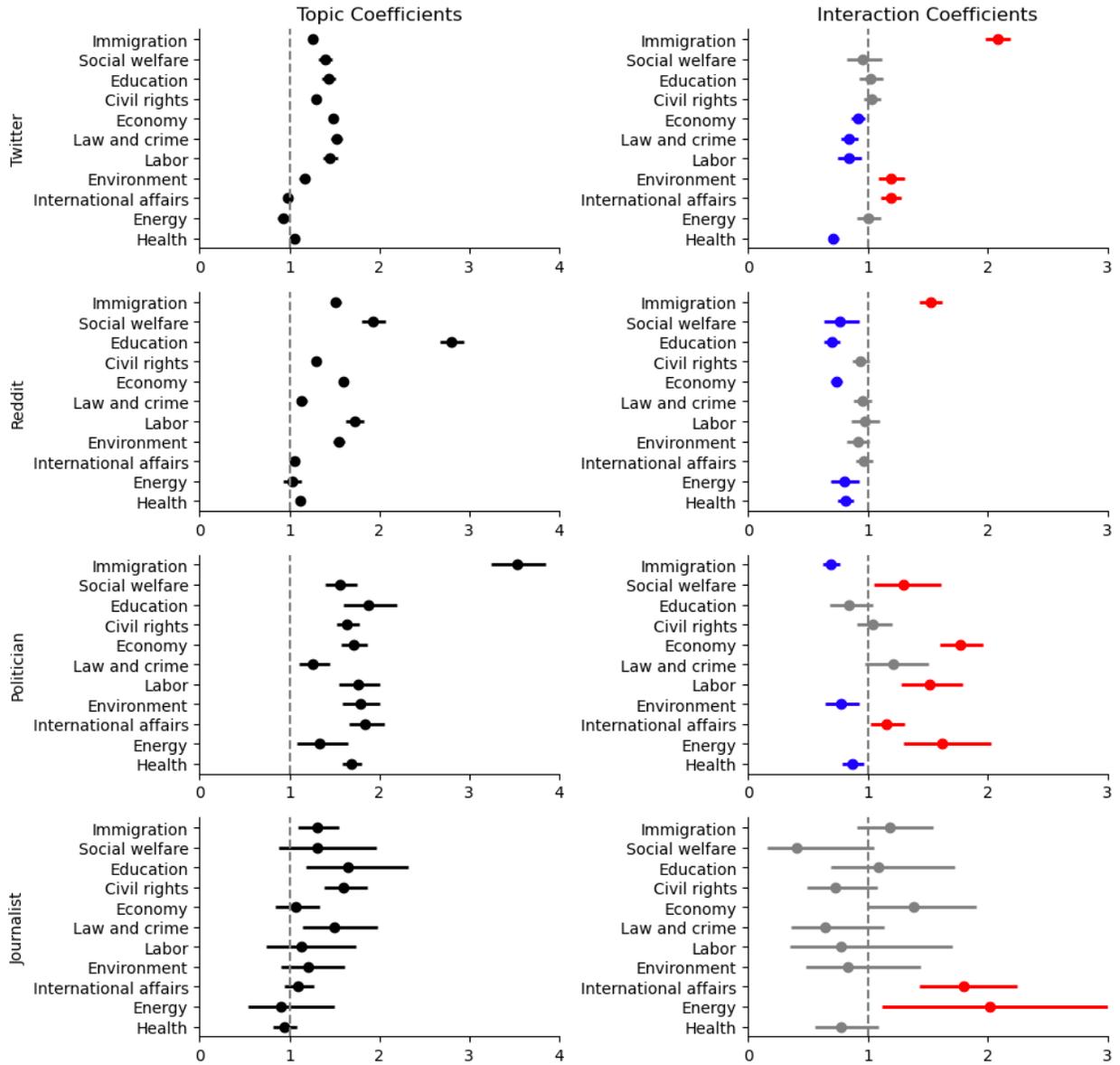

*Figure S26: Re-creating our topic chart with GPT-2 as the base model*